\documentclass[onecolumn,twoside]{IEEEtran}
\usepackage{mathpple}
\usepackage{times}
\usepackage{capt-of}

\usepackage{amsmath}  % Define \boldsymbol (in amsbsy too) and align
\usepackage{amssymb}  % Define \mathbb (in amsfonts too)
\usepackage{mathrsfs} % Define \mathscr, a script font
\usepackage{kantlipsum,cuted}
\usepackage{multirow}% http://ctan.org/pkg/multirow
\usepackage{tabu}

\usepackage{lipsum}

\usepackage{theorem}  % Helps in rendering theorems, etc.
\usepackage{cite}     % Gives multiple references as intervals
\usepackage{comment}  % Comments out with \begin{comment} ... \end{comment}

\usepackage{upref}
\usepackage{amsfonts}

\usepackage{verbatim}

\usepackage[dvipsnames,usenames]{color}
\usepackage{enumerate}
\usepackage{multicol}
\usepackage{blkarray, bigstrut}
\usepackage{graphicx}
\usepackage{subfigure}

\usepackage{latexsym}
\usepackage{hyperref}

\usepackage[all]{xy}

\usepackage{algorithmicx}
\usepackage{algorithm}
\usepackage[noend]{algpseudocode}
\algrenewcommand\alglinenumber[1]{\scriptsize #1:}
\algrenewcommand\algorithmicindent{1em}%

\usepackage{wrapfig}
\usepackage[shortlabels]{enumitem}

\makeatletter
\def\BState{\State\hskip-\ALG@thistlm}

\makeatother

\newcommand{\code}{\ttfamily\bfseries}

\newcommand{\be}[1]{\begin{equation}\label{#1}}
	\newcommand{\ee}{\end{equation}}

\newcommand{\bc}{\begin{center}}
	\newcommand{\ec}{\end{center}}

\newcommand{\ceil}[1]{\lceil{#1}\rceil}

\newcommand{\cA}{{\cal A}}

\newcommand{\cC}{{\cal C}}
\newcommand{\cD}{{\cal D}}

\newcommand{\cI}{{\cal I}}
\newcommand{\cJ}{{\cal J}}

\newcommand{\cL}{{\cal L}}

\newcommand{\cP}{{\cal P}}

\newcommand{\cR}{{\cal R}}
\newcommand{\cS}{{\cal S}}

\newcommand{\bfa}{{\boldsymbol a}}
\newcommand{\bfb}{{\boldsymbol b}}
\newcommand{\bfc}{{\boldsymbol c}}

\newcommand{\bfr}{{\boldsymbol r}}

\newcommand{\bfw}{{\boldsymbol w}}
\newcommand{\bfx}{{\boldsymbol x}}
\newcommand{\bfy}{{\boldsymbol y}}
\newcommand{\bfz}{{\boldsymbol z}}

\DeclareMathOperator*{\argmax}{arg\,max}

\renewcommand{\le}{\leqslant}
\renewcommand{\leq}{\leqslant}
\renewcommand{\ge}{\geqslant}
\renewcommand{\geq}{\geqslant}

\newcommand{\N}{\mathbb{N}}

\newcommand{\Cref}[1]{Co\-rol\-la\-ry\,\ref{#1}}

\theoremstyle{plain} \theorembodyfont{\normalfont\slshape}

\newtheorem{thm}{Theorem$\!$}
\newenvironment{theorem}{\begin{thm}\hspace*{-1ex}{\bf.}}{\end{thm}}

\newtheorem{prop}[thm]{Proposition$\!$}

\newtheorem{lem}[thm]{Lemma$\!$}
\newenvironment{lemma}{\begin{lem}\hspace*{-1ex}{\bf.}}{\end{lem}}

\newtheorem{cor}[thm]{Corollary$\!$}
\newenvironment{corollary}{\begin{cor}\hspace*{-1ex}{\bf.}}{\end{cor}}

\newtheorem{prob}[thm]{Problem$\!$}

\newtheorem{cl}[thm]{Claim$\!$}
\newenvironment{claim}{\begin{cl}\hspace*{-1ex}{\bf.}}{\end{cl}}

\newtheorem{ob}[thm]{Observation$\!$}

\newtheorem{defi}[thm]{Definition$\!$}
\newenvironment{definition}{\begin{defi}\hspace*{-1ex}{\bf.}}{\end{defi}}

\theorembodyfont{\normalfont}

\newtheorem{exam}{Example$\!$}
\newenvironment{example}{\begin{exam}\hspace*{-1ex}{\bf .}}{\end{exam}}

\newtheorem{remrk}{Remark$\!$}

\newcommand{\argmin}{\arg\!\min}

\definecolor{Codecolor}{named}{White}  %{Tan}
	\newcommand{\Copen}{\mbox{\{\kern-5.50pt\{}}
	\newcommand{\Cclose}{\mbox{\}\kern-5.50pt\}}}
	\newcommand{\Cslash}{\mbox{$\backslash\kern-6.02pt\backslash$}}

\newcommand{\RN}[1]{%
	\textup{\uppercase\expandafter{\romannumeral#1}}%
}

\newtheorem{innercustomgeneric}{\customgenericname}
\providecommand{\customgenericname}{}
\newcommand{\newcustomtheorem}[2]{%
	\newenvironment{#1}[1]
	{%
		\renewcommand\customgenericname{#2}%
		\renewcommand\theinnercustomgeneric{##1}%
		\innercustomgeneric
	}
	{\endinnercustomgeneric}
}
\newcustomtheorem{customproposition}{Proposition}
\newcustomtheorem{customtheorem}{Theorem}
\newcustomtheorem{customlemma}{Lemma}
\newcustomtheorem{customclaim}{Claim}

\newcommand{\pr}{\ensuremath{\mathsf{Pr}}}
\newcommand{\ch}{\ensuremath{\mathsf{S}}}
\newcommand{\bsc}{\ensuremath{\mathsf{BSC}}}

\newcommand{\zc}{\ensuremath{\mathsf{Z}}}
\newcommand{\del}{\ensuremath{\mathsf{Del}}}
\newcommand{\ins}{\ensuremath{\mathsf{Ins}}}
\newcommand{\emb}{\ensuremath{\mathsf{Emb}}}
\newcommand{\perr}{\ensuremath{\mathsf{P_{err}}}}
\newcommand{\pfail}{\ensuremath{\mathsf{P_{fail}}}}
\newcommand{\prun}{\ensuremath{\mathsf{P_{run}}}}
\newcommand{\palt}{\ensuremath{\mathsf{P_{alt}}}}
\newcommand{\cp}{\ensuremath{\mathsf{Cap}}}
\newcommand{\ent}{\ensuremath{\mathsf{H}}}
\newcommand{\sups}{\ensuremath{\cS\mathsf{CS}}}
\newcommand{\subs}{\ensuremath{\cL\mathsf{CS}}}
\newcommand{\ML}{\ensuremath{\mathsf{ML}}}

\newcommand{\e}[1]{\textcolor{red}{#1}}

\begin{document}
\title{\textbf{On The Decoding Error Weight of One or Two Deletion Channels}}		

\author{\large Omer~Sabary, %~\IEEEmembership{Student Member,~IEEE}, 
Daniella~Bar-Lev, %~\IEEEmembership{Student Member,~IEEE},
Yotam~Gershon, %~\IEEEmembership{Student Member,~IEEE}, 
Alexander~Yucovich, and Eitan~Yaakobi. %~\IEEEmembership{Senior Member,~IEEE} 
	\thanks{This work was presented in part at the IEEE International Symposium on Information Theory (ISIT), Los Angeles, CA, June 2020 (reference~\cite{SYY20}) and at the IEEE International Symposium on Information Theory (ISIT), Melbourne, Victoria, Australia, July 2021 (reference~\cite{BGSY21}).}
		
		\thanks{O. Sabary, A. Yucovich, and E. Yaakobi are with the Department of Computer Science, Technion --- Israel Institute of Technology, Haifa 3200003, Israel (e-mail: \{omersabary,yucovich,yaakobi\}@cs.technion.ac.il).}
        \thanks{D. Bar-Lev is with the Center for Memory and Recording Research, University of California San Diego, La Jolla, CA\,92093, USA (e-mail: dbarlev@ucsd.edu).}
		\thanks{Y. Gershon is with the Department of Electrical and Computer Engineering, Technion --- Israel Institute of Technology, Haifa 3200003, Israel (e-mail: yotamgr@campus.technion.ac.il).}
		%\thanks{This work was partially funded by the Technion Machine Learning and Intelligent Systems Center under Grants 86703057 and 86703062.} 
		%Part of the results in the paper were presented at the IEEE International Symposium on Information Theory, Aachen, Germany, June 25 -- 30, 2017 (reference~\cite{AY17}).}
		%\\\@date
	}

\begin{comment}
\author{\textbf{Omer Sabary}\IEEEauthorrefmark{4}, \textbf{Daniella Bar-Lev}\IEEEauthorrefmark{2},    \textbf{Yotam Gershon}\IEEEauthorrefmark{3}, \textbf{Alexander Yucovich}\IEEEauthorrefmark{2}, 
        and \textbf{Eitan Yaakobi}\IEEEauthorrefmark{2}\\[0.5mm]
\IEEEauthorblockA{\IEEEauthorrefmark{2} \small Department of Computer Science,
Technion --- Israel Institute of Technology, Haifa, 3200003 Israel\\[0.5mm]}
\IEEEauthorblockA{\IEEEauthorrefmark{3} \small  Department of Electrical Engineering,
Technion --- Israel Institute of Technology, Haifa, 3200003 Israel \\[1mm]}
\IEEEauthorblockA{\IEEEauthorrefmark{4} \small Department of Electrical and Computer Engineering,
University of California San Diego, La Jolla, CA\,92093, USA\\[0.5mm]}
{Emails: \code  \{daniellalev,yaakobi\}@cs.technion.ac.il, \code  yotamgr@campus.technion.ac.il, \code osabary@ucsd.edu} 
%Part of the results in the paper were presented at the IEEE International Symposium on Information Theory, Los Angeles, CA, July 21 -- 26, 2020~\cite{SYY20}.
\end{comment}
%}

		\maketitle
\begin{abstract}
This paper tackles two problems that fall under the study of coding for insertions and deletions. These problems are motivated by several applications, among them is reconstructing strands in DNA-based storage systems. Under this paradigm, a word is transmitted over some fixed number of identical independent channels and the goal of the decoder is to output the transmitted word or some close approximation of it. The first part of the paper studies optimal decoding for a special case of the deletion channel, referred by the \emph{$k$-deletion channel}, which deletes exactly $k$ symbols of the transmitted word uniformly at random. In this part, the goal is to understand how an optimal decoder operates in order to minimize the expected normalized distance. A full characterization of an efficient optimal decoder for this setup, reffered to as \emph{the maximum likelihood* (ML*) decoder}, is given for a channel that deletes one or two symbols. For $k=1$ it is shown that when the code is the entire space, the decoder is the \emph{lazy decoder} which simply returns the channel output. Similarly, for $k=2$ it is shown that the decoder acts as the lazy decoder in almost all cases and when the longest run is significantly long (roughly $(2-\sqrt{2})n$ when $n$ is the word length), it prolongs the longest run by one symbol. 
The second part of this paper studies the deletion channel that deletes a symbol with some fixed probability $p$, while focusing on two instances of this channel. Since operating the maximum likelihood (ML) decoder, in this case, is computationally unfeasible, we study a slightly degraded version of this decoder for two channels and study its \emph{expected normalized distance}. We observe that the dominant error patterns are deletions in the same run or errors resulting from alternating sequences. Based on these observations, we derive lower bounds on the expected normalized distance of the degraded ML decoder for any transmitted $q$-ary sequence of length $n$ and any deletion probability $p$. We further show that as the word length approaches infinity and the channel's deletion probability $p$ approaches zero, these bounds converge to approximately $\frac{3q - 1}{q - 1} p^2$. These theoretical results are verified by corresponding simulations.

% We also study the cases when the transmitted word belongs to the Varshamov Tenengolts (VT) code or the shifted VT code. Additionally, the insertion channel is studied as well as the case of two insertion channels.  

%This paper studies the problem of reconstructing a word given several of its noisy copies. This setup is motivated by several applications, among them is reconstructing strands in DNA-based storage systems. Under this paradigm, a word is transmitted over some fixed number of identical independent channels and the goal of the decoder is to output the transmitted word or some close approximation to it.

%Next, the deletion channel that deletes a symbol with some fixed probability $p$ is studied while focusing on two instances of this channel. While operating the ML decoder in this case is computationally unfeasible we study a slight degraded version of this decoder for two channels and study its expected normalized distance. We observe that the dominant error patterns are deletions in the same run or errors resulting from alternating sequences. Based on these observations, it is derived that the error probability of the degraded ML decoder is roughly $\frac{3q-1}{q-1}p^2$, when the transmitted word is any $q$-ary sequence and $p$ is the channel's deletion probability. We also study the cases when the transmitted word belongs to the Varshamov Tenengolts (VT) code or the shifted VT code. Lastly, the insertion channel is studied as well as the case of two insertion channels.  These theoretical results are verified by corresponding simulations. 
\end{abstract}


\section{Introduction} \label{sec:intro}

Codes correcting insertions/deletions have attracted considerable attention in the past decade due to their relevance to the special error behavior in DNA-based data storage~\cite{Betal16,HMG18,KC14,OAC17,Retal13,SOSAYY19,YGM16,Yetal14}. These codes are relevant for other applications in communications models. For example, insertions/deletions happen during the synchronization of files and symbols of data streams \cite{SalaSchoeny-Sync_TransComm} or due to over-sampling and under-sampling at the receiver side~\cite{DolecekAnan_Sync_2007}. The algebraic concepts of codes correcting insertions/deletions date back to the 1960s when Varshamov and Tenengolts designed a class of binary codes, nowadays called \emph{VT codes}~\cite{VT}. These codes were originally designed to correct a single asymmetric error and later were proven to correct a single insertion/deletion~\cite{L66}. Extensions for multiple deletions were recently proposed in several studies; see e.g.~\cite{BGZ16,GS17,SB19,SRB18}.
%VT codes are asymptotically optimal length-$n$ single-insertion/deletion correcting codes of redundancy $\log(n+1)$. As a generalization of VT codes, Tenengolts presented $q$-ary single-insertion/deletion correcting codes in~\cite{T84}. In~\cite{BGZ16}, Brakensiek \emph{et al.} presented binary multiple-insertion/deletion correcting codes with small asymptotic redundancy. For an explicit small number of deletions, their construction however needs redundancy $c \log n$ where $c$ is a large constant. The recent parallel works by Gabrys et al.~\cite{GS17} and Sima et al.~\cite{SRB18} have presented constructions to correct two deletions with redundancy $8 \log(n)+O(\log( \log (n)))$ \cite{GS17} and $7\log (n) + o(\log (n))$ \cite{SRB18}, respectively. Sima and Bruck \cite{SB19} generalized their construction to correct any $t$ insertions/deletions with redundancy $8t\log (n) + o(\log (n))$.
However, while codes correcting substitution errors were widely studied and efficient capacity-achieving codes both for small and large block lengths are used conventionally, much less is known for codes correcting insertions/deletions. 
%Codes correcting a single deletion or insertion were found by Levenshtein in~\cite{L66} using the Varshamaov and Tenengolts construction for asymmetric errors~\cite{VT65}. However, the task of constructing codes correcting multiple deletions and insertions is extremely more challenging due to the difficulty of this problem. There are several works who tackle this problem, especially from the past 2 year~\cite{AVDG19,AVG18,BM18,BGZ16,CS19,DM00,GS17,GK18,GL19,GW17,HS17,H16,Hag16,HK18,HK19,HE17,HE19,HF02,KM13,KK19,LN16,M09,N19,PSWFC11,R03,RM00,SZ99,SB19,SRB18,S00,TPFV19,TAB19,TTVM17,TFVL17,TFV18,W11}, but many of these works are either not rate efficient, not easy to implement, or do not guarantee successful decoding in the worst case. 
More than that, even the deletion channel capacity is far from being solved~\cite{BGH17,CK15,C19,D11,M09,ED06,RD15, RC23, AT23}. 

In the same context, \emph{reconstruction of sequences} refers to a large class of problems in which there are several noisy copies of the information and the goal is to decode the information, either with small or zero error probability. The first example is the \emph{sequence reconstruction problem} which was first studied by Levenshtein and others~\cite{L011,L012,gabrys2018sequence, yaakobi2013sequence, yaakobi2012uncertainty, sala2015three, levenshtein2009error,  levenshtein2008reconstruction}. Another example, which is also one of the more relevant models to the discussion in the first part of this paper, is the \emph{trace reconstruction problem}~\cite{BKK04, PZ17, NP17, HMP08, HPP18}, where it is assumed that a sequence is transmitted through multiple deletion channels, and each bit is deleted with some fixed probability~$p$. Under this setup, the goal is to determine the minimum number of traces, i.e., channels, required to reconstruct the sequence with high probability.  One of the dominant motivating applications of the sequence reconstruction problems is DNA storage~\cite{CGK12, OAC17, YGM16, GHP15, Anavy433524, BO21}, where every DNA strand has several noisy copies. Several new results on the trace reconstruction problem have been recently studied in~\cite{TrellisBMA, CDR21, SB21, D21, G21, hanna22, hanna23, MS24}.

%Recently, several new coding schemes to support the sequence reconstruction for DNA storage system have been proposed. Kiah et al. presented in~\cite{kiah2016codes} an asymmetric coding technique to eliminate synthesis and sequencing errors. Abroshan et al. presented in~\cite{AVDG19} a new coding scheme, based on the Varshamov Tenegolts (VT) code~\cite{VT}.}

Many of the reconstruction problems are focused on studying the minimum number of channels required for successful decoding. However, in many cases, the number of channels is fixed and then the goal is to find the best code construction that is suitable for this channel setup. Motivated by this important observation, the first part of this paper also studies the error probability of maximum-likelihood decoding when a word is transmitted over two deletion or insertion channels. We should note that we study a degraded version of the maximum likelihood decoder, which allows the decoder to output words of shorter length than the code length. This flexibility of the decoder is useful especially in cases where the same symbol is deleted in both of the channels, or when the code does not have deletion-correcting capabilities. This study is also motivated by the recent works of Srinivasavaradhan \textit{et al.}~\cite{srinivasavaradhan2018maximum,srinivasavaradhan2019symbolwise}, where reconstruction algorithms that are based on the maximum-likelihood approach have been studied. Abroshan \textit{et al.} presented in~\cite{AVDG19} a new coding scheme for sequence reconstruction which is based on the Varshamov Tenengolts (VT) code~\cite{VT} and in~\cite{ISIT20} it was studied how to design codes for the worst case, when the number of channels is given. 

When a word is transmitted over the deletion channel, the channel output is necessarily a subsequence of the transmitted word. Hence, when transmitting the same word over multiple deletion channels, the possible candidate words for decoding are the so-called \emph{common supersequences} of all of the channels' outputs. Hence, an important part of the decoding process is to find the set of all possible common supersequences and in particular the \emph{shortest common supersequences} (\emph{SCS})~\cite{itoga1981string}. Even though this problem is in general NP hard~\cite{B93} for an arbitrary number of sequences, for two words a dynamic programming algorithm exists with quadratic complexity; see~\cite{itoga1981string} for more details and further improvements and approximations for two or more sequences~\cite{Irving1994,hirschberg1975linear,tronicek1999problems, ukkonen1990linear}. The case of finding the \emph{longest common subsequences} (\emph{LCS}) is no less interesting and has been extensively studied in several previous works; see e.g.~\cite{hirschberg1977algorithms, apostolico1992fast, hsu1984computing, masek1980faster, sankoff1972matching, chen2006fast}. Most of these works focused on improving the complexity of the dynamic programming algorithm suggested in~\cite{apostolico1992fast} and presented heuristics and approximations for the LCS.

Back to a single instance of a channel with deletion errors, there are two main models which are studied for this type of errors. While in the first one, the goal is to correct a fixed number of deletions in the worst case, for the second one, which corresponds to the channel capacity of the deletion channel, one seeks to construct codes which correct a fraction $p$ of deletions with high probability~\cite{BBW18,CK15,CS19,DG06,DM07,FD10,hanna2019list,KD10,ED06,TPFV19,VTR13}. The second part of this paper considers a combination of these two models. In this channel, referred as the \emph{$k$-deletion channel}, $k$ symbols of the length-$n$ transmitted word are deleted uniformly at random; see e.g.~\cite{TFA2017, DBLP:journals/corr/abs-1802-00703}. Consider for example the case of $k=1$, i.e., one of the $n$ transmitted symbols is deleted, each with the same probability. In case the transmitted word belongs to a single-deletion-correcting code then clearly it is possible to successfully decode the transmitted word. However, if such error correction capability is not guaranteed in the worst case, two approaches can be of interest. In the first, one may output a list of all possible transmitted words, that is, \emph{list decoding} for deletion errors as was studied recently in several works; see e.g.~\cite{guruswami2020optimally,guruswami2017deletion,haeupler2018synchronization,hanna2019list,hayashi2018list,liu2019list,wachter2017list}. The second one, which is taken in the present work, seeks to output a word that minimizes the expected normalized distance between the decoder's output and the transmitted word. This channel was also studied in several previous works. In~\cite{G15}, the author studied the maximal length of words that can be uniquely reconstructed using a sufficient number of channel outputs of the $k$-deletion channel and calculated this maximal length explicitly for $n-k \le 6$. In~\cite{DBLP:journals/corr/abs-1802-00703}, the goal was to study the entropy of the set of the potentially channel input words given a corrupted word, which is the output of a channel that deletes either one or two symbols. The minimum and maximum values of this entropy were explored. In~\cite{TFA2017,TPFV19}, the authors presented a polar coding solution in order to correct deletions in the $k$-deletion channel.

Mathematically speaking, assume $\ch$ is a channel that is characterized by a conditional probability $\pr_\ch\{  \bfy\textmd{ rec. }| \bfx \textmd{ trans.}\},$
for every pair $(\bfx,\bfy)\in(\Sigma_q^*)^2$. A decoder for a code $\cC$ with respect to the channel $\ch$ is a function $\cD:\Sigma_q^*\rightarrow \cC$. Its \emph{average decoding failure probability} is the probability that the decoder output is not the transmitted word. %defined by $\pfail(\ch,\cC,\cD) = \frac{\sum_{\bfc\in\cC}\pfail(\bfc)}{|\cC|}$, where 
%$\pfail(\bfc) =   \sum_{\bfy:\cD(\bfy) \neq  \bfc} \pr_\ch\{  \bfy\textmd{ rec. }| \bfc \textmd{ trans.}\}.$
The \emph{maximum-likelihood} (\emph{ML}) \emph{decoder} for $\cC$ with respect to $\ch$, denoted by $\cD_{\ML}$, outputs a codeword $\bfc\in\cC$ that maximizes the probability $\pr_\ch\{  \bfy\textmd{ rec. }| \bfc \textmd{ trans.}\}$. 
%That is, for  $\bfy\in \Sigma_{q}^*$,  $\cD_{\ML}(\bfy) =\argmax_{\bfc\in\cC}\left\{\pr_\ch\{  \bfy\textmd{ rec. }| \bfc \textmd{ trans.}\}\right\}$. 
This decoder minimizes the average decoding \emph{failure} probability and thus it outputs only codewords. However, if one seeks to minimize the \emph{expected normalized distance}, then the decoder should consider non-codewords as well. The \emph{expected normalized distance} is the average normalized distance between the transmitted word and the decoder's output, where the distance function depends upon the channel of interest. %Hence, for a distance function $d$ and a decoder $\cD$, we let $\perr(\ch,\cC,\cD,d) = \frac{\sum_{\bfc\in\cC}\perr(\bfc,d)}{|\cC|}$, where \vspace{-2ex}
%$$\perr(\bfc,d) =   \sum_{\bfy:\cD(\bfy) \neq  \bfc} \frac{d(\cD(\bfy),\bfc)}{|\bfc|}\cdot \pr_\ch\{  \bfy\textmd{ rec. }| \bfc \textmd{ trans.}\}.\vspace{-1ex}$$
 In this work we study the \emph{ML$^*$ decoder}, which outputs words that minimize the expected normalized distance.%, for the $t$-deletion channel.

The rest of the paper is organized as follows. Section~\ref{sec:defs} presents the formal definition of channel transmission and maximum likelihood decoding in order to minimize the expected normalized distance. Section~\ref{sec:del_ch} introduces the deletion channel, the insertion channel, and the $k$-deletion channel.  Section~\ref{sec:1-del} studies the 1-deletion channel. It introduces two types of decoders. The first one, referred as the \emph{embedding number decoder}, maximizes the so-called \emph{embedding number} between the channel output and all possible codewords. The second one is called the \emph{lazy decoder} which simply returns the channel output. The main result of this section states that if the code is the entire space then the ML$^*$ decoder is the lazy decoder. Similarly, Section~\ref{sec:2-del} studies the 2-deletion channel where it is shown that in almost all cases the ML$^*$ decoder should act as the lazy decoder and in the rest of the cases it returns a length-$(n-1)$ word which maximizes the embedding number.

In Section~\ref{sec:two deletions}, we present our main results for the case of two deletion channels. We consider the expected normalized distance of a degraded version of the ML decode when the code is the entire space. Among our results, it is shown that when the code is the entire space and the code length $n$ approaches infinity, the expected normalized distance is lower bounded by roughly $\frac{3q-1}{q-1}p^2$, when $q$ is the alphabet size and $p$ is the channel's deletion probability, which approaches zero. We observe that the dominant error patterns are deletions from the same run or errors resulting from alternating sequences. %Based on these observations, it is derived that the error probability of the degraded ML decoder is roughly $\frac{3q-1}{q-1}p^2$, when the transmitted word is any $q$-ary sequence and $p$ is the channel's deletion probability. We also study the cases when the transmitted word belongs to the Varshamov Tenengolts (VT) code or the shifted VT code. 
%Lastly, the insertion channel is studied as well as the case of two insertion channels. 
These theoretical results are verified by corresponding simulations. Section~\ref{sec:conc} concludes the paper and discusses open problems. %Due to the lack of space, some of the proofs in the paper are omitted. 

\section{Definitions and Preliminaries}\label{sec:defs}		
\begin{comment}
For an integer $q\geq 2$, let $\Sigma_q$ denote the $q$-ary alphabet $\{0,1,\ldots,q-1\}$  and $\Sigma_q^* \triangleq \bigcup_{\ell=1}^n \Sigma_q^\ell $. 
We use $\len{\bfx}$ to denote the length of $\bfx$. For $\bfx = (x_1,\ldots, x_\dim) \in \Sigma_q^n$, we let $\run{\bfx}$ denote the number of runs in $\bfx$, that is, \mbox{$\run{\bfx} := 1 + |\{1 \leq i < \dim : x_i \neq x_{i+1}\}|.$} For $\bfx,\bfy \in \Sigma_q^*$, the notation $\bfx\bfy$ denotes the concatenation of $\bfx$ and $\bfy$, where $\len{\bfx\bfy} = \len{\bfx}+\len{\bfy}$. 
\end{comment}
	
We denote by $\Sigma_q =\{0,\ldots,q-1\}$ the alphabet of size $q$ and $\Sigma_q^* \triangleq \bigcup_{\ell=0}^\infty \Sigma_q^\ell, \Sigma_q^{\leq n} \triangleq \bigcup_{\ell=0}^n \Sigma_q^\ell, \Sigma_q^{\geq n} \triangleq \bigcup_{\ell=n}^\infty \Sigma_q^\ell$. 
%Let $\bfx \in \Sigma_q^n\triangleq \{0,1,\ldots,q-1\}^n$ be a word of length $n$ over the alphabet $\Sigma_q$. 
The length of $\bfx\in \Sigma^n$ is denoted by $|\bfx|=n$. % and the length-$m$ prefix of $\bfx$ is denoted by $\bfx_{1:m}$. 
The \emph{Levenshtein distance}  between two words $\bfx,\bfy \in \Sigma_q^*$, denoted by $d_L(\bfx,\bfy)$, is the minimum number of insertions and deletions required to transform $\bfx$ into $\bfy$, and
%Similarly, for two strings $\bfx,\bfy\in \Sigma^*$, $d_E(\bfx,\bfy)$ denotes the \emph{edit} distance between $\bfx$ and $\bfy$, which is the minimum number of insertions, deletions and substitutions required to transform $\bfx$ into $\bfy$. 
$d_H(\bfx,\bfy)$ denotes the \emph{Hamming distance} between $\bfx$ and $\bfy$, when $|\bfx| = |\bfy|$. 
%The \emph{radius-$t$ deletion ball} obtained after exactly $t$ deletions in $\bfx$ is denoted by $\D(\bfx,t)$ and its size is $\VD(\bfx,t)$. The \emph{global deletion ball of $\bfx$} is defined by $\D^*(\bfx) = \bigcup_{t=1}^{|\bfx|} \D(\bfx,t) $ and its size is $\VD^*(\bfx) = |\D^*(\bfx)|$. 
A word $\bfx\in\Sigma_q^*$ will be referred to as an \emph{alternating sequence} if it cyclically repeats all symbols in $\Sigma_q$ in the same order. For example, for $\Sigma_2=\{0,1\}$, the two alternating sequences are $010101\cdots$ and $101010\cdots$, and in general there are $q!$ alternating sequences.  For $n\geq 1$, the set $\{1,\ldots,n\}$ is abbreviated by $[n]$ and for $0\le i< j$ $[i,j]$ denotes the set $\{i,i+1,\ldots, j \}$. 

For a word $\bfx \in \Sigma_q^*$ and a set of indices $I\subseteq [|\bfx|]$, the word $\bfx_I$ is the \emph{projection} of $\bfx$ on the indices of $I$ which is the subsequence of $\bfx$ received by the symbols in the entries of $I$. A word $\bfx \in \Sigma^*$ is called a \emph{supersequence} of $\bfy \in \Sigma^*$, if $\bfy$ can be obtained by deleting symbols from $\bfx$, that is, there exists a set of indices $I\subseteq [|\bfx|]$ such that $\bfy = \bfx_I$. In this case, it is also said that $\bfy$ is a \emph{subsequence} of $\bfx$. Furthermore, $\bfx$ is called a \emph{common supersequence} (\emph{subsequence}) of some words $\bfy_1,\ldots,\bfy_t$ if $\bfx$ is a supersequence (subsequence) of each one of these $t$ words. 
The set of all common supersequences of $\bfy_1,\ldots,\bfy_t\in \Sigma_q^*$ is denoted by $\sups(\bfy_1,\ldots,\bfy_t)$ and $\mathsf{SCS}(\bfy_1,\dots,\bfy_t)$ is the \emph{length of the shortest common supersequence} (\emph{SCS)} of $\bfy_1,\dots,\bfy_t$, that is, 
$\mathsf{SCS}(\bfy_1,\dots,\bfy_t) \triangleq \min_{\bfx\in \sups(\bfy_1,\ldots,\bfy_t)}\{|\bfx|\}$. Similarly, $\subs(\bfy_1,\ldots,\bfy_t)$ is the set of all subsequences of $\bfy_1,\dots,\bfy_t$ and $\mathsf{LCS}(\bfy_1,\dots,\bfy_t)$ is the \emph{length of the longest common subsequence} (\emph{LCS)} of $\bfy_1,\dots,\bfy_t$, that is, $\mathsf{LCS}(\bfy_1,\dots,\bfy_t) \triangleq \max_{\bfx\in \subs(\bfy_1,\ldots,\bfy_t)}\{|\bfx|\}$.

The \emph{radius-$r$ insertion ball} of a word $\bfx\in\Sigma_q^*$, denoted by $I_r(\bfx)$, is the set of all supersequences of $\bfx$ of length $|\bfx|+r$. From~\cite{L66} it is known that $I_r(\bfx)= \sum_{i=0}^r \binom{|\bfx|+r}{i} (q-1)^i$. Similarily,  the \emph{radius-$r$ deletion ball} of a word $\bfx\in\Sigma_q^*$, denoted by $D_r(\bfx)$, is the set of all subsequences of $\bfx$ of length $|\bfx|-r$. %The set of all subsequences, supersequences of $\bfx$ is denoted by $\del^*(\bfx) \triangleq \bigcup_{r\geq 0}\del_r(\bfx), \ins^*(\bfx) \triangleq \bigcup_{r\geq 0}\ins_r(\bfx)$, respectively.

We consider a channel $\ch$ that is characterized by a conditional probability $\pr_\ch$, and is defined by 
$$\pr_\ch\{  \bfy\textmd{ rec. }| \bfx \textmd{ trans.}\},$$
for every pair $(\bfx,\bfy)\in(\Sigma_q^*)^2$ , when the channel is clear from the context, we use the shortened notation of $ p(\bfy | \bfx)$ to denote this probability. %, where $\Sigma_{in}, \Sigma_{out}$ is the input, output alphabet, respectively. In case $\Sigma_{in}  = \Sigma_{out}$, we simply write $(\bfx,\bfy)\in(\Sigma^*)^2$. 
Note that it is not assumed that the lengths of the input and output words are the same as we consider also deletions and insertions of symbols, which are the main topic of this work. As an example, it is well known that if $\ch$ is the \emph{binary symmetric channel} (\emph{BSC}) with crossover probability $0\leq p \leq 1/2$, denoted by $\bsc(p)$, it holds that 
$$\pr_{\bsc(p)}\{  \bfy\textmd{ rec. }| \bfx \textmd{ trans.}\} = p^{d_H(\bfy,\bfx)}(1-p)^{n-d_H(\bfy,\bfx)},$$
for all $(\bfx,\bfy)\in (\Sigma_2^n)^2$, and otherwise (the lengths of $\bfx$ and $\bfy$ is not the same) this probability equals 0. Similarly, for the $Z$-channel, denoted by $\zc(p)$, it is assumed that only a 0 can change to a 1 with probability $p$ and so 
$$\pr_{\zc(p)}\{  \bfy\textmd{ rec. }| \bfx \textmd{ trans.}\} = p^{d_H(\bfy,\bfx)}(1-p)^{n-d_H(\bfy,\bfx)},$$
for all $(\bfx,\bfy)\in (\Sigma_2^n)^2$ such that for any $1\le i \le n$, $\bfx_i \leq \bfy_i$, and otherwise this probability equals 0. 

In the \emph{deletion channel} with deletion probability $p$, denoted by $\del(p)$, every symbol of the word $\bfx$ is deleted with probability $p$. Similarly, in the \emph{insertion channel} with insertion probability $p$, denoted by $\ins(p)$, a symbol is inserted in each of the possible $|\bfx|+1$ positions of the word $\bfx$ with probability $p$, while the probability to insert each of the symbols in $\Sigma_q$ is the same and equals $\frac{p}{q}$. 
Another variation of the deletion channel, studied in this work in Sections~\ref{sec:1-del} and~\ref{sec:2-del}, is the \emph{$k$-deletion channel}, denoted by $k\textrm{-}\mathsf{Del}$, where exactly $k$ symbols are deleted from the transmitted word. The $k$ symbols are selected randomly from the $\binom{n}{k}$ options. This channel was studied in~\cite{DBLP:journals/corr/abs-1802-00703}, where the authors studied the words that maximize and minimize the entropy of the set of the possible transmitted words, given a channel output.  In~\cite{TFA2017}, a polar codes based coding solution that corrects deletions from the $k$-deletion channel was presented.

A decoder for a code $\cC$ with respect to the channel $\ch$ is a function $\cD:\Sigma_q^*\rightarrow \cC$. 
\begin{definition}
\textbf{\emph{Average decoding failure probability.}} 
The average decoding failure probability of a decoder $\cD$, with respect to a channel $\ch$ and a code $\cC$, is denoted by $\pfail(\ch,\cC,\cD)$ and defined as $ \pfail(\ch,\cC,\cD)\triangleq \frac{\sum_{\bfc\in\cC}\pfail(\bfc)}{|\cC|}$, where 
$$\pfail(\bfc) \triangleq   \sum_{\bfy:\cD(\bfy) \neq  \bfc} \pr_\ch\{  \bfy\textmd{ rec. }| \bfc \textmd{ trans.}\}.$$
\end{definition}

We will also be interested in the \emph{expected normalized distance} which is the average normalized distance between the transmitted word and the decoder's output. The distance will depend upon the channel of interest. For example, for the BSC we will consider the Hamming distance, while for the deletion and insertion channels, the Levenshtein distance will be of interest. Formal definition of the expected normalized distance is given below.  

\begin{definition}
\textbf{\emph{The expected normalized distance.}} The expected normalized distance of a decoder $\cD$, with respect to a channel $\ch$, a code $\cC$, and a distance function $d$ is denoted by $\perr(\ch,\cC,\cD,d)$.
Its value is defined as $$\perr(\ch,\cC,\cD,d) \triangleq \frac{\sum_{\bfc\in\cC}\perr(\bfc,d)}{|\cC|},$$ where 
$$\perr(\bfc,d) \triangleq   \sum_{\bfy:\cD(\bfy) \neq  \bfc} \frac{d(\cD(\bfy),\bfc)}{|\bfc|}\cdot \pr_\ch\{  \bfy\textmd{ rec. }| \bfc \textmd{ trans.}\}.$$
\end{definition}
Next, we define the maximum likelihood decoder. 
%In case the distance function will be clear from the context we will simply remove it from these notations. 
\begin{definition} \textbf{\emph{The maximum-likelihood decoder}.}  
The \emph{maximum-likelihood} (\emph{ML}) \emph{decoder} for a code $\cC$ with respect to a channel $\ch$, denoted by $\cD_{\ML}$, outputs a codeword $\bfc\in\cC$ that maximizes the probability $\pr_\ch\{  \bfy\textmd{ rec. }| \bfc \textmd{ trans.}\}$. That is, for  $\bfy\in \Sigma_{q}^*$, 
$$\cD_{\ML}(\bfy) \triangleq \argmax_{\bfc\in\cC}\left\{\pr_\ch\{  \bfy\textmd{ rec. }| \bfc \textmd{ trans.}\}\right\}.$$
%That is, the decoder outputs a codeword $\bfc\in\cC$ that maximizes the probability $\pr_\ch\{  \bfy\textmd{ rec. }| \bfc \textmd{ trans.}\}$.
\end{definition}

It should be noted that, in the analysis presented in this paper, for all the presented decoders, unless stated otherwise explicitly, if there is more than one possible word that satisfies the condition of the decoder's output,  the decoder chooses one of them arbitrarily.

It is well known that for the BSC, the ML decoder simply chooses the closest codeword with respect to the Hamming distance. 
%We will also be interested in the \emph{average decoding error probability} which is the average distance between the transmitted word and the decoder's output. The distance will depend upon the channel of interest. For example, for the BSC we will consider the Hamming distance, while for the deletion and insertion channels, the Levenshtein distance will be of interest. Hence, for a channel $\ch$, distance function $d$, and a decoder $\cD:\Sigma_{q}^*\rightarrow \Sigma_{q}^*$, we let $\perr(\ch,\cC,\cD,d) = \frac{\sum_{\bfc\in\cC}\perr(\bfc,d)}{|\cC|}$, where  $$\perr(\bfc,d) =   \sum_{\bfy:\cD(\bfy) \neq  \bfc} d(\bfy,\bfc)\cdot \pr_\ch\{  \bfy\textmd{ rec. }| \bfc \textmd{ trans.}\}.$$ In case the distance function will be clear from the context we will simply remove it from these notations. Note that here the decoder does not necessarily have to decode the output word into a codeword. 
The \emph{channel capacity} is referred to as the maximum information rate that can be reliably transmitted over the channel $\ch$ and is denoted by $\cp(\ch)$. For example, %it is well known that 
$\cp(\bsc(p))=1-\ent(p)$, where $\ent(p)=-p\log(p)-(1-p)\log(1-p)$ is the binary entropy function. 

The conventional setup of channel transmission is extended to the case of more than a single instance of the channel. Assume a word $\bfx$ is transmitted over some $t$ identical channels of $\ch$ and the decoder receives all channel outputs $\bfy_1,\ldots,\bfy_t$. Unless stated otherwise, it is assumed that all channels are independent and thus this setup is characterized by the conditional probability 
$$\pr_{(\ch,t)}\{ \bfy_1,\ldots,\bfy_t \textmd{ rec.}| \bfx \textmd{ trans.}\} = \prod_{i=1}^t\pr_{\ch}\{ \bfy_i\textmd{ rec.}| \bfx \textmd{ trans.}\}.$$
The definitions of a decoder, the ML decoder, and the error probabilities are extended similarly. The input to the ML decoder is the words $\bfy_1,\ldots,\bfy_t$ and the output is the codeword $\bfc$ which maximizes the probability $\pr_{(\ch,t)}\{ \bfy_1,\ldots,\bfy_t \textmd{ rec.}| \bfc \textmd{ trans.}\}$. That is, 
$$\cD_{\ML}(\bfy_1,\ldots,\bfy_t) \triangleq \argmax_{\bfc\in\cC} \hspace{-.5ex}\left\{ \hspace{-.5ex} \pr_{(\ch,t)}\{ \bfy_1,\ldots,\bfy_t \textmd{ rec.}| \bfc \textmd{ trans.} \hspace{-.5ex} \}\right \}\hspace{-.75ex} .$$
Since the outputs of all channels are independent, the output of the ML decoder is defined to be,
$$ \cD_{\ML}(\bfy_1,\ldots,\bfy_t) \triangleq \argmax_{\bfc\in\cC}\bigg\{\prod_{i=1}^t  {\pr_\ch\{  \bfy_i\textmd{ rec. }| \bfc \textmd{ trans.} \} }\bigg\}.$$ 
The average decoding failure probability, the expected normalized distance is generalized in the same way and is denoted by  $\pfail(\ch,t,\cC,\cD)$, $\perr(\ch,t,\cC,\cD,d)$, respectively. The capacity of this channel is denoted by $\cp(\ch,t)$, so $\cp(\ch,1) = \cp(\ch)$.

The case of the BSC was studied by Mitzenmacher in~\cite{M06}, where he showed that 
%\begin{small}
\begin{align*}
\cp(\bsc(p),t))  =1 \hspace{-0.5ex}+\hspace{-0.5ex}\sum_{i=0}^t\hspace{-0.5ex}\binom{t}{i} \hspace{-1ex} \left( p^i(1\hspace{-0.25ex}-\hspace{-0.25ex}p)^{t-i}\log \frac{p^i(1-p)^{d-i}}{p^i(1-p)^{t-i}+p^{t-i}(1-p)^i} \right)\hspace{-0.5ex}.&
\end{align*}
%\end{small}

\noindent On the other hand, the $Z$ channel is significantly easier to solve and it is possible to verify that  $\cp(\zc(p),t) = \cp(\zc(p^t))$.
%For the binary erasure channel we similarly get that $\cp(\bec(p),t) = \cp(\bec(p^t))$.
It is also possible to calculate the expected normalized distance and the average decoding failure probability for the BSC and $Z$ channels. For example, when $\cC=\Sigma_2^n$, one can verify that 
$$\perr(\zc(p),t,\Sigma_2^n,\cD_{\ML},d_H) = p^t,$$
and if $t$ is odd then
$$\perr(\bsc(p),t,\Sigma_2^n,\cD_{\ML},d_H) = \sum_{i=0}^{\frac{t-1}{2}} \binom{t}{i} p^{t-i}(1-p)^i.$$
Similarly, $\pfail(\zc(p),t,\Sigma_2^n,\cD_{\ML}) = 1- (1-p^t)^n$ for odd $t$, and $\pfail(\bsc(p),t,\Sigma_2^n,\cD_{\ML}) \hspace{-0.25ex}=1-\hspace{-0.25ex} (1 \hspace{-0.25ex}-\hspace{-0.25ex}   \sum_{i=0}^{\frac{t-1}{2}} \binom{t}{i} p^{t-i}(1\hspace{-0.25ex}-\hspace{-0.25ex}p)^i )^n$.
However, calculating these probabilities for the deletion and insertion channels is a far more challenging task. 

We note that the capacity of several deletion channels has been studied in~\cite{HM}, where it was shown that for some $t > 0$ deletion channels with deletion probability $p$, the capacity under a random codebook satisfies
$$\cp(\del(p),t)=1-A(t)\cdot p^t\log(1/p)-O(p^t),$$
where $A(t) = \sum_{j=1}^{\infty}2^{-j-1}tj^t$. For example, when $t=2$, the capacity is $1-6\cdot p^2\log(1/p)-O(p^2)$. 
One of the goals of this paper, which is discussed in Section~\ref{sec:two deletions}, is to study in depth the special case of $t=2$ and estimate the average error and failure probabilities, when the code is the entire space, the Varshamov Tenengolts (VT) code~\cite{VT}, and the shifted VT (SVT) code~\cite{7837631}.

%This model is closely connected to several related problems. In the \emph{reconstruction problem} studied by Levenshtein~\cite{L011,L012}, it was assumed that the word is transmitted over several noisy channels and the goal of the decoder is to decode the transmitted word in the worst case, assuming that all channels' outputs are different from each other. Several extensions of these problems have been studied; see e.g.~\cite{gabrys2018sequence, yaakobi2013sequence, yaakobi2012uncertainty, sala2015three, levenshtein2009error,  levenshtein2008reconstruction}, 
%however in all of them the goal is to find the number of channels that guarantees unique decoding in the worst case. The most relevant case of the reconstruction problem to our work is the one studied in~\cite{8294216}, where it was shown how the shifted VT codes can be used for the two single-deletion channels case. In another work~\cite{ISIT20}, the dual problem is studied where the number of channels is given and then the goal is to find the best code which guarantees successful decoding in the worst case. Hence, the problem studied in this paper can be regarded as the probabilistic variant of the dual problem of the reconstruction problem. Yet another highly related problem is the one of the \emph{trace reconstruction problem}~\cite{BKK04,DOS17,duda2016fundamental,HPP18,HMP08,NP17,PZ17}. The most relevant works to our study are the recent ones~\cite{srinivasavaradhan2018maximum,srinivasavaradhan2019symbolwise}, where decoding algorithms for maximum likelihood are presented for a fixed number of channels. 

\section{Properties of the Deletion and Insertion Channels under ML Decoding}\label{sec:del_ch}
In this section, we establish several basic results for the deletion channels with one or multiple instances. For these cases, the most relevant distance metric is the Levenshtein distance. Thus, unless stated otherwise explicitly, for the rest of the paper, the Levenshtein distance between $\bfx, \bfy \in \Sigma_q^*$ will be denoted shortly by $d (\bfx, \bfy) \triangleq d_L(\bfx, \bfy)$. We continue with several useful definitions. 
% The main goal of this section is studying the error decoding probability for the deletion channel with two copies, i.e., $t=2$. 
For two words $\bfx,\bfy\in\Sigma_q^*$, the number of different ways in which $\bfy$ can be received as a subsequence of $\bfx$ is called the \emph{embedding number of $\bfy$ in $\bfx$} and is defined by
$$\emb(\bfx;\bfy) \triangleq |\{ I\subseteq [|\bfx|] \ | \ \bfx_I=\bfy\}|.$$
Note that if $\bfy$ is not a subsequence of $\bfx$ then $\emb(\bfx;\bfy)=0$. The embedding number has been studied in several previous works; see e.g.~\cite{DBLP:journals/corr/abs-1802-00703,elzinga2008algorithms} and in~\cite{srinivasavaradhan2018maximum} it was referred to as the \emph{binomial coefficient}. In particular, this value can be computed with quadratic complexity~\cite{elzinga2008algorithms}.

While the calculation of the conditional probability $\pr_\ch\{  \bfy\textmd{ rec. }| \bfx \textmd{ trans.}\}$ is a rather simple task for many of the known channels, it is not straightforward for channels that introduce insertions or deletions. The following basic claim is well known and was also stated in~\cite{srinivasavaradhan2018maximum}. It will be used in our derivations to follow.% however it is presented here for the completeness of the results in the paper and since it will be used in our derivations to follow.
\begin{claim}\label{cl:emb}
For all $(\bfx,\bfy)\in(\Sigma_q^*)^2$, it holds that 
$$\pr_{\del(p)}\{  \bfy\textmd{ rec. }| \bfx \textmd{ trans.}\} = p^{|\bfx|-|\bfy|} (1-p)^{|\bfy|}  \cdot \emb(\bfx;\bfy),$$
$$\hspace{-.5ex}\pr_{\ins(p)}\{  \bfy\textmd{ rec. }| \bfx \textmd{ trans.}\} \hspace{-.5ex}=\hspace{-.5ex} \left(\frac{p}{q}\right)^{|\bfy|-|\bfx|}\hspace{-.6ex}(1\hspace{-.25ex}-\hspace{-.25ex}p)^{|\bfx|\hspace{-.25ex}+\hspace{-.25ex}1\hspace{-.25ex}-\hspace{-.25ex}(|\bfy|\hspace{-.25ex}-\hspace{-.25ex}|\bfx|)}\hspace{-.5ex}\cdot\hspace{-.5ex}\emb(\bfy;\bfx).$$
\end{claim}
According to Claim~\ref{cl:emb}, it is  possible to explicitly characterize the ML decoder for the deletion and insertion channels as described also in~\cite{srinivasavaradhan2018maximum}. The proof is added for completeness.
\begin{claim}
	\label{cl:EmbML}
Assume $\bfc\in\cC\subseteq \Sigma_q^n$ is the transmitted word and $\bfy \in \Sigma_q^{\leq n}$ is the output of the deletion channel $\del(p)$, then
$$\cD_{\ML}(\bfy) = \argmax_{\bfc\in\cC}\{ \emb(\bfc;\bfy)\}.$$
Similarly, for the insertion channel $\ins(p)$, and $\bfy \in \Sigma_q^{\geq n}$,
$$\cD_{\ML}(\bfy) = \argmax_{\bfc\in\cC}\{ \emb(\bfy;\bfc)\}.$$
\end{claim}
\begin{IEEEproof}
It can be verified that
\begin{align*}
\cD_{\ML}(\bfy) &\overset{\mathrm{(a)}}{=} \argmax_{\bfc\in\cC}\left\{\pr_\ch\{  \bfy\textmd{ rec. }| \bfc \textmd{ trans.}\}\right\}
\\ & \overset{\mathrm{(b)}}{=}\argmax_{\bfc\in\cC}\left\{ p^{|\bfc|-|\bfy|} (1-p)^{|\bfy|}  \cdot \emb(\bfc;\bfy)  \}\right\}
\\ & \overset{\mathrm{(c)}}{=}\ \argmax_{\bfc\in\cC}\{ \emb(\bfc;\bfy)\},
\end{align*}
where (a) is the definition of the ML decoder, (b) follows from Claim~\ref{cl:emb}, and (c) holds since the value $p^{|\bfc|-|\bfy|} (1-p)^{|\bfy|}$ is the same for every codeword in $\cC$. 
The proof for the insertion channel is similar. 
\begin{comment}
Note that $\cD_{\ML}(\bfy) = \bfc,$ where $\bfc\in\cC$ maximizes the probability $$\pr_{\del(p)}\{  \bfy\textmd{ rec. }| \bfc \textmd{ trans.}\} =  p^{|\bfc|-|\bfy|}\cdot \emb(\bfc;\bfy).$$
Since the value $|\bfc|-|\bfy|$ is the same for every codeword, we get that $\cD_{\ML}(\bfy) = \argmax_{\bfc\in\cC}\{ \emb(\bfc;\bfy)\}.$ A similar proof follows for the insertion channel. 
\end{comment}
\end{IEEEproof}

In case there is more than a single instance of the deletion/insertion channel, the following claim follows.
\begin{claim}
Assume $\bfc\in\cC\subseteq \Sigma_q^n$ is the transmitted word and $\bfy_1,\ldots,\bfy_t \in \Sigma_q^{\leq n}$ are the output words from $t$ instances of the deletion channel $\del(p)$, then
$$\cD_{\ML}(\bfy_1,\ldots,\bfy_t) = \argmax_{\substack{\bfc\in\cC \\ \bfc\in \sups(\bfy_1,\ldots,\bfy_t)}}\bigg\{\prod_{i=1}^t \emb(\bfc;\bfy_i)\bigg\},$$
and for the insertion channel $\ins(p)$, and  $\bfy_1,\ldots,\bfy_t \in \Sigma_q^{\geq n}$, 
$$\cD_{\ML}(\bfy_1,\ldots,\bfy_t) = \argmax_{\substack{\bfc\in\cC \\ \bfc \in \subs(\bfy_1,\ldots,\bfy_t)}}\bigg\{\prod_{i=1}^t \emb(\bfy_i;\bfc)\bigg\}.$$
\end{claim}
\begin{IEEEproof}
It holds that
\begin{align*}
\cD_{\ML}(\bfy_1,\ldots,\bfy_t)& \hspace{-0.5ex}\overset{\mathrm{(a)}}{=}\hspace{-0.5ex}  \argmax_{\bfc\in\cC} \hspace{-0.5ex} \left\{ \hspace{-0.5ex} \pr_{(\ch,t)}\{ \bfy_1,\ldots,\bfy_t \textmd{ rec.}| \bfc \textmd{ trans.} \}\right \}
\\ & \hspace{-0.5ex}\overset{\mathrm{(b)}}{=} \argmax_{\bfc\in\cC}\bigg\{\prod_{i=1}^t  {\pr_\ch\{  \bfy_i\textmd{ rec.}| \bfc \textmd{ trans.} \} }\bigg\}
\\ & \hspace{-0.5ex}\overset{\mathrm{(c)}}{=}\hspace{-2ex} \argmax_{\substack{\bfc\in\cC \\ \bfc\in \sups(\bfy_1,\ldots,\bfy_t)}} \hspace{-1ex} \bigg\{\prod_{i=1}^t  {\pr_\ch\{  \bfy_i\textmd{ rec.} | \bfc \textmd{ trans.} \} }\bigg\}
\\ &\hspace{-0.5ex} \overset{\mathrm{(d)}}{=}\hspace{-2ex}\argmax_{\substack{\bfc\in\cC \\ \bfc\in \sups(\bfy_1,\ldots,\bfy_t)} } \bigg\{\prod_{i=1}^t \emb(\bfy_i;\bfc)\bigg\},
\end{align*}
\noindent where (a) is the definition of the ML decoder, (b) holds since the channels' outputs are independent, (c) follows from the fact that the conditional probability $\pr_\ch\{  \bfy_i\textmd{ rec. }| \bfc \textmd{ trans.}\}$ equals 0 when $\bfc$ is not a supersequnce of $\bfy_i$, for $1 \le i \le t$. Lastly, (d) holds from Claim~\ref{cl:emb} and from the fact that the value $\prod_{i=1}^t p^{|\bfc|-|\bfy_i|} (1-p)^{|\bfy_i|}$ is the same for every codeword in $\cC$. The proof for the insertion channel is similar.
\begin{comment}
Note that $\cD_{\ML}(\bfy) = \bfc,$ where $\bfc\in\cC$ maximizes the probability $$\pr_{\del(p)}\{  \bfy\textmd{ rec. }| \bfc \textmd{ trans.}\} =  p^{|\bfc|-|\bfy|}\cdot \emb(\bfc;\bfy).$$
Since the value $|\bfc|-|\bfy|$ is the same for every codeword, we get that $\cD_{\ML}(\bfy) = \argmax_{\bfc\in\cC}\{ \emb(\bfc;\bfy)\}.$ A similar proof follows for the insertion channel. 
\end{comment}
\end{IEEEproof}
\begin{comment}
\begin{IEEEproof}
Every candidate to be considered in the ML decoder is a common supersequence of $\bfy_1,\ldots,\bfy_t$.  
Hence, $\cD_{\ML}(\bfy) = \bfc,$ where $\bfc\in\cC\cap \sups(\bfy_1,\ldots,\bfy_t)$ maximizes the probability
\textcolor{red}{$$\pr_{\del(p)}\{  \bfy_1,\ldots,\bfy_t \textmd{ rec. }| \bfc \textmd{ trans.}\} =  \prod_{i=1}^tp^{|\bfc|-|\bfy_i|}(1-p)^{|\bfy_i|}\cdot \emb(\bfc;\bfy_i).$$}
Since $\prod_{i=1}^tp^{|\bfc|-|\bfy_i|}$ is the same for all candidates $\bfc$, the statement holds. A similar proof holds for the insertion channel.
\end{IEEEproof}
\end{comment}

Since the deletion (insertion) channel affects the length of its output, it is possible that the length of the shortest (longest) common supersequence (subsequence) of a given channels' outputs will be smaller (larger) than the code length. If the goal is to minimize the average decoding \emph{failure} probability then clearly the decoder's output should be a codeword as there is no point in outputting a non-codeword. However, if one seeks to minimize the \emph{expected normalized distance}, then the decoder should consider non-codewords as well. Therefore, we present here the ML$^*$ decoder, which is an alternative definition of the ML decoder that takes into account non-codewords and in particular words with different length than the code length. That is, the ML$^*$ decoder does not necessarily return a codeword.

\begin{definition} \textbf{\emph{The maximum-likelihood$^*$ (ML$^*$) decoder.}}
The \emph{maximum-likelihood$^*$} (\emph{ML$^*$}) \emph{decoder} for a code $\cC$ with respect to a channel $\ch$, denoted by $\cD_{\ML^*}$, is a decoder that outputs words  
that minimize the expected normalized distance
$\perr(\ch,\cC,\cD,d)$. 
\end{definition}
For every channel output $\bfy\in\Sigma_q^*$, denote the value $\sum_{\bfc:\cD(\bfy) \neq  \bfc}  \frac{d(\cD(\bfy),\bfc)}{|\bfc|}\pr_\ch\{  \bfy\textmd{ rec. }| \bfc \textmd{ trans.}\}$ by $f_{\bfy}(\cD(\bfy))$ (and if $\cD(\bfy)$ is some arbitrary value $\bfx$ then this value is denoted by $f_{\bfy}(\bfx)$). The next claim is used to characterize the output of the ML$^*$ decoder.

\begin{claim} 
Let $\cC$ be a code.  For  any $\bfx \in \Sigma_q^*$, we have  $f_{\bfy}(\bfx) \triangleq \sum_{\substack{\bfc \in \cC}}  \frac{d(\bfx,\bfc)}{|\bfc|}\pr_\ch\{  \bfy\textmd{ rec. }| \bfc \textmd{ trans.}\}$. It holds that,  
$$\cD_{\ML^*}(\bfy)  \triangleq \argmin_{\bfx\in \Sigma_q^*} \{ f_{\bfy}(\bfx) \}.$$
  
\end{claim}
\begin{IEEEproof}
From the definition of the ML$^*$ decoder, we have that it minimizes $\perr(\ch,\cC,\cD = \cD_{\ML^*} ,d)$. Therefore, we have that, 

%\hspace{-3ex}
%\begin{small}
\begin{align*}
\perr(\ch,\cC,\cD,d) &\triangleq \frac{1}{|\cC|}\sum_{\bfc\in\cC}\perr(\bfc,d) \\
& \overset{\mathrm{(a)}}{=} \frac{1}{|\cC|} \sum_{\bfc\in\cC} \sum_{\bfy:\cD(\bfy) \neq  \bfc}  \frac{d(\cD(\bfy),\bfc)}{|\bfc|}\cdot \pr_\ch\{  \bfy\textmd{ rec. }| \bfc \textmd{ trans.}\}
\\ & \overset{\mathrm{(b)}}{=} \frac{1}{|\cC|} \sum_{\bfy\in \Sigma_q^*} \sum_{\bfc:\cD(\bfy) \neq  \bfc}  \frac{d(\cD(\bfy),\bfc)}{|\bfc|}\pr_\ch\{  \bfy\textmd{ rec. }| \bfc \textmd{ trans.}\},
\end{align*}

%\end{small}
where (a) is the definition of the expected normalized distance and in (b) we changed the order of summation, while taking into account all possible channel's outputs. This conclude the statement in the claim. 
\end{IEEEproof}

% because our code in the whole space. 
%Hence, the ML$^*$ decoder is defined to be 

%The ML* deocder for a single deletion channel is discussed in Section~\ref{sec:1-del}. When there are multiple channel outputs $\bfy_1,\ldots, \bfy_t$ the ML* decoder is defined similarly as
%$$ \cD_{\ML^*}(\bfy_1, \ldots ,\bfy_t) = \argmin_{\bfx\in \Sigma_q^*} \{ f(\cD(\bfy_1, \ldots, \bfy_t)) \}$$ 
%$ \frac{\sum_{\bfc\in\cC}\perr(\bfc,d)}{|\cC|}$, where 
%$$\perr(\bfc,d) =   \sum_{\bfy:\cD(\bfy) \neq  \bfc} \frac{d(\cD(\bfy),\bfc)}{|\bfc|}\cdot \pr_\ch\{  \bfy\textmd{ rec. }| \bfc \textmd{ trans.}\}.$$
%Note that since there is more than a single channel, when the goal is to minimize the average decoding error probability, the ML decoder does not necessarily have to output a codeword but any word that minimizes the average decoding error probability. Thus, for the rest of the paper, when discussing the average decoding error probability it is assumed that the ML decoder can output \emph{any} word and not necessarily a codeword from $\cC$. Thus, we get the following claim. 
%In case the ML decoder does not have to output a codeword but only the closest one, we get the following claim. This decoder is denoted by $\cD_{\ML}^*$. 
For the deletion and insertion channels, the ML$^*$ decoder can be characterized as follows. 
\begin{claim}\label{cl:ML*}
Assume $\bfc\in\cC\subseteq \Sigma_q^n$ is the transmitted word and $\bfy \in \Sigma_q^{\leq n}$ is the output word from the deletion channel $\del(p)$, then
$$\cD_{\ML^*}(\bfy) = \argmin_{\bfx\in \Sigma_q^*} \left\{\sum_{\bfc \in \cC}  {d_L(\bfx,\bfc) \emb(\bfc ; \bfy)}\right\},$$ 
%\argmax_{\bfx\in\sups(\bfy_1,\ldots,\bfy_t)}\bigg\{\prod_{i=1}^t \emb(\bfx;\bfy_i)\bigg\},$$
and for the insertion channel $\ins(p)$, and  $\bfy \in \Sigma_q^{\geq n}$, 
$$\cD_{\ML^*}(\bfy) = \argmin_{\bfx\in \Sigma_q^*} \left\{\sum_{\bfc \in \cC}  {d_L(\bfx,\bfc) \emb(\bfy ; \bfc)}\right\},$$ 
%$$\cD_{\ML^*}(\bfy_1,\ldots,\bfy_t) =$$ %\argmax_{\bfx\in\subs(\bfy_1,\ldots,\bfy_t)}\bigg\{\prod_{i=1}^t \emb(\bfy_i;\bfx)\bigg\}.$$
\end{claim}
\begin{IEEEproof}
The following equations hold
\begin{align*}
\cD_{\ML^*}(\bfy) &= \argmin_{\bfx\in \Sigma_q^*} \{ f_{\bfy}(\bfx) \} \\
& \overset{\mathrm{(a)}}{=} \argmin_{\bfx\in \Sigma_q^*} \left\{\sum_{\bfc:\bfx \neq  \bfc}  \frac{d_L(\bfx,\bfc)}{|\bfc|}\pr_\ch\{  \bfy\textmd{ rec. }| \bfc \textmd{ trans.}\} \right\} \\
& \overset{\mathrm{(b)}}{=} \argmin_{\bfx\in \Sigma_q^*} \left\{\sum_{\bfc:\bfx \neq  \bfc}  \frac{d_L(\bfx,\bfc)}{|\bfc|} p^{(|\bfc|-|\bfy|)}(1-p)^{|\bfy|} \emb(\bfc ; \bfy) \right\} \\
%& = \frac{p^{(|c|)}}{|\bfc|}  \argmin_{\bfx\in \Sigma_q^*} {\sum_{\bfc:\bfx \neq  \bfc}  {d_L(\bfx,\bfc) (\frac{1}{p} -1)^{|\bfy|} \emb(\bfc ; \bfy)}} \\
%& =  \argmin_{\bfx\in \Sigma_q^*} {\sum_{\bfc:\bfx \neq  \bfc}  {d_L(\bfx,\bfc) \emb(\bfc ; \bfy)}} \\
& \overset{\mathrm{(c)}}{=}  \argmin_{\bfx\in \Sigma_q^*} \left\{\sum_{\bfc \in \cC}  {d_L(\bfx,\bfc) \emb(\bfc ; \bfy)}\right\},
%& =  \argmin_{\bfx\in \Sigma_q^*} {\min_{m: |\bfx|=m}\sum_{\bfc \in \cC}  {d_L(\bfx,\bfc) \emb(\bfc ; \bfy)}}
\end{align*}
where (a) follows from the definition of the ML$^*$ decoder, (b) follows from Claim~\ref{cl:emb}, and (c) holds since for every $\bfx\in\Sigma_q^*$, the values of $|\bfc|$, $|\bfy|$, and $p$ are fixed. The proof for the insertion channel is similar. 
\end{IEEEproof}
The definition of the ML$^*$ decoder can be easily generalized to the case of multiple channel outputs.
Recall that the definition of the expected normalized distance $\perr(\ch,t,\cC,\cD,d)$ for multiple channels states that 

\hspace{-4ex}
\begin{small}
\begin{align*}
\perr(\ch,t,\cC,\cD,d)  %= \frac{1}{|\cC|}\sum_{\bfc\in\cC}\perr(\bfc,d) \\
& = \frac{1}{|\cC|} \sum_{\bfc\in\cC} \sum_{\substack{\bfy_1, \ldots, \bfy_t \in \Sigma_q^*}} \hspace{-2.5ex} \frac{d(\cD(\bfy_1, \ldots, \bfy_t ),\bfc)}{|\bfc|}\cdot \pr_\ch\{  \bfy_1, \ldots, \bfy_t \textmd{ rec. }| \bfc \textmd{ trans.}\}
\\ & = \frac{1}{|\cC|}\hspace{-0.5ex} \sum_{\substack{\bfy_1, \ldots, \bfy_t \in \Sigma_q^*}} \sum_{\bfc:\cD(\bfy_1, \ldots, \bfy_t ) \neq  \bfc}  \hspace{-5ex}\frac{d(\cD(\bfy_1, \ldots, \bfy_t ),\bfc)}{|\bfc|} \hspace{-0.75ex} \prod_{i=1}^t \pr_\ch\{  \bfy_i\textmd{ rec. }| \bfc \textmd{ trans.}\}.
\end{align*}
\end{small} \hspace{-1.8ex} %where (a) is the definition of the average decoding error probability and in (b) we switched the summation order, while taking into account all possible channel's outputs. % because our code in the whole space. 
%For every $\bfy\in\Sigma_q^*$, denote the value $\sum_{\bfc:\cD(\bfy) \neq  \bfc}  \frac{d(\cD(\bfy),\bfc)}{|\bfc|}\pr_\ch\{  \bfy\textmd{ rec. }| \bfc \textmd{ trans.}\}$ by $f_{\bfy}(\cD(\bfy))$ and if $\cD(\bfy)$ is some arbitrary value $\bfx$ then this value is denoted by $f_{\bfy}(\bfx)$. Hence, the ML$^*$ decoder is defined to be 
%$$\cD_{\ML^*}(\bfy) = \argmin_{\bfx\in \Sigma_q^*} \{ f_{\bfy}(\bfx) \}.$$ 
In this case, we let
\begin{align*}
 f_{\bfy_1, \ldots, \bfy_t }(\cD(\bfy_1, \ldots, \bfy_t)) 
 & \triangleq  \hspace{-2ex}\sum_{\bfc:\cD(\bfy_1, \ldots, \bfy_t) \neq  \bfc}  \frac{d(\cD(\bfy_1, \ldots, \bfy_t),\bfc)}{|\bfc|} \prod_{i=1}^t \pr_\ch\{  \bfy_i\textmd{ rec. }| \bfc \textmd{ trans.}\}, 
\end{align*}
where $\bfy_1, \ldots, \bfy_t$ are the $t$ channel outputs. Then, the ML$^*$ decoder is defined to be 
$$\cD_{\ML^*}(\bfy_1, \ldots, \bfy_t) \triangleq \argmin_{\bfx\in \Sigma_q^*} \{ f_{\bfy_1, \ldots, \bfy_t}(\bfx) \}.$$ 
The following claim solves this setup for the case of deletions or insertions. 
\begin{claim}\label{cl:ML*multiple}
Assume $\bfc\in\cC\subseteq \Sigma_q^n$ is the transmitted word and $\bfy_1,\ldots,\bfy_t \in \Sigma_q^{\leq n}$ are the output words from $t$ deletion channels $\del(p)$. Then,
\begin{align*}
\cD_{\ML^*}(\bfy_1, \ldots, \bfy_t) 
&= \argmin_{\bfx\in \Sigma_q^*} \left\{\sum_{\substack {\bfc \in \cC \\ \bfc \in \sups (\bfy_1, \ldots, \bfy_t )}}  {d_L(\bfx,\bfc) \prod_{i=1}^t  \emb(\bfc ; \bfy_i)}\right\}
\end{align*}
%\argmax_{\bfx\in\sups(\bfy_1,\ldots,\bfy_t)}\bigg\{\prod_{i=1}^t \emb(\bfx;\bfy_i)\bigg\},$$
and for the insertion channel $\ins(p)$, for  $\bfy_1, \ldots, \bfy_t \in \Sigma_q^{\geq n}$, 
\begin{align*}
\cD_{\ML^*}(\bfy_1, \ldots, \bfy_t)
  = \argmin_{\bfx\in \Sigma_q^*} \left\{\sum_{\substack {\bfc \in \cC \\ \bfc \in \subs (\bfy_1, \ldots, \bfy_t )}}  {d_L(\bfx,\bfc) \prod_{i=1}^t \emb(\bfy_i ; \bfc)}\right\} .
\end{align*}
%$$\cD_{\ML^*}(\bfy_1,\ldots,\bfy_t) =$$ %\argmax_{\bfx\in\subs(\bfy_1,\ldots,\bfy_t)}\bigg\{\prod_{i=1}^t \emb(\bfy_i;\bfx)\bigg\}.$$
\end{claim}
\begin{IEEEproof}
The following equations hold
\begin{align*}
\cD_{\ML^*}(\bfy_1, \ldots, \bfy_t)  &= \argmin_{\bfx\in \Sigma_q^*} \{ f_{\bfy_1, \ldots, \bfy_t}(\bfx) \} \\
& \overset{\mathrm{(a)}}{=} \argmin_{\bfx\in \Sigma_q^*} \left\{\sum_{\bfc:\bfx \neq  \bfc}  \frac{d_L(\bfx,\bfc)}{|\bfc|} \prod_{i=1}^t \pr_\ch\{  \bfy_i \textmd{ rec. }| \bfc \textmd{ trans.}\} \right\} \\
& \overset{\mathrm{(b)}}{=} \argmin_{\bfx\in \Sigma_q^*}\hspace{-0.5ex} \left\{\hspace{-0.5ex}\sum_{\bfc:\bfx \neq  \bfc}  \frac{d_L(\bfx,\bfc)}{|\bfc|} \hspace{-0.5ex}\prod_{i=1}^t p^{(|\bfc|-|\bfy_i|)}(1-p)^{|\bfy_i|} \emb(\bfc ; \bfy_i)\hspace{-0.5ex} \right\} \\
%& = \frac{p^{(|c|)}}{|\bfc|}  \argmin_{\bfx\in \Sigma_q^*} {\sum_{\bfc:\bfx \neq  \bfc}  {d_L(\bfx,\bfc) (\frac{1}{p} -1)^{|\bfy|} \emb(\bfc ; \bfy)}} \\
%& =  \argmin_{\bfx\in \Sigma_q^*} {\sum_{\bfc:\bfx \neq  \bfc}  {d_L(\bfx,\bfc) \emb(\bfc ; \bfy)}} \\
& \overset{\mathrm{(c)}}{=}  \argmin_{\bfx\in \Sigma_q^*} \left\{\sum_{\bfc \in \cC}  {d_L(\bfx,\bfc) \prod_{i=1}^t \emb(\bfc ; \bfy_i)}\right\}
\\ & \overset{\mathrm{(d)}}{=}  \argmin_{\bfx\in \Sigma_q^*} \left\{\sum_{\substack { \bfc \in \cC \\ \bfc \in \sups (\bfy_1, \ldots, \bfy_t )}}  {d_L(\bfx,\bfc) \prod_{i=1}^t \emb(\bfc ; \bfy_i)}\right\},
%& =  \argmin_{\bfx\in \Sigma_q^*} {\min_{m: |\bfx|=m}\sum_{\bfc \in \cC}  {d_L(\bfx,\bfc) \emb(\bfc ; \bfy)}}
\end{align*}
where (a) follows from the definition of the ML$^*$ decoder, (b) follows from Claim~\ref{cl:emb}, (c) holds since for every $\bfx\in\Sigma_q^*$, the values of $|\bfc|$, $|\bfy_i|$, and $p$ are fixed, and (d) holds since $\prod_{i=1}^t \emb(\bfc; \bfy_i) = 0$ for every $\bfc \in \cC$ such that $\bfc \notin \sups (\bfy_1, \ldots, \bfy_t)$. The proof for the insertion channel is similar. 
\end{IEEEproof}

In the rest of the paper, we primarily focus on two versions of the deletion channel, the probabilistic channel $\del(p)$, and the combinatorial channel $k\textrm{-}\mathsf{Del}$, both defined in Section~\ref{sec:defs}. The $k\textrm{-}\mathsf{Del}$ channel is studied in Section~\ref{sec:1-del} and Section~\ref{sec:2-del}, where we study, analyze, and characterize the ML$^*$ decoder for $k=1$ and $k=2$. In Section~\ref{sec:two deletions}, we focus on the deletion channel $\del(p)$ and study the case of two instances of this channel. While computing the ML$^*$ decoder, in this case, can be computationally impractical (see Section~\ref{sec:two deletions} for details), we instead analyze a degraded version of this decoder and study its expected normalized distance.

\section{The $1$-Deletion Channel}\label{sec:1-del}
%\subsection{The 1-Deletion Channel}\label{subsec:one-del}
In the following two sections, we consider the $k$-deletion channel. Remember that in the $k$-deletion channel, which was denoted by $k\textrm{-}\mathsf{Del}$, exactly $k$ symbols are deleted from the transmitted word. The $k$ symbols are selected uniformly at random out of the $\binom{n}{k}$ symbol positions, where $n$ is the length of the transmitted word. This channel was studied in~\cite{TFA2017, DBLP:journals/corr/abs-1802-00703}.  As mentioned earlier, given a word $\bfx$, its radius-$r$ deletion ball, denoted by $D_r(\bfx)$, is defined as the set of all words that can be obtained from $\bfx$ by deleting exactly $r$ symbols. Note that the set $D_r(\bfx)$ consists of all words of length $|\bfx| - r$ that are subsequences of the word $\bfx$. Hence, given a word $\bfx$, the set of all possible outputs of the $k$-deletion channel of a word $\bfx$ is  $D_k(\bfx)$.

Recall that, the embedding number of $\bfy$ in $\bfx$, denoted by $\emb (\bfx ;  \bfy )$, is defined as the number of different ways in which $\bfy$  can be received as a subsequence of $\bfx$.   Since the $k$ deleted symbols are selected randomly out of the $\binom{n}{k}$ options, the conditional probability of the $k$-deletion channel is, 
$$\pr_{k\textrm{-}\mathsf{Del}}\{  \bfy\textmd{ rec. }| \bfx \textmd{ trans.}\}=\dfrac{\emb (\bfx; \bfy )}{\binom{n}{k}}.$$
\begin{example}
Assume the word $\bfx=01001$ is transmitted through the $k$-deletion channel, for $k=2$. Then, the set of all possible outputs is the radius-$2$ deletion ball of $\bfx$, which is $D_2(\bfx) = \{ 000, 001, 010,  011, 100, 101 \}$. We denote the word $000$ by $\bfy_1$, and $001$ by $\bfy_2$. Note that $\emb (\bfx; \bfy_1) = 1$ and  $\emb (\bfx; \bfy_2) = 3$, and hence, $\pr_{2\textrm{-}\mathsf{Del}}\{  \bfy_1\textmd{ rec. }| \bfx \textmd{ trans.}\}= \dfrac{1}{\binom{6}{2}}, \pr_{2\textrm{-}\mathsf{Del}}\{  \bfy_2\textmd{ rec. }| \bfx \textmd{ trans.}\}=\dfrac{3}{\binom{6}{2}}.$
\end{example}

In~\cite{DBLP:journals/corr/abs-1802-00703}, it was shown that for any $\bfy \in \Sigma_2^{n-k}$ it holds that $\sum_{\bfx\in \Sigma_2^n} \emb(\bfy;\bfx) = \binom{n}{k}2^{k}.$ This implies that any channel output $\bfy \in \Sigma_2^{n-k}$, obtained from the channel, has the same probability which equals to $\frac{1}{2^{n-k}}$, as shown in the next lemma.
\begin{lemma}
    Let $\cC = \Sigma_2^n$ and $\ch = k\textrm{-}\mathsf{Del}$. For any channel output $\bfy\in\Sigma_2^{n-k}$, it holds that,
    $$\pr_\ch\{  \bfy\textmd{ rec.} \} = \frac{1}{2^{n-k}}. $$
\end{lemma}
\begin{IEEEproof}
From~\cite{DBLP:journals/corr/abs-1802-00703}, it is known that $\sum_{\bfc\in \Sigma_2^n} \emb(\bfc;\bfy) = \binom{n}{k}2^{k}$. Therefore, we have that
    \begin{align*}
    \pr_\ch\{  \bfy\textmd{ rec.} \} &= \sum_{\bfc \in \cC} \pr_\ch\{  \bfy\textmd{ rec.} | \bfc \textmd{ trans.} \} \pr_\ch\{ \bfc \textmd{ trans. }\} 
    \\& = \frac{1}{2^n} \sum_{\bfc \in \cC} \pr_\ch\{  \bfy\textmd{ rec.} | \bfc \textmd{   trans.} \} 
    \\& =  \frac{1}{2^n} \sum_{\bfx \in \Sigma_2^n} \frac{\emb(\bfc;\bfy)}{\binom{n}{k}}   =  \frac{1}{2^n}  \frac{\binom{n}{k}2^{k}}{\binom{n}{k}}  =  \frac{1}{2^{n-k}}.
    \end{align*}
\end{IEEEproof}

In the rest of the section the $1$-deletion channel which deletes one symbol randomly is considered. Note that this is a special case of the $k$-deletion channel where $k=1$. Given a single-deletion-correcting code, any channel output can be easily decoded, and therefore for the rest of this section we assume that the given code is not a single-deletion-correcting code. We start by examining two types of decoders for this channel which are defined next.
\begin{definition} \textbf{\emph{The embedding number decoder.}}
The \emph{embedding number decoder}, denoted by $\cD_{EN}$, is a decoder that for any channel output $\bfy$ returns the codeword $\cD_{EN}(\bfy)$ which is a codeword in the code $\cC$ that maximizes the embedding number of $\bfy$ in $\cD_{EN}(\bfy)$. That is, \vspace{-1ex}
$$\cD_{EN}(\bfy) \triangleq \argmax_{\bfc\in\cC}\{ \emb(\bfc;\bfy)\},\vspace{-1ex}$$
where, if there is more than one such a codeword, the decoder chooses one of them arbitrarily. 
\end{definition}

\begin{definition} \textbf{\emph{The lazy decoder.}}
The \emph{lazy decoder}, denoted by $\cD_{Lazy}$, is a decoder that for any channel output $\bfy$ simply returns $\bfy$ as its output, i.e., $\cD_{Lazy}(\bfy) \triangleq \bfy$. 
\end{definition}
%Note that the lazy decoder does not return a codeword. Additionally, $d_L(\cD_{Lazy}(\bfy),\bfc)=1$ since $\bfy \in D_1(\bfc)$ and hence, the expected normalized distance of the lazy decoder is $\frac{1}{n}$, when $n$ is the code length (see Lemma~\ref{Lem:LazyError}).

\subsection{The ML$^*$ Decoder.} 
In the main result of this section, presented in Theorem~\ref{th:lazy_optimal}, we prove for $\ch=1\textrm{-}\mathsf{Del}$ and $\cC=\Sigma_2^n$, that $\cD_{Lazy}$ performs at least as good as any other decoder, and hence ${\cD_{Lazy} = \cD_{\ML^*}}$.% is the ML$^*$ decoder for the 1-deletion channel. 

For the rest of this section it is assumed that $\cC \subseteq \Sigma_2^n$ and $\ch = 1\textrm{-}\mathsf{Del}$. Under this setup, the Levenshtein distance between the lazy decoder's output $\bfy$ and the transmitted word $\bfc$ is always $d_L(\bfy,\bfc)=1$, since $\bfy \in D_1(\bfc)$. Hence, the following lemma follows immediately. 
\begin{lemma}\label{Lem:LazyError}
The expected normalized distance of the lazy decoder $\cD_{\mathrm{Lazy}}$ under the 1-deletion channel $1\textrm{-}\mathsf{Del}$ is
	\begin{equation*}
	P_{\mathrm{err}}(1\textrm{-}\mathsf{Del},\cC,\cD_{\mathrm{Lazy}},d_L) = \frac{1}{n}.
	\end{equation*}
\end{lemma}

\begin{IEEEproof}
	The expected normalized distance of the lazy decoder for each codeword $\bfc$ is calculated as follows.
	\begin{align*}
	\label{Eq:AvgErr}
	P_{\mathrm{err}}(\bfc,d_L) & = \sum_{\bfy: \cD_{\mathrm{Lazy}}(\bfy)\neq\bfc} \frac{d_L\left( \cD_{\mathrm{Lazy}}(\bfy),\bfc \right) }{|\bfc|} p(\bfy | \bfc) & \\ 
	& = \sum_{\bfy\in D_1(\bfc)} \frac{1}{n} p(\bfy | \bfc) = \frac{1}{n}. &
	\end{align*}
	Since this is true for every $\bfc \in \cC$, we get that
	\begin{equation*}
	P_{\mathrm{err}}(1\textrm{-}\mathsf{Del},\cC,\cD_{\mathrm{Lazy}},d_L) = \frac{1}{n}\cdot |\cC| \cdot \frac{1}{|\cC|} = \frac{1}{n}.
	\end{equation*}
\end{IEEEproof}

We can now show the main result of this section, which claims that the lazy decoder is preferable, with respect to the expected normalized distance, over any decoder that outputs a word of the same length as its input.
\begin{theorem} \label{th:lazy_optimal}
    Let $\cD$ be a decoder and let $\cC = \Sigma_2^n$. Then, it holds that, 
\begin{equation*}
P_{\mathrm{err}}(1\textrm{-}\mathsf{Del},\cC,\cD,d_L) \ge  P_{\mathrm{err}}(1\textrm{-}\mathsf{Del},\cC,\cD_{\mathrm{Lazy}},d_L) = \frac{1}{n}.
\end{equation*}
\end{theorem}

\begin{IEEEproof}
Recall the definition of the expected normalized distance, where $\ch = 1\textrm{-}\mathsf{Del}$, $\cC=\Sigma_2^n$, and $d=d_L$. 
\begin{align*}
\perr(\ch,\cC,\cD,d) & \triangleq \frac{1}{|\cC|}\sum_{\bfc\in\cC}\sum_{\bfy:\cD(\bfy) \neq  \bfc} \frac{d(\cD(\bfy),\bfc)}{|\bfc|}\cdot \pr_\ch\{  \bfy\textmd{ rec. }| \bfc \textmd{ trans.}\}
\\ & = \frac{1}{|\cC|}\sum_{\bfc\in \Sigma_2^n}\sum_{\bfy:\cD(\bfy) \neq  \bfc} \frac{d(\cD(\bfy),\bfc)}{|\bfc|}\cdot \pr_\ch\{  \bfy\textmd{ rec. }| \bfc \textmd{ trans.}\}
\\ & = \frac{1}{n  |\cC|}\sum_{\bfy \in \Sigma_2^{n-1}}\sum_{ \bfc\in \Sigma_2^n} d(\cD(\bfy),\bfc)\cdot \pr_\ch\{  \bfy\textmd{ rec. }| \bfc \textmd{ trans.}\}
\\ & {=} \frac{1}{n  |\cC|} \left( \sum_{\substack{\bfy \in \Sigma_2^{n-1}, \\ |\cD(\bfy)| \ne n}}\sum_{ \bfc\in \Sigma_2^n} d(\cD(\bfy),\bfc)\cdot \pr_\ch\{  \bfy\textmd{ rec. }| \bfc \textmd{ trans.}\} + \sum_{\substack{\bfy \in \Sigma_2^{n-1}, \\ |\cD(\bfy)| = n}}\sum_{ \bfc\in \Sigma_2^n} d(\cD(\bfy),\bfc)\cdot \pr_\ch\{  \bfy\textmd{ rec. }| \bfc \textmd{ trans.}\} \right).
\end{align*}
Let us define $K \triangleq \{ \bfy: |\cD(\bfy)| = n\}$. 
We start by deriving a lower bound on $\sum_{\substack{\bfy \notin K}}\sum_{ \bfc\in \Sigma_2^n} d(\cD(\bfy),\bfc)\cdot \pr_\ch\{  \bfy\textmd{ rec. }| \bfc \textmd{ trans.}\}$. 
Observe that for any $\bfy \in \Sigma_2^{n-1} \setminus K$, we have that, 
\begin{align*} 
\sum_{\bfc \in \Sigma_2^n} \pr_\ch\{  \bfy\textmd{ rec. }| \bfc \textmd{ trans.} \} &= \sum_{\bfc \in \Sigma_2^n} \frac{\pr_\ch \{ \bfc \text{ tran. and  } \bfy \text{ rec. }\}}{\pr_\ch\{ \bfc \text{ trans. }\}} \\&= \sum_{\bfc \in \Sigma_2^n} \frac{\pr_\ch \{ \bfc \text{ tran. and  } \bfy \text{ rec. }\}}{1/2^n}
\\&=2^n \sum_{\bfc \in \Sigma_2^n} \pr_\ch \{ \bfc \text{ tran. and  } \bfy \text{ rec.}\} \\&= 2^n \pr_\ch \{  \bfy \textmd{ rec.} \} \\&=  \frac{2^n}{2^{n-1}} = 2. 
\end{align*}
Therefore, since $d(\cD(\bfy),\bfc)>1$, we get that, 
\begin{align} \label{ineq:ynotinK}
%\sum_{\substack{\bfy \notin K}}\sum_{ \bfc\in \Sigma_2^n} d(\cD(\bfy),\bfc)\cdot \pr_\ch\{  \bfy\textmd{ rec. }| \bfc \textmd{ trans.}\} \ge \sum_{\substack{\bfy \notin K}} 1 \cdot \frac{1}{2^{n-1}}  =  \frac{2^{n-1} - |K| }{2^{n-1}}.
%\\ 
\sum_{\substack{\bfy \notin K}}\sum_{ \bfc\in \Sigma_2^n} d(\cD(\bfy),\bfc)\cdot \pr_\ch\{  \bfy\textmd{ rec. }| \bfc \textmd{ trans.}\} \ge \sum_{\substack{\bfy \notin K}} 1 \cdot 2  =  2\left( 2^{n-1} - |K|  \right).
\end{align}
Next, we consider channel outputs $\bfy \in K$. Note that if  $\cD(\bfy)$ is not the transmitted word, its Levenshtein distance is at least $2$. This is due to the fact that at least one insertion and one deletion are required to transform $\cD(\bfy)$ into the transmitted word. On the other hand, if $\cD(\bfy)$ is the transmitted word, then the Levenshtein distance is $0$. Furthermore, we note that, $$\pr_\ch\{ \cD(\bfy) \text{ trans. and } \bfy \text{ rec.}  \} = \pr_\ch\{ \cD(\bfy) \text{ trans.} \} \pr_\ch\{ \bfy \text{ rec. } | \cD(\bfy) \text{ trans.}  \} \le \pr_\ch\{\cD(\bfy) \text{ trans.} \} = \frac{1}{2^n}.$$
This implies that $$\sum_{\bfc \ne \cD(\bfy)} \pr_\ch\{ \bfc \text{ trans. and } \bfy \text{ rec.}  \} = \pr_\ch \{ \bfy \text{ trans.} \}  - \pr_\ch \{ \cD(\bfy) \text{ trans. and } \bfy \text{ rec.} \} \ge \frac{1}{2^{n-1}} - \frac{1}{2^n}=\frac{1}{2^n}.$$
Thus, 
\begin{align}
    \notag \sum_{\substack{\bfy \in K}}\sum_{ \bfc\in \Sigma_2^n} d(\cD(\bfy),\bfc)\cdot \pr_\ch\{  \bfy\textmd{ rec. }| \bfc \textmd{ trans.}\}
    & = \sum_{\substack{\bfy \in K}}\sum_{ \bfc\in \Sigma_2^n} d(\cD(\bfy),\bfc)\cdot \frac{\pr_\ch\{ \bfc \text{ trans. and } \bfy \text{ rec. } \} }{\pr_\ch\{\bfc \text{ trans. } \} }
    \notag \\& = 2^n \sum_{\substack{\bfy \in K}}\sum_{ \bfc\in \Sigma_2^n} d(\cD(\bfy),\bfc)\cdot \pr_\ch\{ \bfc \text{ trans. and } \bfy \text{ rec. } \} 
    \notag \\& \ge 2^{n+1}  \sum_{\substack{\bfy \in K}}\sum_{ \bfc\ne \cD(\bfy)}  \pr_\ch\{\bfc \text{ trans. and } \bfy \text{ rec. } \}
    \notag \\& \label{ineq:yinK} \ge  2^{n+1}  \sum_{\substack{\bfy \in K}} \frac{1}{2^{n}} = 2^{n+1} \frac{|K|}{2^{n}} = 2|K|.
\end{align}
Combining the results in (\ref{ineq:ynotinK}) and in (\ref{ineq:yinK}), we get that, 

\begin{align*}
\perr(\ch,\cC,\cD,d) &  = \frac{1}{n  |\cC|} \left( \sum_{\substack{\bfy: \cD(\bfy) \neq  \bfc, \\ |\cD(\bfy)| \ne n}}\sum_{ \bfc\in \Sigma_2^n} d(\cD(\bfy),\bfc)\cdot \pr_\ch\{  \bfy\textmd{ rec. }| \bfc \textmd{ trans.}\} + \sum_{\substack{\bfy: \cD(\bfy) \neq  \bfc, \\ |\cD(\bfy)| = n}}\sum_{ \bfc\in \Sigma_2^n} d(\cD(\bfy),\bfc)\cdot \pr_\ch\{  \bfy\textmd{ rec. }| \bfc \textmd{ trans.}\} \right)
\\ & \ge  \frac{1}{n |\cC|} \left( 2^n-2|K|+2|K| \right) = \frac{2^n}{n |\cC|} = \frac{1}{n}.
\end{align*}

\end{IEEEproof}

\subsection{The Embedding Number Decoder}
In this section, we characterize and study the performance of the embedding number decoder.  Our main result in this section is  Theorem~\ref{Lem:ENover_n}, which states that the embedding number decoder minimizes the expected normalized distance amongst all other decoders that output words of the code's length. In the previous section, in Theorem~\ref{th:lazy_optimal}, it was shown that the lazy decoder optimizes the expected normalized distance. However, this decoder outputs words which are not of the code's length. Therefore, in this section, we complete these results and show optimality for the case where the decoder output is of the code's length. 
Next, it is shown that a decoder that prolongs an arbitrary run of maximal length within the decoder's input word (i.e., the channel output) is equivalent to the embedding number decoder.
\begin{lemma}
	\label{Lem:MaxEmb}
	Given $\bfy \in \Sigma_2^{n-1}$, the word $\widehat{\bfx} \in \Sigma_2^{n}$ obtained by prolonging a run of maximal length in $\bfy$ satisfies
	\begin{equation*}
	\emb(\widehat{\bfx} ; \bfy) = \max_{\bfx \in \Sigma_2^{n}} \{\emb(\bfx ; \bfy)\}.
	\end{equation*}
\end{lemma}
\begin{IEEEproof}
	Let $\bfy$ be a word with $n_r$ runs of lengths $r_1,r_2,\dots,r_{n_r}$. Let $\bfx_0$ be any word obtained from $\bfy$ by creating a new run of length one, and so $\emb(\bfx_0 ; \bfy) = 1$. Let $\bfx_i , \ 1\leq i \leq n_r$ be the word obtained from $\bfy$ by prolonging the $i$-th run by one, and so $\emb(\bfx_i ; \bfy) = r_i + 1$. Hence, it follows that 
	\begin{equation*}
	\argmax_{0\leq i \leq n_r} \{\emb(\bfx_i ; \bfy)\} = \argmax_{0\leq i \leq n_r} \{r_i+1\},
	\end{equation*}
	where by definition $r_0\triangleq0$. It should be noted that the union of the words $\bfx_0$ and $\bfx_i$, $1 \le i \le n_r$ comprises all of the words that $\bfy$ can be obtained from, by introducing one deletion, and hence are the only words of length $n-1$ with an embedding number larger than $0$.\\
\end{IEEEproof}

According to Lemma~\ref{Lem:MaxEmb}, we can arbitrarily choose the decoder that prolongs the first run of maximal length as the embedding number decoder.
\begin{definition} \textbf{\emph{Equivalent decoder to the embedding number decoder.}}
The {embedding number decoder} $\cD_{\mathrm{EN}}$ prolongs the first run of maximal length in $\bfy$ by one symbol.
A decoder $\cD$ that prolongs one of the runs of maximal length in $\bfy$ by one symbol is said to be \textit{equivalent} to the embedding number decoder, and is denoted by $\cD \equiv \cD_{\mathrm{EN}}$.
\end{definition}
The rest of this section will focus on the case for which $\cC = \Sigma_2^n$. The following lemmas will be stated for the embedding number decoder for the simplicity of the proofs, but unless stated otherwise they hold for any decoder $\cD$ for which $\cD \equiv \cD_{\mathrm{EN}}$.
\begin{lemma}
	\label{Lem:ENerror1}
	For every codeword $\bfc \in \cC$, the embedding number decoder satisfies
	\begin{equation*}
	P_{\mathrm{err}}(\bfc,d_L) = \frac{2}{n} \cdot  \sum_{\substack{\bfy\in D_1(\bfc) \\ \bfc \ne \cD_{\mathrm{EN}(\bfy)}}} \frac{\emb(\bfc;\bfy) }{n}. %\cdot \mathbb{I}\{\cD_{\mathrm{EN}}(\bfy)\neq\bfc\}.
	\end{equation*}
\end{lemma}
\begin{IEEEproof}
	Let $\bfc \in \cC$ be a codeword and let $\bfy \in D_1(\bfc)$ be a channel output such that $\cD_{\mathrm{EN}}(\bfy) \neq \bfc$. Since $\cD_{\mathrm{EN}}(\bfy)$ can be obtained from a word in $D_1(\bfc)$ by one insertion, it follows that $d_L(\cD_{\mathrm{EN}}(\bfy),\bfc)=2$. Thus,
	\begin{align*}
	\label{Eq:AvgErr}
	P_{\mathrm{err}}(\bfc,d_L)   & = \sum_{\bfy: \cD_{\mathrm{EN}}(\bfy)\neq\bfc} \frac{d_L\left( \cD_{\mathrm{EN}}(\bfy),\bfc \right) }{|\bfc|} p(\bfy | \bfc) \\
	& = \frac{2}{n} \sum_{\bfy\in D_1(\bfc)} p(\bfy | \bfc) \cdot \mathbb{I}\{\cD_{\mathrm{EN}}(\bfy)\neq\bfc\} \\
	& = \frac{2}{n} \cdot  \sum_{\substack{\bfy\in D_1(\bfc) \\ \bfc \ne \cD_{\mathrm{EN}(\bfy)}}} \frac{\emb(\bfc;\bfy) }{n}.
		\end{align*}
\end{IEEEproof}

For $\bfy\in D_1(\bfc)$, we have that $\cD_{\mathrm{EN}}(\bfy)=\bfc$ if and only if the deletion occurred within the run corresponding to the first run of maximal length in $\bfy$. Hence, the embedding number decoder will fail at least for any deletion occurring outside of the first run of maximal length in $\bfc$. This observation will be used in the proof of Lemma~\ref{Lem:ENerror2}. Before presenting this lemma, one more definition is introduced. For a word $\bfx\in\Sigma_2^n$, we denote by $\tau(\bfx)$ the length of its maximal run. For example $\tau(00111010) = 3$ and $\tau(01010101) =1$. For a code $\cC\subseteq \Sigma_2^n$, we denote by $\tau(\cC)$ the average length of the maximal runs of its codewords. That is, 
$$\tau(\cC) = \frac{\sum_{\bfc\in\cC}\tau(\bfc)}{|\cC|}.$$
Furthermore, if $N(r)$, for $1\leq r\leq n$ denotes the number of codewords in $\cC$ in which the length of their maximal run is $r$, then $\tau(\cC) = \frac{\sum_{r=1}^{n}r\cdot N(r)}{|\cC|}$. We are now ready to present a lower bound on the expected normalized distance of the embedding number decoder. 
% on the the proof of Lemma~\ref{Lem:ENerror2}.
\begin{lemma}
	\label{Lem:ENerror2}
	The expected normalized distance of the embedding number decoder $\cD_{\mathrm{EN}}$ satisfies
	\begin{equation*}
	P_{\mathrm{err}}(1\textrm{-}\mathsf{Del},\cC,\cD_{\mathrm{EN}},d_L) \geq \frac{2}{n} \cdot \left(1 - \frac{\tau(\cC)}{n} \right).
	\end{equation*}
%	where $\mathbb{E}[r]$ is the expectation value of the length of the longest run in the codewords $\bfc\in\cC$.
\end{lemma}

\begin{IEEEproof}
	Let $\mathcal{C}_r \subseteq \cC$ be the subset of codewords with maximal run length of $r$, and let its size be denoted by $N(r)$. For any codeword $\bfc$, since the decoder $\cD_{EN}$ prolongs the first run of maximal length, any deletion error that occurs outside of the first run of maximal length will result in a decoding failure. Since the sum 
	\begin{equation*}
	\sum_{\substack{\bfy\in D_1(\bfc) \\ \bfc \ne \cD_{\mathrm{EN}}(\bfy) }} \frac{\emb(\bfc;\bfy)}{n} %\cdot \mathbb{I}\{\cD_{\mathrm{EN}}(\bfy)\neq\bfc\}
	\end{equation*} is equivalent to counting the indices in $\bfc$ in which a deletion will result in a decoding failure (and normalizing it by $n$), using 
	Lemma~\ref{Lem:ENerror1} we get that for every $ \bfc \in \mathcal{C}_r$,
	\begin{equation*}
	P_{\mathrm{err}}(\bfc,d_L) \geq \frac{2}{n}\cdot \frac{n-r}{n},
	\end{equation*}
	and the expected normalized distance becomes
	\begin{align*}
	P_{\mathrm{err}}(1\textrm{-}\mathsf{Del},\cC,\cD_{\mathrm{EN}},d_L) &= \frac{1}{|\mathcal{C}|} \sum_{\bfc\in \cC}  P_{\mathrm{err}}(\bfc,d_L)\\  
	& = \frac{1}{|\mathcal{C}|} \sum_{r=1}^{n} \sum_{\bfc\in \mathcal{C}_r}  P_{\mathrm{err}}(\bfc,d_L) \geq \frac{1}{|\mathcal{C}|} \sum_{r=1}^{n}\sum_{\bfc\in \mathcal{C}_r} \frac{2}{n}\cdot \frac{n-r}{n} \\
	& = \frac{1}{|\mathcal{C}|} \frac{2}{n}  \sum_{r=1}^{n} N(r) \left( 1- \frac{r}{n} \right) = \frac{2}{n} \left( \frac{\sum_{r=1}^{n} N(r)}{|\cC|} -  \frac{\sum_{r=1}^{n} r N(r)}{n |\cC|}\right) \\ 
	& = \frac{2}{n}  \left( 1 - \frac{1}{n}\frac{\sum_{r=1}^{n} r \cdot N(r)}{|\mathcal{C}|} \right) = \frac{2}{n} \cdot \left(1 - \frac{\tau(\cC)}{n} \right).
	\end{align*}
\end{IEEEproof}

For the special case of $\cC = \Sigma_2^n$, the next claim is proved in Appendix~\ref{app:A}.
\begin{claim}\label{cl:tau}
For all $n\geq 1$ it holds that $\tau(\Sigma_2^n) \leq 2\log_2(n)$.
\end{claim}
  
We will now show that the embedding number decoder is preferable over any other decoder that outputs a word of the original codeword length.

\begin{theorem} 
	\label{Lem:ENover_n}
	Let $\cD : \Sigma_2^{n-1} \rightarrow \Sigma_2^{n}$ be a general decoder that prolongs the input length by one. It follows that
	\begin{equation}
	P_{\mathrm{err}}(1\textrm{-}\mathsf{Del},\cC,\cD,d_L) \geq P_{\mathrm{err}}(1\textrm{-}\mathsf{Del},\cC,\cD_{\mathrm{EN}},d_L).
	\end{equation}
	and equality is obtained if and only if $\cD \equiv \cD_{\mathrm{EN}}$.
\end{theorem}
\begin{IEEEproof}
	We have the following sequence of equalities and inequalities 
	\begin{align*}
	P_{\mathrm{err}}( 1\textrm{-}\mathsf{Del},\cC,\cD,d_L) &= \frac{1}{|\mathcal{C}|} \sum_{\bfc \in \cC} \sum_{\bfy: \cD(\bfy)\neq\bfc} \frac{d_L(\cD(\bfy),\bfc)}{|\bfc|} p(\bfy|\bfc) \\
	& \overset{\mathrm{(a)}}{=} \frac{1}{|\mathcal{C}|} \sum_{\bfy \in \Sigma_2^{n-1}} \sum_{\bfc \in I_1(\bfy)} \frac{d_L(\cD(\bfy),\bfc)}{|\bfc|} p(\bfy|\bfc) \\
	& \overset{\mathrm{(b)}}{\geq} \frac{1}{|\mathcal{C}|} \sum_{\bfy \in \Sigma_2^{n-1}} \frac{2}{n} \left( \left( \sum_{\bfc \in I_1(\bfy)} p(\bfy|\bfc) \right) - p(\bfy|\cD(\bfy)) \right) \\
	& = \frac{2}{n|\mathcal{C}|} \sum_{\bfy \in \Sigma_2^{n-1}} \sum_{\bfc \in I_1(\bfy)} p(\bfy|\bfc) - \frac{2}{n|\mathcal{C}|} \sum_{\bfy \in \Sigma_2^{n-1}} p(\bfy|\cD(\bfy)) \\
	& \overset{\mathrm{(c)}}{=} \frac{2}{n|\mathcal{C}|} \sum_{\bfy \in \Sigma_2^{n-1}} \sum_{\bfc \in I_1(\bfy)} p(\bfy|\bfc)%\\
	%&\ \ \
	 - \frac{2}{n^2|\mathcal{C}|} \sum_{\bfy \in \Sigma_2^{n-1}} \emb(\cD(\bfy) ; \bfy) \\
	& \overset{\mathrm{(d)}}{\geq} \frac{2}{n|\mathcal{C}|} \sum_{\bfy \in \Sigma_2^{n-1}} \sum_{\bfc \in I_1(\bfy)} p(\bfy|\bfc) %\\
	%&\ \ \ 
	- \frac{2}{n^2|\mathcal{C}|} \sum_{\bfy \in \Sigma_2^{n-1}} \max_{\bfc\in\cC}\{\emb(\bfc ; \bfy)\} \\
	& \overset{\mathrm{(e)}}{\geq} \frac{2}{n|\mathcal{C}|} \sum_{\bfy \in \Sigma_2^{n-1}} \sum_{\bfc \in I_1(\bfy)} p(\bfy|\bfc) %\\
	%&\ \ \ 
	- \frac{2}{n^2|\mathcal{C}|} \sum_{\bfy \in \Sigma_2^{n-1}} \emb(\cD_{EN}(\bfy) ; \bfy) \\
	& = P_{\mathrm{err}}(1\textrm{-}\mathsf{Del},\cC,\cD_{\mathrm{EN}},d_L),
	\end{align*}
	where (a) is a result of replacing the order of summation, (b) holds since for every $\bfc$ such that $\cD(\bfy) \neq \bfc$ we have that $d_L(\cD(\bfy),\bfc) \geq 2$, and for $ \bfc^* = \cD(\bfy)$ $d_L(\cD(\bfy),\bfc^*) = 0$. The equality (c) is obtained by the definition of the $1\textrm{-}\mathsf{Del}$ channel, and in (d) we simply choose the word that maximizes the value of $\emb(\bfc ; \bfy)$, which is the definition of the ML decoder as derived in step (e). From steps (b) and (e) it also follows that equality is obtained if and only if $\cD \equiv \cD_{\mathrm{EN}}$.
\end{IEEEproof}

\section{The $2$-Deletion Channel}\label{sec:2-del}

In this section, we consider the case of a single $2$-deletion channel over a code which is the entire space, i.e., $\cC = \Sigma_2^n$. In this setup, a word $\bfx \in \Sigma_2^n$ is transmitted over the channel $2\textrm{-}\mathsf{Del}$, where exactly 2 symbols from $\bfx$ are selected and deleted, resulting in the channel output $\bfy \in \Sigma_2^{n-2}$. We construct a decoder that is based on the lazy decoder and on a variant of the embedding number decoder and prove that it minimizes the expected normalized distance, that is, we explicitly find the ML$^*$ decoder for the $2\textrm{-}\mathsf{Del}$ channel. 

Recall that the expected normalized distance of a decoder $\cD$ over a single $2$-deletion channel is defined as
\begin{align*}
\perr(\cD)& =\frac{1}{|\cC|} \sum_{\bfc \in \cC}   P_{\mathrm{err}}(\bfc) %& \\
%&
= \frac{1}{|\cC|\cdot |\bfc|} \sum_{\bfc \in \cC} \sum_{\bfy: \cD(\bfy) \neq c }  d_L(\cD(\bfy),\bfc)\cdot p (  \bfy| \bfc ). &
\end{align*}
We can rearrange the sum as follows
\begin{align*}
\perr(\cD)=\frac{1}{|\cC|\cdot |\bfc|} \sum_{\bfy \in \Sigma_2^{n-2} }  \sum_{\bfc \in \cC}  d_L(\cD(\bfy),\bfc)\cdot  p (  \bfy| \bfc ).
\end{align*}
As mentioned before, we denote $\sum_{\bfc:\cD(\bfy) \neq  \bfc}  \frac{d_L(\cD(\bfy),\bfc)}{|\bfc|}p (  \bfy| \bfc )$ by $f_{\bfy}(\cD(\bfy))$. Recall that, a decoder that minimizes $f_{\bfy}(\cD(\bfy))$ for any channel output $\bfy \in \Sigma_2^{n-2}$, also minimizes the expected normalized distance.  Hence, if for two decoders $\cD_1$ and $\cD_2$, we have that for any $\bfy \in \Sigma_q^{n-2}, f_{\bfy}(\cD_1(\bfy)) \le f_{\bfy}(\cD_2(\bfy))$ then the we have that the expected normalized distance of $\cD_1$ is smaller to equal to the one of $\cD_2$. Therefore, when comparing the two decoders, showing that for any $\bfy$, $f_{\bfy}(\cD_1(\bfy)) \le f_{\bfy}(\cD_2(\bfy))$ is a sufficient condition to show that $\cD_1$ has smaller (or equal) expected normalized distance.

Before we continue, two more families of decoders are introduced.
\begin{definition} \textbf{\emph{The {maximum likelihood* decoder of length $m$}.}}
The \emph{maximum likelihood* decoder of length $m$}, denoted by $\cD_{\ML^*}^{m}$, is the decoder that for any given channel output $\bfy$ returns a word $\bfx$ of length $m$ that minimizes $f_{\bfy}(\bfx)$. That is, 
$$\cD_{\ML^*}^m(\bfy) = \argmin_{\bfx\in\Sigma_2^m}\{ f_{\bfy}( \bfx)\}.$$
\end{definition}
\begin{definition}
\textbf{\emph{The {embedding number decoder of length $m$}.}}
The \emph{embedding number decoder of length $m$}, denoted by $\cD_{EN}^{m}$, is the decoder that for any given channel output $\bfy$ returns a word $\bfx$ of length $m$ that maximizes the embedding number of $\bfy$ in $\bfx$. That is, 
$$\cD_{EN}^m(\bfy) = \argmax_{\bfx\in\Sigma_2^m}\{ \emb(\bfx;\bfy)\}.$$
\end{definition}

%In these decoders' definitions, and unless stated otherwise, if there is more than one word $\bfx$ that optimizes these expressions, the decoder chooses one of them arbitrarily. 

Similarly to the analysis of the $1\textrm{-}\mathsf{Del}$ channel in Section~\ref{sec:1-del}, any embedding number decoder prolongs existing runs in the word $\bfy$. The following lemma proves that any embedding number decoder of length $n-1$ prolongs at least one of the longest runs in $\bfy$ by at least one symbol. 
\begin{lemma} \label{lem:endec1}
Let $\bfy \in \Sigma_2^{n-2}$ be a channel output.  The decoder $\cD_{EN}^{n-1}$ prolongs one of the longest runs of $\bfy$ by at least one symbol.
\end{lemma}
\begin{IEEEproof}
Assume that the number of runs in $\bfy$ is $\rho(\bfy)=r$ and let $r_j$ denote the length of the $j$-th run for $1\le j\le r$. We further assume that the $i$-th run is of longest length in the word $\bfy$, and that its length is denoted by $r_i$. Assume to  the contrary that none of the longest runs in $\bfy$ was prolonged. Furthermore, let $i'$ be one of the indices of the runs in $\bfy$, such that the $i'$-th run of $\bfy$ was prolonged by the decoder $\cD_{EN}^{m}$, and note that $r_i > r_{i'}$. Thus, by editing the decoder to prolong the $i$-th run instead of the $i'$-th run (while maintaining the number of symbols that are added to the run), we get a decoder output with a strictly larger embedding number, in contradiction to the definition of the decoder. 
%Furthermore, assume that the first longest run in $\bfy$ is the $i$-th run, for $1\le i\le t$. Let the $i'$-th run be the first longest run that was prolonged by the decoder $\cD_{EN}^{m}$. Since the $i$-th run is the first longest run of $\bfy$, it holds that $i<i'$ and $r_i\ge r_{i'}$. Hence the decoder can prolong the $i$-th run with the same number of symbols as it prolonged the $i'$-th run without decreasing the embedding number. 
\end{IEEEproof}
For simplicity, we assume that in the case where there are two or more longest runs in $\bfy$, the embedding number decoder $\cD_{EN}^{m}$ for $m>|\bfy|$ necessarily chooses to prolong the first ones. Moreover, if there is more than one option that maximizes the embedding number, the embedding number decoder $\cD_{EN}^{m}$ will choose the one that prolongs the least number of runs, where the runs are chosen as the first runs in $\bfy$. 

In the following lemma, a useful property regarding $\cD_{EN}^{n}$, the embedding number decoder of length $n$, is given. 
\begin{lemma} \label{lem:endec2}
Let $\bfy \in \Sigma_2^{n-2}$ be a channel output. Assume that the number of runs in $\bfy$ is $\rho(\bfy)=r$ and let $r_i$ denote the length of the $i$-th run for $1\le i\le r$. In addition, let the $i$-th and the $j$-th runs be the first two longest runs in $\bfy$, such that  $r_i \geq r_j$, and let $a$ be the length of the longest alternating segment in $\bfy$.
The decoder $\cD_{EN}^{n}$ operates as follows.
\begin{enumerate}
    \item If $a \ge 2(r_i +1)(r_j+1)$ and $a \ge (r_i+2)(r_i+1)$, the decoder prolongs the (first) longest alternating segment by two symbols.
	\item Otherwise, if $r_i \geq 2 r_j$, the decoder prolongs the $i$-th run by two symbols.
	\item Otherwise, if $r_i < 2 r_j$, the decoder prolongs the $i$-th and the $j$-th runs, each by one symbol.
\end{enumerate}
\end{lemma}
\begin{IEEEproof} 
First, it should be noted that for any decoder $\cD$ that prolongs the alternating segment, we have that $\emb(\bfy; \cD(\bfy)) = \lfloor \frac{a+2}{2} \rfloor.$ Therefore, the embedding number decoder has three options. The first one is to prolong one of the longest alternating segments by two symbols (i.e., introducing two new runs of length one), the second one is to prolong one of the runs in $\bfy$ by two symbols, and the second is to prolong two runs in $\bfy$, each by one symbol. We ignore the option of creating new runs that are not part of the longest alternating segment since it won't increase the embedding number. Thus, the maximum embedding number value is given by
\begin{align*}
& \max \left\lbrace \max_{1\le s<\ell\le r} \left\lbrace \left( \begin{matrix} r_s+1 \\ 1 \end{matrix} \right) \cdot \left( \begin{matrix} r_\ell+1 \\ 1 \end{matrix} \right) \right\rbrace  , \max_{1\le s\le r} \left( \begin{matrix} r_s+2 \\ 2 \end{matrix} \right), \lfloor \frac{a+2}{2} \rfloor \right\rbrace \\
& = \max \left\lbrace \max_{1\le s<\ell\le r}  \left\lbrace (r_s+1)(r_\ell+1) \right\rbrace , \max_{1\le s\le r} \left\lbrace \frac{(r_s+1)(r_s+2)}{2}  \right\rbrace ,  \frac{a+2}{2} -1 \right\rbrace \\
& = \max \left\lbrace (r_i+1)(r_j+1) , \frac{(r_i+1)(r_i+2)}{2}, \frac{a}{2}\right\rbrace.
\end{align*}
Finally, to determine the option that maximizes the embedding number, it is left to compare between $ \frac{a}{2}$, $(r_i+1)(r_j+1)$, and $\ \frac {(r_i+1)(r_i+2)}{2}$. Thus, given our assumption that the decoder prefers to create and prolong the least umber of runs, the decoder $\cD_{EN}^{n}$ chooses the first option, i.e., prolonging the longest alternating segment by two symbols, only if $a \ge 2(r_i +1)(r_j+1)$ and $a \ge (r_i+2)(r_i+1)$. Otherwise, it decides to prolong the longest run with two symbols, if and only if $\frac{r_i+2}{2}  \ge (r_j+1) $ which is equivalent to $r_i \ge 2r_j$. 
\end{IEEEproof}
%To conclude, the embedding number decoder of length $n-1$ prolongs the first longest run in $\bfy$, and the embedding number decoder of length $n+1$ prolongs up to the three longest runs in $\bfy$.
%We also prove several properties regarding the lazy decoder, which for every given channel output simply returns it, and regarding the embebdding number decoder, which 

In the rest of this section we prove several properties on $\cD_{\ML^*}$, the ML$^*$ decoder for a single $2$-deletion channel and lastly in Theorem~\ref{Th:2-del} we construct this decoder explicitly.
 %decoder $\cD_{2-Del}$, an decoder that minimizes $f_{\bfy}(\cD(\bfy))$ for a single $2$-deletions channel. 
Unless specified otherwise, we assume that  $\cD_{\ML^*}$ returns a word with minimum length that minimizes $f_{\bfy}(\cD(\bfy))$.  
%The following lemmas prove several properties and explain the construction of $\cD_{2-Del}$. \textcolor{red}{In these proofs we use the definition of insertion ball.  The \emph{radius-$r$ insertion ball} of a word $\bfx\in\Sigma_q^*$, denoted by $\ins_r(\bfx)$ or shortly $I_r(\bfx)$, is the set of all supersequences of $\bfx$ of length $|\bfx|+r$. From~\cite{} it is know that $I_r(\bfx)= \sum_{i=0}^r \binom{|\bfx|+r}{i}$. Similarily,  the \emph{radius-$r$ deletion ball} of a word $\bfx\in\Sigma_q^*$, denoted by $\del_r(\bfx)$ or shortly $D_r(\bfx)$, is the set of all subsequences of $\bfx$ of length $|\bfx|-r$. The set of all subsequences, supersequences of $\bfx$ is denoted by $\del^*(\bfx) \triangleq \bigcup_{r\geq 0}\del_r(\bfx), \ins^*(\bfx) \triangleq \bigcup_{r\geq 0}\ins_r(\bfx)$, respectively.} 
%In the rest of this section, in order to construct a decoder that minimizes $f_{\bfy}(\cD(\bfy))$, we prove several lemmas and properties consider the \emph{almost lazy decoders}, denoted by $\cD_{AL}$. The almost lazy decoders work as follows, for any $\bfy' \in\Sigma^{n-2}$ they return $\cD(\bfy')=\bfy'$, the same output as the lazy decoder, with the exception of $\bfy$. The output of $\cD_{AL}$ on $\bfy$ varies in its length. Each lemma of the following lemmas examine one specific decoder among the almost lazy deocders, and compare it to the lazy decoder. Following these lemmas and their proofs we construct an optimal decoder for a single $2$-deletions channel and prove in Theorem~\ref{Th:2-del} that it minimizes the expected normalized distance. 

\begin{lemma} \label{lem:2deldeclength}
For any channel output $\bfy \in \Sigma_2^{n-2}$, it holds that
$$n-2 \le |\cD_{\ML^*}(\bfy)| \le n+1.$$
 \end{lemma}  

\begin{IEEEproof}
Let $\bfy\in\Sigma_2^{n-2}$ be a channel output and assume to the contrary that  $|\cD_{\ML^*}(\bfy)|\geq n+2$ or $|\cD_{\ML^*}(\bfy)|\leq n-3$.  In order to show a contradiction, we prove that 

\vspace{-1ex}
%\begin{small}
\begin{align*}
f_{\bfy}(\cD_{\ML^*}(\bfy))= \sum_{\bfc \in I_2(\bfy)}  \frac{d_L(\cD_{\ML^*}(\bfy),\bfc)}{|\bfc|}\cdot p (  \bfy| \bfc ) \geq 
\sum_{\bfc \in I_2(\bfy)}  \frac{d_L(\cD_{Lazy}(\bfy),\bfc)}{|\bfc|}\cdot p (  \bfy| \bfc ) = f_{\bfy}(\cD_{\text{Lazy}}(\bfy)),
\end{align*}
%\end{small}

and equality can be obtained only in the case $|\cD_{\ML^*}(\bfy)| =n+2$. If $|\cD_{\ML^*}(\bfy)| \leq n-3$ or $|\cD_{\ML^*}(\bfy)| \ge n+3$, then $d_L(\cD_{\ML^*}(\bfy),\bfc) \geq 3$ and since $d_L(\cD_{Lazy}(\bfy),\bfc)=2$ a strict inequality holds for each $\bfy$. In case $|\cD_{\ML^*}(\bfy)| = n+2$, $d_L(\cD_{\ML^*}(\bfy),\bfc) \geq 2$ and the inequality holds. Recall that $\cD_{\ML^*}(\bfy)$ returns a word with minimum length which implies that $|\cD_{\ML^*}(\bfy)|\le n+1$.
 \end{IEEEproof}

For $\bfy \in \Sigma_2^{n-2}$, Lemma~\ref{lem:2deldeclength} implies that $m = |\cD_{\ML^*}(\bfy)| \in \{n-2, n-1, n, n+1\}$.  In the following lemmas, we show that for any $m \in \{n-2, n-1, n\}$,
\begin{align*}
 \cD_{\ML^*}^{m}=\cD_{EN}^{m}. 
\end{align*}
%\textcolor{red}{We will deal with the case of $m=n+1$ later on.}
\begin{lemma} \label{lem:decoder_EN}
It holds that for $\cC = \Sigma_2^n$
$$ \cD_{\ML^*}^{n-2} = \cD_{EN}^{n-2} = \cD_{Lazy}.$$
\end{lemma}
\begin{IEEEproof}
Let $\bfy\in\Sigma_2^{n-2}$ be a channel output. Each $\bfy'\in\Sigma_2^{n-2}$ such that $\bfy'\ne \bfy$ satisfies ${\emb(\bfy';\bfy) =0}$. Hence ${\cD_{EN}^{n-2}(\bfy) = \bfy}$, which implies that  $\cD_{EN}^{n-2}=\cD_{Lazy}$. 

In order to show that $\cD_{Lazy} = \cD_{\ML^*}^{n-2}$, let us consider any decoder $\cD $ that outputs words of length $n-2$ such that $\cD~\neq~\cD_{\mathrm{Lazy}}$, i.e., there exists $\bfy \in \Sigma_2^{n-2}$ such that $\cD(\bfy) = \bfy' \neq \bfy$. Since $\bfy' \neq \bfy$ it holds that $I_2(\bfy') \neq I_2(\bfy)$ and hence, without the loss of the generality, there exists a codeword $\bfc \in \Sigma_2^n$ such that $\bfc \in I_2(\bfy)$ and $\bfc \notin I_2(\bfy')$. Equivalently, $\bfy \in D_2(\bfc)$,  $\bfy' \notin D_2(\bfc)$ and therefore $d_L(\bfc,\bfy') \geq 4$ (at least one more deletion and one more insertion are needed in addition to the two insertions needed for every word in the deletion ball). Hence, 

\vspace{-1ex}
%\begin{small}
\begin{align*}
f_{\bfy}(\cD(\bfy)) & =   \sum_{\bfc' \in \Sigma_2^n}  \frac{d_L(\cD(\bfy),\bfc')}{|\bfc'|} p(\bfy|\bfc') 
 \\ & =   \sum_{\substack {\bfc' \in \Sigma_2^n \\ \bfc' \ne \bfc} }  \frac{d_L(\cD(\bfy),\bfc')}{|\bfc'|} p(\bfy|\bfc')  +   \frac{d_L(\cD(\bfy),\bfc)}{|\bfc|} p(\bfy|\bfc)
  \\ &  \ge \sum_{\substack {\bfc' \in \Sigma_2^n \\ \bfc' \ne \bfc} }  \frac{2}{|\bfc'|} p(\bfy|\bfc')  +   \frac{d_L(\cD(\bfy),\bfc)}{|\bfc|} p(\bfy|\bfc)
    \\ &  \ge \sum_{\substack {\bfc' \in \Sigma_2^n \\ \bfc' \ne \bfc} }  \frac{2}{|\bfc'|} p(\bfy|\bfc')  +   \frac{4}{|\bfc|} p(\bfy|\bfc)
      \\ &  > \sum_{\substack {\bfc' \in \Sigma_2^n } }  \frac{2}{|\bfc'|} p(\bfy|\bfc')   = f_{\bfy}(\cD_{\mathrm{Lazy}}(\bfy))  =  f_{\bfy}(\cD_{EN}^{n-2}(\bfy)).
\end{align*}
%\end{small}

These inequalities state that $\cD_{Lazy}$ is the decoder that minimizes $f_{\bfy}(\cD(\bfy))$ for any $\bfy \in \Sigma_2^{n-2}$ among all decoders that return words of length $n-2$. Hence, we deduce that the ML$^*$ decoder of length $n-2$ is $\cD_{Lazy}$. 
\begin{comment}
	
	Hence, it is derived that
	\begin{align*}
	P_{\mathrm{err}}(\bfc,d_L) & = \sum_{\bfz \in D_2(\bfc)} \frac{d_L(\cD(\bfz),\bfc)}{n} p(\bfy | \bfz) \\ 
	& \geq \sum_{\substack {\bfz \in D_2(\bfc) \\ \bfz \ne \bfy}} \frac{2}{n} p(\bfz | \bfc)  + \frac{d_L(\cD(\bfy)=\bfy',\bfc)}{n} \cdot p(\bfy | \bfc) \\
		& > \sum_{\bfz \in D_2(\bfc)} \frac{2}{n} p(\bfz | \bfc) = \frac{2}{n}.
	\end{align*}
	If $\cC = \Sigma_2^n$ it must hold that $\bfc \in \cC$, and so
	\begin{equation*}
	P_{\mathrm{err}}(1\textrm{-}\mathsf{Del},\Sigma_2^n,\cD,d_L) \geq \frac{|\cC|-1}{|\cC|}\cdot \frac{1}{n} + \frac{1}{|\cC|}P_{\mathrm{err}}(\bfc,d_L) > \frac{1}{n}.
	\end{equation*}
	Combining with Lemma~\ref{Lem:LazyError} again completes the proof.
\end{comment}
\end{IEEEproof}

Based on the discussion at the beginning of this section, when comparing two decoders $\cD_1$ and $\cD_2$, we can deduce that $\cD_1$ has higher expected normalized distance by evaluating the sufficient condition that for any $\bfy \in \Sigma_2^{n-2}$, $f_\bfy(\cD_1(\bfy)) \ge f_{\bfy}(\cD_2(\bfy))$. Next, we show that the above condition holds for a word $\bfy \in \Sigma_2^{n-2}$, if and only if, 
 \vspace{-1ex}
 %\begin{small}
 \begin{align} \label{inequlaity_avg_dec}
%&  \sum_{\bfc\in I_2(\bfy)}  d_L(\cD(\bfy),\bfc) \emb(\bfc; \bfy)-  \sum_{\bfc\in I_2(\bfy)}  d_L(\cD_{EN}^{n-1}(\bfy),\bfc) \emb(\bfc; \bfy) \\ 
& \sum_{\bfc\in I_2(\bfy)}  \emb(\bfc; \bfy)  \Big(  d_L(\cD_1(\bfy),\bfc) -  d_L(\cD_2(\bfy),\bfc) \Big) \ge 0. 
\end{align}

The equivalency of $f_\bfy(\cD_1(\bfy)) \ge f_{\bfy}(\cD_2(\bfy))$ and inequality~\ref{inequlaity_avg_dec} follows from the following equations.   Given two decoders $\cD_1$ and $\cD_2$, we have that, 
\begin{align*}
& f_\bfy(\cD_1(\bfy)) - f_{\bfy}(\cD_2(\bfy)) \\& =  \sum_{\bfc:\cD_1(\bfy) \neq  \bfc}  \frac{d_L(\cD_1(\bfy),\bfc)}{|\bfc|} p (  \bfy| \bfc ) -  \sum_{\bfc:\cD_2(\bfy) \neq  \bfc}  \frac{d_L(\cD_2(\bfy),\bfc)}{|\bfc|} p (  \bfy| \bfc ) \\
& = \frac{1}{|\bfc|} \left( \sum_{\bfc\in \Sigma_2^n}  d_L(\cD_1(\bfy),\bfc) p (  \bfy| \bfc ) -  \sum_{\bfc\in\Sigma_2^n}  d_L(\cD_2(\bfy),\bfc) p (  \bfy| \bfc ) \right) \\ 
& = \frac{1}{|\bfc|} \sum_{\bfc\in \Sigma_2^n} p (  \bfy| \bfc ) \Big(  d_L(\cD_1(\bfy),\bfc) -  d_L(\cD_2(\bfy),\bfc) \Big) \\
& = \frac{1}{\binom{n}{2} |\bfc|}  \sum_{\bfc\in \Sigma_2^n}  \emb(\bfc; \bfy)  \Big( d_L(\cD_1(\bfy),\bfc) -  d_L(\cD_2(\bfy),\bfc) \Big) \\
& = \frac{1}{\binom{n}{2} |\bfc|}  \sum_{\bfc\in I_2(\bfy)}  \emb(\bfc; \bfy)  \Big( d_L(\cD_1(\bfy),\bfc) -  d_L(\cD_2(\bfy),\bfc) \Big),
\end{align*}
where the last equality holds since for any $\bfc \in \Sigma_2^n$ such that $\bfc \notin I_2(\bfy)$ it holds that $\emb(\bfc;\bfy)=0.$ 
Hence when comparing the expected normalized distance of two decoders $\cD_1$ and $\cD_2$, inequality~\ref{inequlaity_avg_dec} is a sufficient condition. 
 
%\end{small}
\begin{lemma} \label{lem:decoder_EN_n-1}
It holds that
$$ \cD_{\ML^*}^{n-1} = \cD_{EN}^{n-1}.$$
\end{lemma}
\begin{IEEEproof}
By similar arguments to those presented in Lemma~\ref{lem:endec1}, for any channel output $\bfy$, $\cD_{EN}^{n-1}(\bfy)$ is obtained from $\bfy$ by prolonging the first longest run of $\bfy$ by one symbol. Let $\bfy$ be the channel output and let $\cD$ be a decoder such that $|\cD(\bfy)| = n-1$. Our goal is to prove that the inequality stated in~(\ref{inequlaity_avg_dec}) holds when $\cD_1=\cD$ and $\cD_2 = \cD_{EN}^{n-1}$. This completes the lemma's proof. The latter will be verified in the following claims. 

\begin{claim} \label{D_n-1_run}
For any decoder $\cD$ such that $\cD(\bfy) \ne \cD_{EN}^{n-1}(\bfy)$ and $|\cD(\bfy)|=n-1$, where $\cD(\bfy)$ is obtained from $\bfy$ by prolonging one of the runs in $\bfy$, the inequality stated in~(\ref{inequlaity_avg_dec}) holds and thus $f_{\bfy} (\cD(\bfy)) \ge f_{\bfy}{(\cD_{EN}^{n-1}(\bfy))}$. 
\end{claim}
\begin{IEEEproof}
Assume that the number of runs in $\bfy$ is $\rho(\bfy)=r$, let $r_j$ denote the length of the $j$-th run for $1\le j\le r$, and let the $i$-th run of $\bfy$ be the first longest run of $\bfy$. Assume that $\cD(\bfy)$ is obtained by prolonging the $j$-th run of $\bfy$ by one symbol. Since $\cD(\bfy) \ne\cD_{EN}^{n-1}(\bfy) $ it holds that $j\ne i$. Note that $$|I_1(\cD(\bfy))\cap I_1(\cD_{EN}^{n-1}(\bfy))| = 1$$ since the only word in this set is the word that is obtained by prolonging the $i$-th and $j$-th runs of $\bfy$. It holds that for ${\bfc \in I_1(\cD(\bfy))\cap I_1(\cD_{EN}^{n-1}(\bfy))}$, $d_L(\cD(\bfy),\bfc)=d_L(\cD_{EN}^{n-1}(\bfy),\bfc)=1$ and hence this word can be eliminated from inequality~(\ref{inequlaity_avg_dec}). Similarly for words $\bfc$ such that $\bfc\notin I_1(\cD(\bfy))$ and $\bfc\notin I_1(\cD_{EN}^{n-1}(\bfy))$, we get that $d_L(\cD(\bfy),\bfc) = d_L(\cD_{EN}^{n-1}(\bfy), \bfc) = 3$ and therefore these words can also be eliminated from inequality~(\ref{inequlaity_avg_dec}). Note that from~\cite{L66}, the number of such words is
\begin{small}
\begin{align*}
& \left|I_2\left(\bfy\right)\right|  - \left|I_1\left(\cD_{EN}^{n-1}(\bfy)\right)\right| - \left|I_1\left(\cD(\bfy)\right)\right| + \left|I_1\left(\cD(\bfy)\right)\cap I_1\left(\cD_{EN}^{n-1}(\bfy)\right)\right| & \\
& =  \binom{n}{2} + n + 1 -2(n+1) + 1 = \binom{n}{2} - n.&
\end{align*}
\end{small}
 Let us consider the remaining $2n$ words in $I_2(\bfy)$, which are not in the intersection $I_1(\cD(\bfy))\cap I_1(\cD_{EN}^{n-1}(\bfy))$.
\begin{enumerate}
\item \label{case1_run} $\bfc\in I_1(\cD_{EN}^{n-1}(\bfy))$ and $\bfc\notin I_1(\cD(\bfy))$: Since the embedding number decoder prolongs a run in $\bfy$, $I_1({\cD_{EN}^{n-1}}(\bfy)) \subseteq I_2(\bfy)$. Therefore, there are
%\begin{small}
 \begin{align*}
 |I_1(\cD_{EN}^{n-1}(\bfy))| - |I_1(\cD(\bfy))\cap I_1(\cD_{EN}^{n-1}(\bfy))| = n+1-1 = n 
  \end{align*}
  %\end{small}
such words and for each one of them,
 \begin{align*}
d_L(\cD(\bfy),\bfc)) = 3 \text{ and } d_L(\cD_{EN}^{n-1}(\bfy), \bfc) = 1.
\end{align*}
We consider three possible options for the word $\bfc$ in this case. If $\bfc$ is the word obtained by prolonging the $i$-th run of $\bfy$ by two symbols, then $\emb(\bfc; \bfy) = \binom{r_i +2}{2}$. Let $\bfc = \bfc_h$ be the word obtained by prolonging the $i$-th and the $h$-th run for $h \ne i, j$. Since there are $r-2$ runs other than the $i$-th and the $j$-th run,  the number of such words is $r-2$, while $\emb (\bfc_h ; \bfy) = (r_i+1) (r_h+1) $.  Lastly, if $\bfc$ is obtained by prolonging the $i$-th run and creating a new run in $\bfy$ then $\emb(\bfc; \bfy) = r_i+1$, and the number of such words is $n-r+1$. Thus, 
%\vspace{-1ex}
%\begin{footnotesize}
\begin{align*}
& \sum_{\substack{\bfc\in I_1(\cD_{EN}^{n-1}(\bfy)) \\ \bfc\notin I_1(\cD(\bfy))}}   \emb(\bfc; \bfy) \left( d_L(\cD(\bfy),\bfc) -  d_L(\cD_{EN}^{n-1}(\bfy),\bfc) \right)
 \\ & = 2  \left( \binom{r_i+2}{2} + \sum_{\substack{h=1 \\ h \ne j,i }}^{r} (r_h+1) (r_i+1) + (n-r+1) (r_i+1)\right)
 \\& = 2 \left( \binom{r_i+2}{2}  + (r_i+1)(n-2-r_j-r_i+r-2) +  (n-r+1) (r_i+1)\right)
  \\& = 2 \left(\binom{r_i+2}{2} + (r_i+1)(n-r_j-r_i+r-4) +  (n-r+1) (r_i+1)\right).
 \end{align*}
%\end{footnotesize}
\item $\bfc\notin I_1(\cD_{EN}^{n-1}(\bfy))$ and $\bfc\in I_1(\cD(\bfy))$:  The decoder $\cD$ prolongs a run in $\bfy$, and therefore $I_1(\cD(\bfy)) \subseteq I_2(\bfy)$. Similarily to Case~\ref{case1_run}, there are $n$ such words, and
%\vspace{-1ex}
%\begin{footnotesize}
\begin{align*}
& \sum_{\substack{\bfc\notin I_1(\cD_{EN}^{n-1}(\bfy)) \\ \bfc\in I_1(\cD(\bfy))}}   \emb(\bfc; \bfy) \left( d_L(\cD(\bfy),\bfc) -  d_L(\cD_{EN}^{n-1}(\bfy),\bfc) \right)
 \\ & = 2  \left( \binom{r_j+2}{2} + \sum_{\substack{h=1 \\ h \ne j,i }}^{r} (r_h+1) (r_j+1) + (n-r+1) (r_j+1)\right)
 \\& = 2 \left( \binom{r_j+2}{2}  + (r_j+1)(n-2-r_j-r_i+r-2) +  (n-r+1) (r_j+1)\right)
  \\& = 2 \left(\binom{r_j+2}{2} + (r_j+1)(n-r_j-r_i+r-4) +  (n-r+1) (r_j+1)\right).
 \end{align*}
% \end{footnotesize}
\end{enumerate}
Thus, 
%\begin{footnotesize}
\begin{align*}
 & \sum_{\bfc\in I_2(\bfy)}   \emb(\bfc; \bfy)  \Big(  d_L(\cD_1(\bfy),\bfc) -  d_L(\cD_2(\bfy),\bfc) \Big)  
\\ & \ \ \ =  2 \left(\binom{r_i+2}{2} + (r_i+1)(n-r_j-r_i+r-4) +  (n-r+1) (r_i+1)\right)
\\ &\ \ \  -2 \left(\binom{r_j+2}{2} + (r_j+1)(n-r_j-r_i+r-4) +  (n-r+1) (r_j+1)\right)
\\ &  \ \ \ \ge 0,
\end{align*}
%\end{footnotesize}
where the last inequality holds since $r_i \ge r_j$.
\end{IEEEproof}

\begin{claim} 
For any decoder $\cD$ such that $\cD(\bfy) \ne \cD_{EN}^{n-1}(\bfy)$ and $|\cD(\bfy)|=n-1$, where $\cD(\bfy)$ is obtained from $\bfy$  by creating a new run of one symbol in $\bfy$, the inequality stated in~(\ref{inequlaity_avg_dec}) holds and thus $f_{\bfy} (\cD(\bfy)) \ge f_{\bfy}{(\cD_{EN}^{n-1}(\bfy))}$. 
\end{claim}

\begin{IEEEproof}
Assume that the number of runs in $\bfy$ is $\rho(\bfy)=r$, let $r_j$ denote the length of the $j$-th run for $1\le j\le r$, and let the $i$-th run of $\bfy$ be the first longest run of $\bfy$.
As in Claim~\ref{D_n-1_run}, if $\bfc \in \left(I_1(\cD(\bfy))\cap I_1(\cD_{EN}^{n-1}(\bfy))\right)$, then $\bfc$ can be eliminated from (\ref{inequlaity_avg_dec}). It should be noted that, if the new run which is created in $\bfy$ by $\cD$ is in the begging or the end of $\bfy$, or if it is  adjacent to the $i$-th run of $\bfy$, or if it is splitting the $i$-th run of $\bfy$, than $\left|\left(I_1(\cD(\bfy))\cap I_1(\cD_{EN}^{n-1}(\bfy))\right)\right| =2 $, otherwise $\left|\left(I_1(\cD(\bfy))\cap I_1(\cD_{EN}^{n-1}(\bfy))\right)\right|=1$. To lower bound the value of inequality (\ref{inequlaity_avg_dec}), we can assume the size of this intersection is one.
Similarly, any word $\bfc$ such that  $\bfc\notin I_1(\cD(\bfy))$ and $\bfc\notin I_1(\cD_{EN}^{n-1}(\bfy))$ can be eliminated from (\ref{inequlaity_avg_dec}). 
Let us consider the remaining $2n$ (or $2n-1$) words in $I_2(\bfy)$:
\begin{enumerate}
\item $\bfc\in I_1(\cD_{EN}^{n-1}(\bfy))$ and $\bfc\notin I_1(\cD(\bfy))$:  From arguments similar to those presented in Claim~\ref{D_n-1_run},  there are $n$ such words, given as follows. The first word is obtained by prolonging the $i$-th run with an additional symbol. The embedding number of this word is $\binom{r_i+2}{2}$. Additionally, there are $r-1$ words obtained by prolonging the $h$-th run in $\bfy$ by an additional symbol, for $1 \le h \le r$, $h \ne i$. These words stratify $\emb(\bfy; \bfc) = (r_i +1)(r_h+1)$. Finally, we have at least $n-r-1$ words that are obtained by creating a new run, which is different than the run created by the decoder $\cD$.  These words have an embedding number of $(r_i+1)$. 
%\vspace{-1ex}
%\begin{footnotesize}
\begin{align*}
& \sum_{\substack{\bfc\in I_1(\cD_{EN}^{n-1}(\bfy)) \\ \bfc\notin I_1(\cD(\bfy))}}   \emb(\bfc; \bfy) \left( d_L(\cD(\bfy),\bfc) -  d_L(\cD_{EN}^{n-1}(\bfy),\bfc) \right)
  \\& \ge 2 \left(\binom{r_i+2}{2} + (r_i+1)(n-r_i+r-1) +  (n-r-1) (r_i+1)\right)
    \\& = 2 \left(\binom{r_i+2}{2} + (r_i+1)(2n-r_i-2)\right)
    \\& = (r_i+1)(-r_i-2+4n).
 \end{align*}
 %\end{footnotesize}
Note that the difference compared to Claim~\ref{D_n-1_run} follows from the fact that the number of runs is different. 

\item $\bfc\notin I_1(\cD_{EN}^{n-1}(\bfy))$ and $\bfc\in I_1(\cD(\bfy))$:  As in Claim~\ref{D_n-1_run}, the number of such words is $n$, and for each of these words, 
\begin{align*}
d_L(\cD(\bfy),\bfc) = 1 \text{ and } d_L(\cD_{EN}^{n-1}(\bfy), \bfc) = 3.
\end{align*}

We consider three possible options for the word $\bfc$ in this case. If $\bfc$ is the word obtained by prolonging the new run of $\cD(\bfy)$ by additional symbol then $\emb(\bfc; \bfy) = 1$. Let $\bfc = \bfc_h$ be the word obtained by prolonging the $h$-th run of $\bfy$  for $h \ne i$ and creating the same  new run of one symbol as in $\cD(\bfy)$. Since there are $r-1$ runs other than the $i$-th run,  the number of such words is $r-1$, while $\emb (\bfc_h ; \bfy) =  (r_h+1) $.  Lastly, if $\bfc$ is obtained by creating an additional new run in $\cD(\bfy)$, then we distinguish two cases; the first case includes two words in which the two additional runs create an alternating segment. Note that there are two such words since the alternating segment can be created by both of its edges. In this case, the length of such alternating segment is at most $r+2$  and $\emb(\bfc; \bfy) = \lfloor \frac{r+2}{2} \rfloor $~\footnote{This value equals $r$ if and only if the inserted two symbols creates alternating segment of length $r$ in $\bfc$, see more details in~\cite{BSG21}.}. The second case includes all the other $n-r-2$ words, and in this case, $\emb(\bfc; \bfy) =1 $ . Hence, 
%\vspace{-1ex}
% \begin{small}
\begin{align*}
& \sum_{\substack{\bfc\notin I_1(\cD_{EN}^{n-1}(\bfy)) \\ \bfc\in I_1(\cD(\bfy))}}   \emb(\bfc; \bfy) \left( d_L(\cD(\bfy),\bfc) -  d_L(\cD_{EN}^{n-1}(\bfy),\bfc) \right)
 \\ & \ge 1  \left(1 + \sum_{\substack{h=1 \\ h \ne i }}^{r} (r_h+1) +  1(n-r)\right) 
 %\\ & \hspace{5ex}
  - 3  \left(1 + \sum_{\substack{h=1 \\ h \ne i }}^{r} (r_h+1) + (n-r-2) +2\lfloor \frac{r+2}{2} \rfloor \right)
   \\& \ge -2 -2 \sum_{\substack{h=1 \\ h \ne i }}^{r} (r_h+1) +(n-r)-3(n-r-2)-3(r+2).
    \\& = -2(n+r-r_i-3)-2(n-r-2)-3(r+2).
 \end{align*}
 %\end{small}
% If $\bfc$ is the word obtained by prolonging the new run of $\cD(\bfy)$ by additional symbol, then $\emb(\bfc; \bfy) = 1$. 
%Else, assume $\bfc$ is a word obtained by prolonging the $h$-th run for $h \ne i$. Since there are $t-1$ runs other than the $i$-th run,  the number of such words is $t-1$ and each of these words satisfies $\emb (\bfc ; \bfy) = (r_h+1) $. 
% Otherwise, $\bfc$ is obtained by creating additional new run in $\cD(\bfy)$ and $\emb(\bfc; \bfy) \le 2$~\footnote{explain the alternate case}, and the number of such words is $n-t$. Hence,
\end{enumerate}
Thus, 
%\begin{small}
\begin{align*}
 & \sum_{\bfc\in I_2(\bfy)}   \emb(\bfc; \bfy)  \Big(  d_L(\cD_1(\bfy),\bfc) -  d_L(\cD_2(\bfy),\bfc) \Big) 
 \\ & \ge (r_i+1)(-r_i-2+4n)  -2(n+r-r_i-3)-2(n-r-2)-3(r+2)
 \\& = -(r_i)^2 + r_i (4 n - 1) - 3 r + 2
  \\& \ge -(r_i)^2 + r_i (4 r_i - 1) - 3 \cdot 1 + 2 
  \\ & =   3(r_i)^2 -r_1 - 1 
\\ & \ge 0,
\end{align*}
%\end{small}
where the last inequality holds for any $1 \le r_i, r \le n$.
\end{IEEEproof}

\begin{claim} \label{D_n-1_nosupersequence}
For any decoder $\cD$ such that $\cD(\bfy) \ne \cD_{EN}^{n-1}(\bfy)$ and $|\cD(\bfy)|=n-1$, where $\cD(\bfy)$ is not a supersequence of $\bfy$, the inequality stated in~(\ref{inequlaity_avg_dec}) holds and thus $f_{\bfy} (\cD(\bfy)) \ge f_{\bfy}{(\cD_{EN}^{n-1}(\bfy))}$. 
\end{claim}

\begin{IEEEproof}
By definition $\cD(\bfy)$ is not a supersequence of $\bfy$ which implies that $\bfy \notin D_1(\cD(\bfy))$.
Note that for any word $\bfc\in I_2(\bfy)$ such that $\bfc \notin I_1 ( \cD (\bfy) )$, it holds that $d_L(\cD(\bfy), \bfc) \ge 3$, while $d_L(\cD_{EN}^{n-1}(\bfy), \bfc) \le 3$. Hence, if $I_2(\bfy) \cap I_1 (\cD(\bfy)) = \emptyset $ then, 
%\vspace{-1ex}
%\begin{small}
\begin{align*}
  \sum_{\bfc\in I_2(\bfy)} &  \emb(\bfc; \bfy)  \Big(  d_L(\cD_1(\bfy),\bfc) -  d_L(\cD_2(\bfy),\bfc) \Big) 
 %\\ & 
 \ge \sum_{\bfc\in I_2(\bfy)}   \emb(\bfc; \bfy)  \Big(3 - 3\Big)   =  0.
\end{align*}
%\end{small}
Otherwise, let $\bfc$ be a word such that ${\bfc \in \Big(I_2 (\bfy) \cap I_1 ( \cD (\bfy) )\Big)}$, let $\rho(\bfc)=r'$ be the number of runs in $\bfc$ and denote by $r'_{j}$ the length of the $j$-th run in $\bfc$. Let the $i$-th run in $\bfc$ be the first longest run in $\bfc$.  Note that $\bfy \in D_2 (\bfc)$ and $\cD (\bfy) \in D_1 (\bfc)$. Consider the following distinct cases. 
\begin{enumerate}
\item \textbf{Case 1: $\bfy$ is obtained from $\bfc$ by deleting two symbols from the same runs.} There exists an index $1\le j \le r'$ such that $\bfy$ is obtained from $\bfc$ by deleting two symbols from the $j$-th run of $\bfc$. In this case, since $\cD(\bfy)$ is not a supersequence of $\bfy$, $\cD(\bfy)$ must be obtained from $\bfc$ by deleting one symbol from the $h$-th run of $\bfc$ for some $h \ne j$. Hence, $\bfc$ is the unique word that is obtained by inserting to $\bfy$ the two symbols that were deleted from the $j$-th run of $\bfc$, that is,
$$ I_2 (\bfy) \cap I_1 ( \cD (\bfy) ) = \{ \bfc \}. $$

Note that, $\emb( \bfc ; \bfy ) = \binom{r'_{j}}{2} \le \binom{r'_{i}} {2}$ and ${d_L(\cD(\bfy),\bfc)=1}$, while $d_L(\cD_{EN}^{n-1}(\bfy),\bfc) \in \{1, 3\}$. If $d_L(\cD_{EN}^{n-1}(\bfy),\bfc)=1$, ~(\ref{inequlaity_avg_dec}) holds (since $\bfc$ is the only word in the intersection). Otherwise $d_L(\cD_{EN}^{n-1}(\bfy),\bfc)=3$ and our goal is to find $\bfc' \in I_2 (\bfy)$ such that 
\begin{align*}
& \sum_{\substack{\bfw \in I_2 (\bfy)}} \emb(\bfw ; \bfy) \left( d_L(\cD(\bfy) , \bfw) - d_L(\cD_{EN}^{n-1}(\bfy), \bfw) \right)
\\ &  = \sum_{\substack{\bfw \in I_2 (\bfy) \\ \bfw \ne \bfc, \bfc'}} \emb(\bfw ; \bfy) \left( d_L(\cD(\bfy) , \bfw) - d_L(\cD_{EN}^{n-1}(\bfy), \bfw) \right)
\\& \ \ \ \ + \emb(\bfc ; \bfy) \left( d_L(\cD(\bfy), \bfc) - d_L(\cD_{EN}^{n-1}(\bfy), \bfc) \right)  
\\ & \ \ \ \ + \emb(\bfc' ; \bfy) \left( d_L(\cD(\bfy), \bfc') - d_L(\cD_{EN}^{n-1}(\bfy), \bfc') \right) \ge 0.
\end{align*}
Since $d_L(\cD(\bfy),\bfw)-  d_L(\cD_{EN}^{n-1}(\bfy),\bfw) \ge 0$ for every $\bfw \ne \bfc $, it is enough to find $\bfc' \in I_2 (\bfy) $ such that, 
\begin{align*}
& \emb(\bfc ; \bfy) \Big( d_L(\cD(\bfy), \bfc) - d_L(\cD_{EN}^{n-1}(\bfy), \bfc) \Big) 
%\\ & 
+ \emb(\bfc' ; \bfy) \Big( d_L(\cD(\bfy), \bfc') - d_L(\cD_{EN}^{n-1}(\bfy), \bfc') \Big) \ge 0.
\end{align*}

Recall that the embedding number decoder prolongs the first longest run in $\bfy$. If the first longest run in $\bfc$, which is the $i$-th run, satisfies $i \ne j$, this run is also the first longest run in $\bfy$. In this case, let $\bfc'$ be the word obtained from $\bfy$ by prolonging this run by two symbols. It holds that, $d_L(\cD_{EN}^{n-1}(\bfy), \bfc')=1$,  $d_L(\cD(\bfy), \bfc')=5$, and $\emb(\bfc'; \bfy) = \binom{r'_{i}+2}{2}$. Recall that $r'_{i} \ge r'_{j}$ and hence, 
\begin{align*}
& -2 \binom{r'_{j}}{2}  +4 \binom{r'_{i}+2}{2}   \ge 0.
\end{align*}

Else, if the first longest run in $\bfc$ is the $j$-th run (i.e., $i=j$) and all the other runs in $\bfc$ are strictly shorter in more than two symbols from the $j$-th run. Then, the $j$-th run is also the first longest run in $\bfy$. In this case $\cD(\bfy) = \cD_{EN}^{n-1} (\bfy)$ which is a contradiction to the definition of $\cD(\bfy)$. Otherwise, the longest run in $\bfc$ is the $j$-th run and there exists $s < j$ such that $r'_s +2 \ge r'_j$, which implies that the $s$-th run is the first longest run in $\bfy$. By Lemma~\ref{lem:endec2}, $D_{EN}^{n-1}$ prolongs the $s$-th run of $\bfy$ by one symbol. Let $\bfc'$ be the word that is  obtained from $\bfy$ by prolonging the $s$-th run by two symbols, it holds that $d_L(\cD_{EN}^{n-1}(\bfy), \bfc')=1$, $d_L(\cD(\bfy), \bfc')=5$ and 
$$\emb(\bfc'; \bfy) = \binom{r'_{s}+2}{2} \ge \binom{r'_{j}}{2} = \emb(\bfc; \bfy).$$ 
Which implies that , 
\begin{align*}
& -2 \binom{r'_{j}}{2}  +4 \binom{r'_{s}+2}{2}   \ge 0.
\end{align*}

\item \textbf{Case 1: $\bfy$ is obtained from $\bfc$ by deleting symbols from two different runs.} There exist $1 \le j < j' \le r'$ such that $\bfy$ is obtained from $\bfc$ by deleting one symbol from the $j$-th run and one symbol from the $j'$-th run. Similarly to the previous case, $\cD(\bfy)$ must be obtained from $\bfc$ by deleting one symbol from the $h$-th run for some $h\neq j , j'$. 
Hence, $\bfc$ is the unique word that is obtained from $\bfy$ by inserting one symbol to the $j$-th run, and one symbol to the $j'$-th run, that is,
$$ I_2 (\bfy) \cap I_1 ( \cD (\bfy) ) = \{ \bfc \}. $$

Note that $\emb(\bfc ; \bfy)=r'_{j}  r'_{j'} $ and that $d_L(\cD(\bfy), \bfc) =1$ and $d_L(\cD_{EN}^{n-1}(\bfy),\bfc) \in \{ 1, 3\}$. Similarly to the previous case we can assume that $d_L(\cD_{EN}^{n-1}(\bfy),\bfc)=3$ and our goal is to find  a word $\bfc' \in I_2 (\bfy)$ such that, 
\begin{align*}
& \emb(\bfc ; \bfy) \left( d_L(\cD(\bfy), \bfc) - d_L(\cD_{EN}^{n-1}(\bfy), \bfc) \right) 
%\\ &  
+ \emb(\bfc' ; \bfy) \left( d_L(\cD(\bfy), \bfc') - d_L(\cD_{EN}^{n-1}(\bfy), \bfc') \right) \ge 0. 
\end{align*}
As in the previous case, if the $i$-th run, which is the first longest run in $\bfc$ satisfies $i \ne j, j'$, the same run is also the first longest run in $\bfy$. Let $\bfc'$ be the word that is obtained from $\bfy$ by prolonging this longest run by two symbols. It holds that $d_L(\cD_{EN}^{n-1}(\bfy), \bfc')=1$, $d_L(\cD(\bfy), \bfc')=5$ and $\emb(\bfc'; \bfy)= \binom{r'_{i}+2}{2}$, and since, $r'_{i} \ge r'_{j}, r'_{j'}$, 
\begin{align*}
& -2 r'_{j}r'_{j'}  +4 \binom{r'_{i}+2}{2}   \ge 0.
\end{align*}

Else, we consider the case in which the first longest run in $\bfc$ is the $j$-th run, or the $j'$-th run (i.e.,  $i\in \{j ,j'\}$), and the same run is also the first longest run in $\bfy$. In this case, it holds that  $d_L(\cD(\bfy), \bfc) = d_L(\cD_{EN}^{n-1} (\bfy), \bfc) =1$ and therefore $\emb(\bfc ; \bfy) \left( d_L(\cD(\bfy), \bfc) - d_L(\cD_{EN}^{n-1}(\bfy), \bfc) \right)=0$. Otherwise, we have that $i\in \{j ,j'\}$, and there exists $s < j, j'$ such that $r'_s +1 \ge r'_j , r'_{j'}$. In other words this run is the first longest run in $\bfy$. By Lemma~\ref{lem:endec2}, $\cD_{EN}^{n-1}$ prolongs this run by one symbol. Assume w.l.o.g. that $r'_j \ge r'_{j'}$ and let $\bfc'$ be the word obtained from $\bfc$ by deleting one symbol from the $j'$-th run and prolonging the $s$-th run by one symbol. In this case $d_L(\cD_{EN}^{n-1}(\bfy), \bfc')=1$, $d_L(\cD(\bfy), \bfc')= 3$ and
\begin{align*}
 \emb(\bfc'; \bfy) = r'_{j} (r'_{s}+1)\ge r'_{j}r'_{j'}   = \emb(\bfc; \bfy).
\end{align*}
Therefore, 
\begin{align*}
& -2 r'_{j}r'_{j'}  +2r'_j({r'_{s}+1})  \ge 0.
\end{align*}
\end{enumerate}
\end{IEEEproof}
Combining the results from the above three claims, we get that $\cD_{\ML^*}^{n-1} = \cD_{EN}^{n-1}$.
\end{IEEEproof}

\begin{lemma}
Let $\bfy \in \Sigma_2^{n-2}$ be a channel output. It holds that, for any $n\ge 5$,
$$|\cD_{\ML^*}(\bfy)|\ne n.$$
\end{lemma}
\begin{IEEEproof}
Assume to the contrary that $|\cD_{\mathsf{ML}^*}(\bfy)|=n$.  We  show that  $$f_{\bfy}(\cD_{\mathsf{ML}^*}(\bfy))\ge f_{\bfy}(\cD_{Lazy}(\bfy)),$$ which is a contradiction to the definition of the ML$^*$ decoder (since the ML$^*$ decoder is defined to return the shortest word that minimizes $f_{\bfy}(\cdot)$). %By Lemma~\ref{lem:decoder_EN_n}, $\cD_{EN}^{n}(\bfy)$ is the decoder that minimizes $f_{\bfy}(\cD(\bfy))$ among all other decoders that return a word of length $n$ for the channel output $\bfy$.  Hence, it is enough to show that $f_{\bfy}(\cD_{EN}^n(\bfy)) - f_{\bfy}(\cD_{Lazy}(\bfy)) \ge 0$.  
%If $|\cD_{AL}(\bfy)|=n$, it holds that:
%	\begin{align*}
%	P_{\mathrm{err}}(2\textrm{-}\mathsf{Del},(\Sigma_2)^{n},\cD_{AL},d_L) \geq P_{\mathrm{err}}(2\textrm{-}\mathsf{Del},(\Sigma_2)^{n},\cD_{\mathrm{Lazy}},d_L).
%	\end{align*}
%\end{lemma}  
%Since for any $\bfy' \neq \bfy$, the decoders $\cD_{AL}$ and $\cD_{Lazy}$ returns the same output, it is enough to show that:

First we note that if $\cD_{\ML^*}(\bfy)$ is not a supersequence of $\bfy$, we have that $d_L(\cD_{\ML^*}(\bfy), \bfc) \ge 3 $, and thus $f_{\bfy}(\cD_{\mathsf{ML}^*}(\bfy))\ge f_{\bfy}(\cD_{Lazy}(\bfy))$. Therefore, let us consider the case in which $\cD_{\mathsf{ML}^*}$ returns a word of length $n$ that is a supersequence of $\bfy$ and therefore any possible output of $\cD_{\mathsf{ML}^*}$ is either of distance $0, 2,$ or $4$ from the transmitted word $\bfc$. Hence, 

\begin{align*}
& f_{\bfy}(\cD_{\mathsf{ML}^*}(\bfy)) - f_{\bfy}(\cD_{Lazy}(\bfy))
 \\ & \ \ \ \ = \sum_{\bfc \in I_2(\bfy)}   \frac{p ( \bfy| \bfc )}{|\bfc|}  \left( {d_L(\cD_{\mathsf{ML}^*}(\bfy),\bfc)}    -   {d_L(\cD_{Lazy}(\bfy),\bfc)} \right)
 \\ & \ \ \ \  \overset{\mathrm{(a)}}{=}  \sum_{\substack{\bfc \in I_2(\bfy)  \\d_L(\cD_{\mathsf{ML}^*}(\bfy),\bfc)=4} }  \frac{p ( \bfy| \bfc )}{|\bfc|}  \left( 4-2 \right)
%\\ & \hspace{5ex}
  +  \sum_{\substack{\bfc \in I_2(\bfy)  \\d_L(\cD_{\mathsf{ML}^*}(\bfy),\bfc)=2} }  \frac{p ( \bfy| \bfc )}{|\bfc|}  \left( 2-2 \right)
 %\\ & \hspace{5ex}  
 + \sum_{\substack{\bfc \in I_2(\bfy)  \\d_L(\cD_{\mathsf{ML}^*}(\bfy),\bfc)=0} }  \frac{p ( \bfy| \bfc )}{|\bfc|}  \left( 0-2 \right)
\\ & \ \ \ \ \overset{\mathrm{(b)}}{=}  \frac{2}{n} \left( \sum_{\substack{\bfc \in I_2(\bfy)  \\d_L(\cD_{\mathsf{ML}^*}(\bfy),\bfc)=4} }  {p ( \bfy| \bfc )} 
  - \sum_{\substack{\bfc \in I_2(\bfy)  \\d_L(\cD_{\mathsf{ML}^*}(\bfy),\bfc)=0} }  {p ( \bfy| \bfc )} \right),
\end{align*}
where (a) holds since $d_L(\cD_{Lazy}(\bfy),\bfc)=2$ for every $\bfc \in I_2(\bfy)$ and (b) holds since $|\bfc|=n$.

Denote,
\begin{align*}
& \cS um_4 \triangleq \sum_{\substack{\bfc \in I_2(\bfy)\\  d_L(\cD_{\mathsf{ML}^*}(\bfy),\bfc)=4 }}   p( \bfy| \bfc ),
\\ & \cP_0 \triangleq \sum_{\substack{\bfc \in I_2(\bfy)\\  d_L(\cD_{\mathsf{ML}^*}(\bfy),\bfc)=0 }}   p( \bfy| \bfc ) = p\left(\bfy | \cD_{\mathsf{ML}^*}(\bfy)\right).
\end{align*}
From the above discussion, our objective is to prove that $\cS um_4 \geq  \cP_0 $. Recall that $|I_2(\bfy)|= \binom{n}{2} +n+1$. Let the $i$-th, $i'$-th run be the first, second longest run of $\bfy$, respectively, and denote their lengths by $r_i \ge r_{i'}$. We will bound the number of possible words $\bfc \in I_2(\bfy)$ such that $d_L(\cD_{\ML^*}(\bfy),\bfc) =4$. \\\\
\textbf{Case 1:} $\cD_{\mathsf{ML}^*}$ prolongs one run of  $\bfy$ by two symbols. We denote the index of the run by $i'$ and its length by $r_{i'}$.  There is one word $\bfc \in I_2(\bfy)$ such that $d_L(\cD_{\mathsf{ML}^*}(\bfy),\bfc) =0$. Note that the set of words $\bfc \in I_2(\bfy)$ such that $d_L(\cD_{\mathsf{ML}^*}(\bfy),\bfc) =2$ consists of words $\bfc$ that can be obtained from $\bfy$ by prolonging the $i'$-th run by exactly one symbol. Consider the word $\bfy'$, which is the word obtained from $\bfy$ by prolonging the $i'$-th run by exactly one symbol. $\bfy'$ is a word of length $n-1$, and the words $\bfc$, such that $d_L(\cD_{\mathsf{ML}^*}(\bfy), \bfc)=2$ are all the words in the radius-$1$ insertion ball centered at $\bfy'$ expect to the word $\cD_{\mathsf{ML}^*} (\bfy)$. The number of such words is
$$I_1(\bfy')-1=n+1-1= n. $$
Hence, there are $\binom{n}{2}$ words $\bfc \in I_2(\bfy)$ for which ${d_L(\cD_{\mathsf{ML}^*}(\bfy),\bfc) =4}$ and the conditional probability of each of these words is $p(\bfy | \bfc) \geq \dfrac{1}{\binom{n}{2}}$.  
Therefore, 
$$\cS um_4 = \sum_{\substack{\bfc \in I_2(\bfy) \\ d_L(\cD_{\mathsf{ML}^*}(\bfy),\bfc)=4 }}  p(\bfy | \bfc) \geq \binom{n}{2} \cdot \dfrac{1}{\binom{n}{2}} =1 .$$  
On the other hand,  $$ \cP_0  =  \dfrac{\binom{r_{i'}+2}{2} } {\binom{n}{2}} \leq 1, $$ which implies $ \cS um_4\geq  \cP_0  $ for every $n>0$ and thus, 
 $$ f_{\bfy}(\\cD_{\mathsf{ML}^*}(\bfy)) - f_{\bfy}(\cD_{Lazy}(\bfy)) \ge 0. $$\\
\textbf{Case 2:} $\cD_{\mathsf{ML}^*} (\bfy)$ prolongs two runs of $\bfy$, each by one symbol. We assume the indices of the runs are given by $i'$ and $j'$ and their corresponding lengths by $r_{i'}$ and $r_{j'}$. %By Lemma~\ref{lem:endec2}, we know that $\cD_{\mathsf{ML}^*} (\bfy)$ prolongs these two runs if and only if $\binom{r_i}{2} \leq r_i  r_{i'}$, and consequently,  $\dfrac{r_i-1}{2} \leq r_{i'} < r_i$. 

The only word $\bfc \in I_2(\bfy)$ that satisfies $d_L(\cD_{\mathsf{ML}^*}(\bfy),\bfc) =0$ is the word $\bfc = \cD_{\mathsf{ML}^*}(\bfy)$. In addition the set of words $\bfc \in I_2(\bfy)$ such that $d_L(\cD_{\mathsf{ML}^*}(\bfy),\bfc) =2$ consists of words $\bfc$ that can be obtained from $\bfy$ by prolonging either the $i'$-th run or the $j'$-run by exactly one symbol. Let $\bfy'$ be the word obtained from $\bfy$ by prolonging the $i'$-th run by one symbol and let $\bfy''$ be the word obtained from $\bfy$ by prolonging the $j'$-th run by one symbol. Similarly to the first case the number of such words is
$$I_1(\bfy')-1+I_1(\bfy'')-1=2n,$$
which implies that the number of words $\bfc \in I_2(\bfy)$ such that $d_L(\cD_{\mathsf{ML}^*}(\bfy),\bfc) =4$ is $\binom{n}{2} -n$ and the conditional probabilities of these words satisfy $ p(\bfy | \bfc) \geq \dfrac{1}{\binom{n}{2}}$.  
Hence, 
$$\cS um_4 = \sum_{\substack{\bfc \in I_2(\bfy)\\ d_L(\cD_{\mathsf{ML}^*}(\bfy),\bfc)=4 }} p(\bfy | \bfc)  \geq  \dfrac{\binom{n}{2}  -n }{\binom{n}{2}}.$$
%The decoder $\cD_{AL}$ maximizes the embedding number of $\cD_{AL}(\bfy)$. Therefore, it implies that $\binom{r_1}{2} \leq r_1 \cdot r_2$. Consequently, we can say that $\dfrac{r_1-1}{2} < r_2 < r_1$ and following that it holds that $r_1 \leq \dfrac{2}{3} \cdot n$. 
On the other hand, 
\begin{align*}
  \cP_0  & = \dfrac{(r_{i'} +1)  (r_{j'} +1)} {\binom{n}{2}} \overset{\mathrm{(a)}}{\leq} \dfrac{(r_{i'}+1)  (n-r_{i'}-1) } {\binom{n}{2}} 
%\\ &
\overset{\mathrm{(b)}}{\leq}  \dfrac{(\frac{n}{2}-1)^2 } {\binom{n}{2}} = \dfrac{\frac{n^2}{4} - n +1 } {\binom{n}{2}}, 
\end{align*}
where (a) holds since $r_{i'} + r_{j'}\leq n-2$ and (b) holds since the maximum of the function $f(x) = x(n-x)$ is achieved for $x=n/2$. Hence, $\cS um_4  \geq  \cP_0 $ when $\frac{n^2}{4} - n +1\leq \binom{n}{2}-n$, which holds for any $n \geq 4$. Thus, for $n \geq 4$, 
 $$ f_{\bfy}(\cD_{\mathsf{ML}^*}(\bfy)) - f_{\bfy}(\cD_{Lazy}(\bfy)) \ge 0. $$

\noindent \textbf{Case 3:} $\cD_{\mathsf{ML}^*}$ prolongs one run of in $\bfy$ by one symbol and creates a new run. We denote the index of the run by $i'$ and its length by $r_{i'}$. The only word $\bfc \in I_2(\bfy)$ that satisfies $d_L(\cD_{\mathsf{ML}^*}(\bfy),\bfc) =0$ is the word $\bfc = \cD_{\mathsf{ML}^*}(\bfy)$. In addition the set of words $\bfc \in I_2(\bfy)$ such that $d_L(\cD_{\mathsf{ML}^*}(\bfy),\bfc) =2$ consists of words $\bfc$ that can be obtained from $\bfy$ by prolonging either the $i'$-th run or by introducing the new run. Let $\bfy'$ be the word obtained from $\bfy$ by prolonging the $i'$-th run by one symbol and let $\bfy''$ be the word obtained from $\bfy$ by introducing the same run as $\cD_{\mathsf{ML}^*}$. Similarly to the previous case the number of such words is
$I_1(\bfy')-1+I_1(\bfy'')-1=2n,$ and   
hence, 
$\cS um_4 = \sum_{\substack{\bfc \in I_2(\bfy)\\ d_L(\cD_{\mathsf{ML}^*}(\bfy),\bfc)=4 }} p(\bfy | \bfc)  \geq  \dfrac{\binom{n}{2}  -n }{\binom{n}{2}}.$ 

Additionally, we have that, 
$
  \cP_0   = \dfrac{(r_{i'} +1)} {\binom{n}{2}} \leq \dfrac{(n-1) } {\binom{n}{2}}$. Thus, for $n \geq 5$, 
  $f_{\bfy}(\cD_{\mathsf{ML}^*}(\bfy)) - f_{\bfy}(\cD_{Lazy}(\bfy)) \ge 0. $

\noindent \textbf{Case 4:} $\cD_{\mathsf{ML}^*}$ creates two new runs in $\bfy$.  In this case, it should be noted that the inserted two symbols can creates an alternating sequence of length which is bounded by $n$.  Thus, from the same arguments as in the previous case we have that for $n \ge 5$,  $f_{\bfy}(\cD_{\mathsf{ML}^*}(\bfy)) - f_{\bfy}(\cD_{Lazy}(\bfy)) \ge 0. $
\end{IEEEproof}

\begin{lemma}
Let $\bfy \in \Sigma_2^{n-2}$ be a channel output. For any decoder $\cD$, such that $\cD(\bfy)$ is not a supersequence of $\bfy$ and $|\cD(\bfy)|~=~n+1$, it holds that 
 \begin{align*}
  f_\bfy(\cD(\bfy)) \ge f_{\bfy}(\cD_{EN}^{n-1}(\bfy)).
\end{align*}

\end{lemma}  
\begin{IEEEproof}
Since $\cD(\bfy)$ is not a supersequence of $\bfy$, it is also not a supersequence of the transmitted word $\bfc$. Therefore, for each $\bfc \in I_2(\bfy)$ it holds that $d_L(\cD(\bfy),\bfc) \ge 3$, while $d_L(\cD_{EN}^{n-1}(\bfy),\bfc) \le 3$. Thus,

\vspace{-1ex}
\begin{small}
\begin{align*}
& f_\bfy(\cD(\bfy)) - f_{\bfy}(\cD_{EN}^{n-1}(\bfy)) \\& =  \sum_{\bfc\in I_2(\bfy)}  \frac{d_L(\cD(\bfy),\bfc)}{|\bfc|} p (  \bfy| \bfc ) -  \sum_{\bfc\in I_2(\bfy)}  \frac{d_L(\cD_{EN}^{n-1}(\bfy),\bfc)}{|\bfc|} p (  \bfy| \bfc ) \\
& = \frac{1}{|\bfc|} \left( \sum_{\bfc\in I_2(\bfy)} d_L(\cD(\bfy),\bfc) p (  \bfy| \bfc ) - \sum_{\bfc\in I_2(\bfy)} d_L(\cD_{EN}^{n-1}(\bfy),\bfc) p (  \bfy| \bfc ) \right) \\ 
& = \frac{1}{|\bfc|} \sum_{\bfc\in I_2(\bfy)} p (  \bfy| \bfc ) \left(  d_L(\cD(\bfy),\bfc) -  d_L(\cD_{EN}^{n-1}(\bfy),\bfc) \right) \\
& \ge  \frac{1}{|\bfc|} \sum_{\bfc\in I_2(\bfy)} p (  \bfy| \bfc ) \left( 3-3 \right) \ge 0.
 \end{align*}
 \end{small}
\end{IEEEproof}

\begin{lemma}
Let $\bfy \in \Sigma_2^{n-2}$ be a channel output. For any decoder $\cD$, such that $\cD(\bfy)$ is a supersequence of $\bfy$ and $|\cD(\bfy)|=n+1$, it holds that 
 \begin{align*}
  f_\bfy(\cD(\bfy)) \ge f_{\bfy}(\cD_{EN}^{n-1}(\bfy)).
\end{align*}

\end{lemma}  
\begin{IEEEproof}
From similar arguments to those presented in Lemma~\ref{lem:decoder_EN_n-1}, our goal is to prove that (\ref{inequlaity_avg_dec}) holds for $\cD(\bfy)$ and $ \cD_{EN}^{n-1}(\bfy)$, i.e., to prove that 
\begin{align*}
\sum_{\bfc\in I_2(\bfy)}  \emb(\bfc; \bfy)  \left(  d_L(\cD(\bfy),\bfc) -  d_L(\cD_{EN}^{n-1}(\bfy),\bfc) \right)  \ge 0.
\end{align*}

\begin{comment}
Using these observations we get that,
\begin{align*}
f_{\bfy}(\cD_{EN}^{n+1}(\bfy)) & =   \sum_{\bfc \in \Sigma_2^n}  \frac{d(\cD(\bfy),\bfc)}{|\bfc|} p(\bfy|\bfc) 
\\ &=  \sum_{\bfc \in I_2(\bfy)}   \frac{d(\cD_{EN}^{n+1}(\bfy),\bfc)}{|\bfc|}\cdot p(\bfy|\bfc)    
 \\ & =  \sum_{\substack{\bfc \in I_2(\bfy) \\ d(d(\cD_{EN}^{n+1}(\bfy),\bfc)= 1}}  \frac{1}{|\bfc|}\cdot p(\bfy|\bfc)   
 \\ &+ \sum_{\substack{\bfc \in I_2(\bfy) \\ d(d(\cD_{EN}^{n+1}(\bfy),\bfc)=3 }}  \frac{3}{|\bfc|}\cdot p(\bfy|\bfc)   
 \\ &+ \sum_{\substack{\bfc \in I_2(\bfy) \\ d(d(\cD_{EN}^{n+1}(\bfy),\bfc)=5 }}  \frac{5}{|\bfc|}\cdot p(\bfy|\bfc)   
 \end{align*}
And from the same reasons, 
\begin{align*}
f_{\bfy}(\cD_{EN}^{n-1}(\bfy)) & =   \sum_{\bfc \in \Sigma_2^n}  \frac{d(\cD(\bfy),\bfc)}{|\bfc|} p(\bfy|\bfc) 
\\ &=  \sum_{\bfc \in I_2(\bfy)}   \frac{d(\cD_{EN}^{n-1}(\bfy),\bfc)}{|\bfc|}\cdot p(\bfy|\bfc)    
 \\ & =  \sum_{\substack{\bfc \in I_2(\bfy)\\ d(d(\cD_{EN}^{n-1}(\bfy),\bfc)= 1}}  \frac{1}{|\bfc|}\cdot p(\bfy|\bfc)   
 \\ &+ \sum_{\substack{\bfc \in I_2(\bfy) \\ d(d(\cD_{EN}^{n-1}(\bfy),\bfc)=3 }}  \frac{3}{|\bfc|}\cdot p(\bfy|\bfc)   
 \end{align*}
 \end{comment}
Assume that the number of runs in $\bfy$ is $\rho(\bfy)=r$, let $r_j$ denote the length of the $j$-th run for $1\le j\le r$, and let the $i$-th run of $\bfy$ be the first longest run of $\bfy$. 
Note that the Levenshtein distance of $\cD_{EN}^{n-1}(\bfy)$ from the transmitted word $\bfc$ can be either $1$ or $3$. 
Similarly, $\cD(\bfy)$  can have distance of $1$, $3$ or $5$ from $\bfc$. Recall that $\cD_{EN}^{n-1}$ prolongs the $i$-th run by one symbol and that $I_1(\cD_{EN}^{n-1}(\bfy)) \subseteq I_2 (\bfy) $. $\cD(\bfy)$ is a supersequence of $\bfy$, and hence $\cD(\bfy)$ is obtained from $\bfy$ by prolonging existing runs or by creating new runs in $\bfy$. 
From the discussion above, for every word $\bfc \in I_2(\bfy)$ such that 
$$\bfc \notin \left( I_1(\cD_{EN}^{n-1}(\bfy)) \cup D_1(\cD(\bfy) )\right),$$ 
it holds that $d_L(\cD_{EN}^{n-1}(\bfy), \bfc)=3$ while $d_L(\cD(\bfy), \bfc) \ge 3$.
 Additionally, every word $\bfc \in I_2(\bfy)$ such that 
 $$\bfc \in \left( I_1(\cD_{EN}^{n-1}(\bfy)) \cap D_1(\cD(\bfy) ) \right),$$ 
 satisfies $d_L(\cD_{EN}^{n-1}(\bfy), \bfc)=d_L(\cD(\bfy), \bfc)=1$. 
 Hence, for these words it holds that $d_L(\cD(\bfy), \bfc)-d_L(\cD_{EN}^{n-1}(\bfy), \bfc)\ge 0$ and they can be eliminated from inequality (\ref{inequlaity_avg_dec}). 
 In order to complete the proof, the words $\bfc \in I_2(\bfy)$ such that $$\bfc \in I_1(\cD_{EN}^{n-1}(\bfy)) \text{ and } \bfc \notin D_1(\cD(\bfy) )$$ and the words $\bfc \in I_2(\bfy)$ such that $$\bfc \notin I_1(\cD_{EN}^{n-1}(\bfy)) \text{ and } \bfc \in D_1(\cD(\bfy) )$$ should be considered. 
For words in the first case it holds that ${d_L(\cD_{EN}^{n-1}(\bfy), \bfc)=1}$ and $d_L(\cD(\bfy), \bfc)\ge 3$, while for words in the second case, $d_L(\cD_{EN}^{n-1}(\bfy), \bfc)=3$ and $d_L(\cD(\bfy), \bfc)\ge 1$. Hence, 
\begin{align*}
&  \sum_{\bfc\in I_2(\bfy)}  \emb(\bfc; \bfy)  \left(  d_L(\cD(\bfy),\bfc) -  d_L(\cD_{EN}^{n-1}(\bfy),\bfc) \right) 
\\&  \ge \sum_{\substack{\bfc \in I_2 (\bfy)  \\ \bfc \in I_1(\cD_{EN}^{n-1}(\bfy)) \\ \bfc \notin D_1(\cD(\bfy) )}} \emb(\bfc ; \bfy) \left( d_L(\cD(\bfy) , \bfc) - d_L(\cD_{EN}^{n-1}(\bfy), \bfc) \right)
\\ & + \sum_{\substack{\bfc \in I_2 (\bfy)  \\ \bfc \notin I_1(\cD_{EN}^{n-1}(\bfy)) \\ \bfc \in  D_1(\cD(\bfy) )}} \emb(\bfc ; \bfy) \left( d_L(\cD(\bfy) , \bfc) - d_L(\cD_{EN}^{n-1}(\bfy), \bfc) \right)
\\&  \ge 2 \sum_{\substack{\bfc \in I_2 (\bfy)  \\ \bfc \in I_1(\cD_{EN}^{n-1}(\bfy)) \\ \bfc \notin D_1(\cD(\bfy) )}}  \emb(\bfc ; \bfy)  - 2\sum_{\substack{\bfc \in I_2 (\bfy)  \\ \bfc \notin I_1(\cD_{EN}^{n-1}(\bfy)) \\ \bfc \in  D_1(\cD(\bfy) )}}  \emb(\bfc ; \bfy).
\end{align*}

We first assume that $\cD(\bfy)$ is obtained from $\bfy$ by prolonging the $i$-th run by exactly one symbol. Let $\bfc \in I_2(\bfy)$ and consider the cases mentioned above. 
\begin{enumerate}
\item {$\bfc \in I_1(\cD_{EN}^{n-1}(\bfy)) \text{ and } \bfc \notin D_1(\cD(\bfy) )$}: Recall that both decoders return supersequences of $\bfy$. By the assumption $\cD(\bfy)$ is obtained from $\bfy$ by prolonging the $i$-th run by one symbol and then performing two more insertions to the obtained word. Since $\bfc \in I_1(\cD_{EN}^{n-1}(\bfy) )$, $\bfc$ must be obtained from $\bfy$ by prolonging the $i$-th run and performing one more insertion. $\bfc \notin D_1(\cD(\bfy))$, and therefore the number of such words equals to 
\begin{align*}
& |I_1(\cD_{EN}^{n-1} (\bfy))| 
\\ & \hspace{5ex}- \left| \left\{ \bfc \in I_2(\bfy) \ : \ \bfc \in I_1(\cD_{EN}^{n-1}(\bfy)) \cap D_1(\cD(\bfy) ) \right \} \right|.
\end{align*}
Note that 
$$\left| \left\{ \bfc \in I_2(\bfy) \ : \ \bfc \in I_1(\cD_{EN}^{n-1}(\bfy)) \cap D_1(\cD(\bfy) ) \right \} \right| \le 2$$
since the words in the latter intersection are the words that obtain from $\bfy$ by prolonging the $i$-th run by one symbol and then performing one of the two other insertions performed to receive $\cD(\bfy)$.  Hence, there are at least $|I_1(\cD_{EN}^{n-1} (\bfy))|-2 = n-1$ such words in this case and for each of them $\emb(\bfc ; \bfy) \ge (r_i+1) $. Recall that these words satisfy $d(\cD_{EN}^{n-1}(\bfy), \bfc)=1$ and $d(\cD(\bfy), \bfc)\ge 3$. 

\item {$\bfc \notin I_1(\cD_{EN}^{n-1}(\bfy)) \text{ and } \bfc \in D_1(\cD(\bfy) )$}:  By the assumption, $\cD$ prolongs the $i$-th run by one symbol and performs two more insertions into the obtained word and $\cD_{EN}^{n-1}$ prolongs the $i$-th run by one symbol. Hence, the words $\bfc \in I_2(\bfy)$ such that $\bfc \notin I_1(\cD_{EN}^{n-1}(\bfy)) \text{ and } \bfc \in D_1(\cD(\bfy) )$ can not be obtained from $\bfy$ by prolonging the $i$-th run. Therefore, it implies that $\bfc$ is the unique word obtained from $\cD(\bfy)$ by deleting the symbol that was inserted to the $i$-th run of $\bfy$.  %Otherwise, $\bfc$ is the word obtained from $\cD(\bfy)$ by deleting one symbol from the prolonged or new runs that were created by $\cD$. Since $\cD$ returns a supersequence of $\bfy$, we have at most three prolonged or new runs and hence,
%$$|\{\bfc | \bfc \in I_2 (\bfy), \bfc \notin I_1(\cD_{EN}^{n-1}(\bfy)), \bfc \in D_1(\cD_{EN}^{n+1}(\bfy) ) \} | \le 3.$$ 
It holds that ${\emb(\bfc;\bfy)   \le (r_{i}+1)^2}$ and $d_L(\cD_{EN}^{n-1}(\bfy), \bfc)=3$ and $d_L(\cD(\bfy), \bfc) = 1$. 
\end{enumerate}
Note that $r_i \le n-2$ since it is the length of the $i$-th run of $\bfy \in \Sigma_2^{n-2}$.
Thus, 
\begin{align*}
&   2 \sum_{\substack{\bfc \in I_2 (\bfy)  \\ \bfc \in I_1(\cD_{EN}^{n-1}(\bfy)) \\ \bfc \notin D_1(\cD(\bfy) )}}  \emb(\bfc ; \bfy)  - 2\sum_{\substack{\bfc \in I_2 (\bfy)  \\ \bfc \notin I_1(\cD_{EN}^{n-1}(\bfy)) \\ \bfc \in  D_1(\cD(\bfy) )}}  \emb(\bfc ; \bfy)
\\ & \ge 2(n-1)(r_i+1) -2 \cdot(r_i+1)^2 
%\\ & 
\ge 2(r_i+1)^2 -2 \cdot(r_i+1)^2   \ge 0.
 \end{align*}
 
 Second we assume that $\cD(\bfy)$ is obtained from $\bfy$ by prolonging the $i$-th run by at least two symbols. In this case, it holds that $\left(D_1(\cD(\bfy)) \cap I_2(\bfy) \right) \subseteq I_1(\cD_{EN}^{n-1}(\bfy))$, which implies that 
$$\left| \left\{ \bfc \in I_2 (\bfy) \ : \ \bfc \notin I_1(\cD_{EN}^{n-1}(\bfy)) \text{ and } \bfc \in D_1(\cD(\bfy) ) \right\} \right| = 0,$$
 and therefore,
  \begin{align*}
&   2 \sum_{\substack{\bfc \in I_2 (\bfy)  \\ \bfc \in I_1(\cD_{EN}^{n-1}(\bfy)) \\ \bfc \notin D_1(\cD(\bfy) )}}  \emb(\bfc ; \bfy)  - 2\sum_{\substack{\bfc \in I_2 (\bfy)  \\ \bfc \notin I_1(\cD_{EN}^{n-1}(\bfy)) \\ \bfc \in  D_1(\cD(\bfy) )}}  \emb(\bfc ; \bfy) \ge 0.
 \end{align*}

Lastly, we assume that $\cD(\bfy)$ is obtained from $\bfy$ by three insertions such that neither of these insertions prolongs the $i$-th run. For this scenario, we first note that  it is possible that the three symbols that are inserted by $\cD$ creates (or prolongs) an alternating sequence which is adjacent to the $i$-th run.  In this case, we have that, $\left| \left\{ \bfc \in I_2(\bfy) \ : \ \bfc \in I_1(\cD_{EN}^{n-1}(\bfy)) \cap D_1(\cD(\bfy) ) \right \} \right| = 2$, where the two words are obtained by either prolonging the alternating sequence by two symbols, or by adding one symbol to the $i$-th run, and one additional symbol. Therefore, the number of words $\bfc \in I_2(\bfy)$ such that $\bfc \in I_1(\cD_{EN}^{n-1}(\bfy)) \text{ and } \bfc \notin D_1(\cD(\bfy) )$ equals to $|I_1(\cD_{EN}^{n-1}(\bfy))|-2 = n-1$.

For any such word $\bfc$ it holds that $\emb (\bfc;\bfy) \ge r_i+1$. Furthermore, $ |D_1(\cD(\bfy))| $ equals to the number of runs in $\cD(\bfy)$~\cite{L66} and any $\bfc \in D_1(\cD(\bfy)) \cap I_2(\bfy)$ is obtained from $\cD(\bfy)$ by deleting one of the three symbols that were inserted into $\bfy$ in order to obtain  $\cD(\bfy)$.  Hence, there are at most three such words, and each is obtained by deleting one of the three inserted symbols. Let $\bfc$ be one of those words. If the two remaining symbols belong to the same run, then $\emb(\bfc; \bfy)=\binom{m}{2}$ where $m$ is the length of this run in $\bfc$ and $m \le r_i+2$. In this case consider the word $\bfc'$ that is obtained by prolonging the $i$-th run of $\bfy$ by two symbols. It holds that, 
 $$\emb (\bfc'; \bfy) = \binom{r_i+2}{2} \ge \binom{m}{2} = \emb(\bfc ;\bfy).$$

Otherwise,  $\emb(\bfc; \bfy)=m_1 m_2$ where $m_1$ and $m_2$ are the lengths of the runs that include the remaining inserted symbols and $m_1, m_2 \le r_i+1$. Let $\bfc'$ be the word that is obtained from $\bfy$ by prolonging the $i$-th run and the run of length $\max \{ m_1 -1, m_2-1\}$ that is prolonged by $\cD$. In this case, 
 $$\emb (\bfc'; \bfy) = m_1({r_i+1}) \ge {m_1}{m_2} = \emb(\bfc ;\bfy).$$
Note that there is at most one such word $\bfc$ that is obtained by prolonging the same run with two symbols, which implies that there is always a selection of words $\bfc'$ such that,
\begin{align*} 
&   2 \sum_{\substack{\bfc \in I_2 (\bfy)  \\ \bfc \in I_1(\cD_{EN}^{n-1}(\bfy)) \\ \bfc \notin D_1(\cD(\bfy) )}}  \emb(\bfc ; \bfy)  - 2\sum_{\substack{\bfc \in I_2 (\bfy)  \\ \bfc \notin I_1(\cD_{EN}^{n-1}(\bfy)) \\ \bfc \in  D_1(\cD(\bfy) )}}  \emb(\bfc ; \bfy) \ge 0.
 \end{align*} 
 We proved that for any decoder $\cD$ such that $\cD(\bfy)$ is a supersequence $\bfy$ and $|\cD(\bfy)|=n+1$,
 \begin{align*} 
&   2 \sum_{\substack{\bfc \in I_2 (\bfy)  \\ \bfc \in I_1(\cD_{EN}^{n-1}(\bfy)) \\ \bfc \notin D_1(\cD(\bfy) )}}  \emb(\bfc ; \bfy)  - 2\sum_{\substack{\bfc \in I_2 (\bfy)  \\ \bfc \notin I_1(\cD_{EN}^{n-1}(\bfy)) \\ \bfc \in  D_1(\cD(\bfy) )}}  \emb(\bfc ; \bfy) \ge 0.
 \end{align*} 
Thus, 
$$ f_\bfy(\cD(\bfy)) - f_{\bfy}(\cD_{EN}^{n-1}(\bfy)) \ge 0.$$

\end{IEEEproof}

From the previous lemmas it holds that for a given channel output $\bfy \in \Sigma_2^{n-2}$, the length of $\cD_{\textsf{ML}^*} (\bfy)$ is either $n-1$ or~$n-2$. Lemma~\ref{lem:decoder_EN_n-1} implies that if $|\cD_{\textsf{ML}^*} (\bfy)|=n-1$, then $\cD_{\textsf{ML}^*} (\bfy)=\cD_{EN}^{n-1}(\bfy)$.  In the following result we define a condition on the length of the longest run in $\bfy$ to decide whether prolonging  it by one symbol can minimize the expected normalized distance. In other words, this result defines a criteria on a given channel output $\bfy$ to define whether using the same output as $\cD_{{Lazy}}$ or using the same output as $\cD_{EN}^{n-1}$ is better in terms of minimizing $f_\bfy(\cD(\bfy))$ (and therefore minimizing the expected normalized distance).  An immediate conclusion of this result is Theorem~\ref{Th:2-del} which determines the ML$^*$ decoder for the case of a single 2-deletion channel.   
\begin{lemma}
Let $\bfy \in \Sigma_2^{n-2}$ be a channel output, such that the number of runs in $\bfy$ is $\rho(\bfy)=r$, and the first longest run in $\bfy$ is the $i$-th run. Denote by $r_j$ the length of the $j$-th for $1\le j \le r$.
It holds that 
 $$f_\bfy(\cD_{EN}^{n-1}(\bfy)) - f_{\bfy}(\cD_{Lazy}(\bfy)) \ge 0$$
 if and only if 
 $$2n^2-4nr_i -6n+r_i^2+3r_i+r+1 \geq 0.$$
\begin{comment}
 \begin{align*}
&\sum_{\bfc \in I_2(\bfy)}  \frac{d(\cD_{EN}^{n-1}(\bfy),\bfc)}{|\bfc|}\cdot \pr_\ch\{  \bfy\textmd{ rec. }| \bfc \textmd{ trans.}\} \geq 
\\ &\sum_{\bfc \in I_2(\bfy)}  \frac{d(\cD_{\mathsf{Lazy}}(\bfy),\bfc)}{|\bfc|}\cdot \pr_\ch\{  \bfy\textmd{ rec. }| \bfc \textmd{ trans.}\}
\end{align*}
\end{comment}
\end{lemma}  

\begin{IEEEproof}
By Lemma~\ref{lem:endec1}, $\cD_{EN}^{n-1}$ prolongs the $i$-th run of $\bfy$ by one symbol. Therefore, the Levenshtein distance of $\cD_{EN}^{n-1}(\bfy)$ from the transmitted word $\bfc$ can be either $1$ or $3$.  Hence, 
\begin{align*}
& f_\bfy(\cD_{EN}^{n-1}(\bfy)) - f_{\bfy}(\cD_{Lazy}(\bfy))
 \\& = \sum_{\bfc \in I_2(\bfy)} \frac{p(\bfy | \bfc)}{|\bfc|} \left(   {d_L(\cD_{EN}^{n-1}(\bfy),\bfc)}  -  {d_L(\cD_{{Lazy}}(\bfy),\bfc)} \right)    
  \\ & =\sum_{\substack{\bfc \in I_2(\bfy)\\  d_L(\cD_{EN}^{n-1}(\bfy),\bfc)= 3}}  \frac{p(\bfy | \bfc)}{|\bfc|} \left(   3-2 \right)  
  %\\ & \hspace{5ex}
  +\sum_{\substack{\bfc \in I_2(\bfy)\\  d_L(\cD_{EN}^{n-1}(\bfy),\bfc)= 1}}  \frac{p(\bfy | \bfc)}{|\bfc|} \left(   1-2 \right)   
    \\ & =\frac{1}{n} \left( \sum_{\substack{\bfc \in I_2(\bfy)\\  d_L(\cD_{EN}^{n-1}(\bfy),\bfc)= 3}}  {p(\bfy | \bfc)}   -\sum_{\substack{\bfc \in I_2(\bfy)\\  d_L(\cD_{EN}^{n-1}(\bfy),\bfc)= 1}}  {p(\bfy | \bfc)}\right).  
\end{align*}
Denote
\begin{align*}
& \cS um_3 \triangleq \sum_{\substack{\bfc \in I_2(\bfy)\\  d_L(\cD_{EN}^{n-1}(\bfy),\bfc)=3 }}   p( \bfy| \bfc ),
\\ & \cS um_1 \triangleq \sum_{\substack{\bfc \in I_2(\bfy)\\  d_L(\cD_{EN}^{n-1}(\bfy),\bfc)=1 }}   p( \bfy| \bfc ).
\end{align*}  Let us prove that $$2n^2-4nr_i -6n+r_i^2+3r_i+r+1 \geq 0$$ is a necessary and sufficient condition for the inequality $\cS um_3\geq \cS um_1$ to hold. First, we count the number of words $\bfc \in I_2(\bfy)$ such that $d_L(\cD_{EN}^{n-1}(\bfy),\bfc)=1$. Each such $\bfc$  is a supersequence of $\cD_{EN}^{n-1}(\bfy)$ and therefore $\bfc$ can be obtained from $\bfy$ only by one of the three following ways. The first way is by prolonging the $i$-th run and the $j$-th of $\bfy$ for $j\ne i$, each by one symbol. The number of such words is $r-1$. The second way is by prolonging the $i$-th run in $\bfy$ by one symbol and creating a new run in $\bfy$. The number of options to create a new run in $\bfy$ is $n-r+1$ and therefore, there are $n-r+1$ such words. The third way is by prolonging the $i$-th run by two symbols and there is only one such word. 
Hence, the total number of words $\bfc \in I_2(\bfy)$ such that ${d_L(\cD_{EN}^{n-1}(\bfy),\bfc)=1}$ is $n+1=| I_1(\cD_{EN}^{n-1}(\bfy))|$.  Among them, the $r-1$ words that are obtained by the first way has an embedding number of $\emb(\bfc; \bfy)={(r_i+1)(r_j+1)}$. Similarly the $n-r+1$ words that are obtained from $\bfy$ using the second way satisfy $\emb(\bfc;\bfy)=r_i+1$.  Lastly, for the word $\bfc$ that is obtained by prolonging the $i$-th run of $\bfy$ by two symbols it holds that $\emb(\bfc;\bfy)=\binom{r_i+2}{2}$.
Hence, 
\begin{align*}
\cS um_1 & = \sum_{\substack{\bfc \in I_2(\bfy)\\ d_L(\cD_{EN}^{n-1}(\bfy),\bfc)=1 }}   p(\bfy |\bfc)  
%\\&
=  \dfrac{\binom{r_i+2}{2}}{\binom{n}{2}} + \sum_{\substack{1 \le j \le r \\ j \ne i}} \dfrac{(r_i +1)(r_j+1)}{\binom{n}{2}} + \sum_{j=1}^{n-r+1} \dfrac{(r_i+1)}{\binom{n}{2}} 
\\& \overset{\mathrm{(a)}}{=} \dfrac{(r_i+2)(r_i+1)}{2\binom{n}{2}}+\dfrac {(n-r_i-2+r-1) (r_i+1)}{\binom{n}{2}} 
%\\& \hspace{5ex}
+ \dfrac {(n-r+1) (r_i+1)}{\binom{n}{2}} 
\\ &= \dfrac {(2n-\frac{r_i}{2}-1) \cdot (r_i+1)}{\binom{n}{2}} = \dfrac {(4n-{r_i}-2) \cdot (r_i+1)}{n\cdot(n-1)},
%\\ &= \dfrac{4(r_i+1)}{n-1} - \dfrac{r_i^2+3r_i+2}{n\cdot(n-1)}.
%\\ & = \dfrac{4(r_t+1)}{n-1} - \dfrac{r_t(r_t-1)-2}{n\cdot(n-1)}
\end{align*}
where (a) holds since $\sum_{\substack{j \ne i}}r_j = n-2-r_i$. 

Next, let us evaluate the summation $\cS um_3$. Note that if $d_L(\cD_{EN}^{n-1}(\bfy), \bfc)=3$ then $\bfc$ is not in a supersequence of $\cD_{EN}^{n-1}(\bfy)$, and hence $\bfc \notin I_1(\cD_{EN}^{n-1}(\bfy))$. %Recall that this set of words includes all the words that are part of the insertion ball of radius 2 of $\bfy$, but are not in the insertion ball of radius 1 of $\cD(\bfy)$. 
The words that contribute to the summation $\cS um_3$ can be divided into three different types of words ${\bfc \in I_2 (\bfy)}$.

\textbf{Case 1:}
Let $\cC_1 \subseteq I_2 (\bfy)$ be the set of words $\bfc \in \cC_1$, such that $\bfc$ includes additional run(s) that does not appear in $\bfy$. Such additional runs can be either one run of length $2$, or two runs of length $1$ each. The number of words such that the length of the new run is two is  $n-r$. And the number of words with two additional runs is $\binom{n-r}{2}$. Additionally, for $\bfc \in \cC_1 $, $\emb(\bfc;\bfy)\ge 1$, which implies, 
\begin{align*}
 \sum_{c\in \cC_1} p(\bfy | \bfc )  &= \sum_{c\in \cC_1} \dfrac{1}{\binom{n}{2}}
%\\ &
= \dfrac{1}{\binom{n}{2}}  \left( \binom{n-r}{2} + n-r \right)
\\ & = \dfrac{2}{n(n-1)}  \left(\frac{(n-r-1)(n-r)}{2} + n-r\right) 
%\\ & 
= \dfrac{(n-r)(n-r+1)}{n(n-1)}.
\end{align*}

\textbf{Case 2:}
Let $\cC_2 \subseteq I_2 (\bfy)$ be the set of words $\bfc \in \cC_2$, such that $\bfc$ is obtained from $\bfy$ by prolonging the $j$-th run and by creating a new run in $\bfy$. Note that the prolonged run cannot be the $i$-th run in order to ensure $d_L(\cD_{EN}^{n-1}(\bfy), \bfc) = 3$, i.e., $j \ne i$.
The number of words in $\cC_2$ is $(r-1)(n-r+1)$, since there are $r-1$ options for the index $j$, and $n-r+1$ ways to create a new run in the obtained word. For such a word $ \bfc \in \cC_2$, it holds that $\emb(\bfc;\bfy) = r_j+1$ and hence,
\begin{align*}
\sum_{\bfc\in \cC_2}p(\bfy | \bfc) & = \sum_{\substack{1 \le j \le r \\ j \ne i}} (n-r+1) \cdot \dfrac{r_j+1}{\binom{n}{2}} 
\\ &=  \frac{(n-r+1)}{\binom{n}{2}}  \sum_{\substack{1 \le j \le r \\ j \ne i}} ({r_j+1}) \\
& = \dfrac{2(n-r+1)}{n(n-1)} (n-r_i +r-3 ).
\end{align*}

\textbf{Case 3:}
Let $\cC_3 \subseteq I_2 (\bfy)$ be the set of words $\bfc \in \cC_3$, such that $\bfc$ is obtained from $\bfy$ by prolonging one or two existing runs in $\bfy$ (other than the $i$-th run). The number of words $\bfc \in \cC_3$ obtained from $\bfy$ by prolonging a single run by two symbols is $r-1$. If the $j$-th run is the prolonged run then ${\emb (\bfc ; \bfy)=\binom{r_j+2}{2}}$. Additionally, there are $\binom{r-1}{2}$ words in $\cC_3$ that are obtained by prolonging the $j$-th and the $j'$-th runs of $\bfy$, each by one symbol. These words satisfy $\emb(\bfc; \bfy)=(r_j+1)(r_{j'}+1)$. Therefore,
\begin{align*}
&\sum_{\bfc\in \cC_3} p(\bfy | \bfc) = \sum_{\substack{1 \le j \le r \\ j \ne i}} \dfrac{\binom{r_j+2}{2}}{\binom{n}{2}} + \sum_{\substack{1 \leq j < j' \leq r \\ j,j' \ne i } } \dfrac{(r_{j'}+1)(r_j+1)}{\binom{n}{2}} 
\\ & = \dfrac{2}{n(n-1)}  \Big( \sum_{\substack{1 \le j \le r \\ j \ne i}} \dfrac{(r_j+2)(r_j+1)}{2}  
%\\ &\hspace{5ex} 
+  \frac{1}{2} \sum_{\substack{1 \leq j \leq r \\ j \ne i }}\sum_{\substack{1 \leq  j' \leq r \\ j' \ne i }} (r_j+1) (r_{j'}+1) -\frac{1}{2} \sum_{\substack{1 \leq j \leq r \\ j\ne i }} (r_{j}+1)^2 \Big) 
\\ & = \dfrac{2}{n(n-1)} \cdot \Big( \dfrac{1}{2}\sum_{\substack{1 \le j \le r \\ j \ne i}} {(r_j^2+3r_j+2)}
%\\& \hspace{5ex} 
+ \dfrac{1}{2}(n-r_i+r-3)^2 - \frac{1}{2} \sum_{\substack{1 \le j \le r \\ j \ne i}} r_j^2 - \sum_{\substack{1 \le j \le r \\ j \ne i}} r_j - \frac{r-1}{2} \Big)
\\ & = \dfrac{(n-r_i +r -3 )(n-r_i+r-2)}{n(n-1)}.
\end{align*} 

Thus, 
\begin{align*}
 \cS um_3 & = \sum_{\substack{\bfc\in I_2(\bfy)\\ d_L(\cD_{EN}^{n-1}(\bfy), \bfc)=3}} p(\bfy | \bfc)
\\ & =  \sum_{\bf c\in \cC_1} p(\bfy | \bfc) + \sum_{\bfc \in \cC_2} p(\bfy | \bfc) + \sum_{\bfc \in \cC_3} p(\bfy | \bfc)  
\\ & \ge \dfrac{(n-r)(n-r+1)}{n(n-1)} 
%\\ & \hspace{5ex} 
+ \dfrac{(n-r_i+r-3)}{n(n-1)} \cdot (3n-r-r_i) 
\\ &  = \dfrac{1}{n(n-1)} \cdot ({4n^2 -4nr_i -8n+ r_i^2 +3r_i +2r}).
\end{align*}

It holds that $\cS um _3 - \cS um_1 \geq 0$ if and only if
\begin{align*}
& {4n^2 -4nr_i -8n+ r_i^2 +3r_i +2r} \geq {4n(r_i+1)}- r_i^2-3r_i-2 
\\&2n^2-4nr_i -6n+r_i^2+3r_i+r+1 \geq 0.
\end{align*}
\end{IEEEproof}

Using this result we can explicitly define the ML$^*$ decoder $\cD_{\mathsf{ML}^*}$. This decoder works as follows. For each word $\bfy$ it calculates the number of runs $r$ and the length of the longest run $r_i$ and then checks if 
\begin{align}\label{eq_cond_2_del}
2n^2-4nr_i -6n+r_i^2+3r_i+r+1 \geq 0.
\end{align}
 If this condition holds, the decoder works as the lazy decoder and  returns the word $\bfy$. Otherwise, it acts like the embedding number decoder of length $n-1$ and prolongs the first longest run by one. The next theorem summarizes this result. %This result is summarized in the following theorem. 
\begin{theorem} \label{Th:2-del}
The ML$^*$ decoder $\cD_{\mathsf{ML}^*}$ for a single 2-deletion channel is a decoder that performs as the lazy decoder if inequality (\ref{eq_cond_2_del}) holds and otherwise it acts like the embedding number decoder of length $n-1$. i.e.,
$$
\cD_{\mathsf{ML}^*}(\bfy) = 
\begin{cases}
\cD_{Lazy}(\bfy) & \text{ inequality (\ref{eq_cond_2_del}) holds }, \\
\cD_{EN}^{n-1}(\bfy) & \text{ otherwise.}
\end{cases}$$
\end{theorem}
\begin{IEEEproof}
Using the previous lemmas, one can verify that $\cD_{\mathsf{ML}^*}$  minimizes the expected normalized distance for any possible channel output $\bfy$ and hence it is the ML$^*$ decoder.
\end{IEEEproof}

The result of Theorem~\ref{Th:2-del} states that if the ML$^*$ decoder chooses the same output as the decoder $\cD_{EN}^{n-1}$ then inequality  (\ref{eq_cond_2_del}) does not hold. It can be shown that this implies that $r_i \geq (2-\sqrt{2})n$ and thus, by Claim~\ref{cl:tau}, in almost all cases the output of the ML$^*$ decoder is the lazy decoder's output. 
\section{Two Deletion Channels}\label{sec:two deletions}

In this section, we shift to alphabet of size $q\ge 2$, and study the case of two instances of the deletion channel, $\del(p)$, where every symbol is deleted with probability $p$. % and prove in Theorem~\ref{th:2ch} an approximation for the expected normalized distance in the form of $$\perr(q,p) \approx \frac{3q-1}{q-1}p^2 + O(p^3).$$
%Complexity wise, it is well known that the time complexity to calculate the SCS length and the embedding numbers of two sequences are both quadratic with their length. However, the number of SCSs can grow exponentially~\cite{itoga1981string, elzinga2008algorithms}.  Thus, given a list of SCSs of size $L$, the complexity of the ML decoder for $t=2$ will be $O(Ln^2)$. The main idea behind these algorithms uses dynamic programming in order to calculate the SCS length and the embedding numbers for all prefixes of the given words. However, when calculating for example the SCS for $\bfy_1$ and $\bfy_2$ it is already known that $\mathsf{SCS}(\bfy_1,\bfy_2) \leq n$. Hence, it is not hard to observe that (see e.g.~\cite{apostolico1992fast}) many paths corresponding to prefixes which their length difference is greater than $d_1+d_2$ can be eliminated, when $d_1,d_2$ is the number of deletions in $\bfy_1,\bfy_2$, respectively. In particular, when $d_1$ and $d_2$ are fixed, then the time complexity is linear. In our simulations we used this improvement when implementing the ML decoder. Other improvements and algorithms of the ML decoder are discussed in~\cite{srinivasavaradhan2018maximum,srinivasavaradhan2019symbolwise}.
Recall that for a given codeword $\bfc \in \cC$ and two channel outputs  $\bfy_1, \bfy_2 \in (\Sigma_q)^{\leq |\bfc|}$, by Claim~\ref{cl:ML*multiple}, the output of the ML$^*$ decoder is
\begin{align*}
\cD_{\ML^*}(\bfy_1, \bfy_2) = \argmin_{\bfx\in \Sigma_q^*} \left\{\sum_{\substack{\bfc \in \cC \\ \bfc \in \sups (\bfy_1, \bfy_2 )}}  {d_L(\bfx,\bfc) \prod_{i=1}^2 \emb(\bfc ; \bfy_i)}\right\}. 
\end{align*}
Since the number of shorterst common supersequences of $\bfy_1$ and $\bfy_2$ can grow exponentially with their lengths~\cite{itoga1981string}, a direct computation of the ML$^*$ decoder might be impractical in this case. Hence, not only that the number of candidates $\bfx$ is large~\cite{itoga1981string}, the number of codewords  $\bfc\in \cC \cap \sups (\bfy_1, \bfy_2 )$ that are evaluated in the summation can be exponential. Therefore, we suggest a suboptimal approach, which is yet very practical. Instead of using the formal definition of the ML$^*$ decoder, in this section a degraded version of the ML$^*$ decoder is used. The decoder is designed with a limitation that may result in producing an output that is not necessarily a codeword, but rather a word of shorter length. This decoder, denoted by $\cD_{\ML^D}$ and referred as the ML$^D$ \emph{decoder}, is defined as follows
%the following definition, that approximates the ML$^*$ decoder. For convenience, we keep it under the same notation of the ML$^*$ decoder. 
\begin{align*}
 \cD_{\ML^D}(\bfy_1, \bfy_2) = \argmax_{\bfx\in \sups(\bfy_1, \bfy_2)} \left\{ \emb(\bfx ; \bfy_1) \emb(\bfx ; \bfy_2)\right\}. 
\end{align*}

For the rest of this section we assume that $\cC$ is $\Sigma_q^n$ and the expected normalized Levenshtein distance between the input and the decoded output is denoted by $\perr(n,q,p)$. This value provides an upper bound on the corresponding expected normalized distance (and the error probabilities) of the ML$^*$ decoder. Note that a lower bound on this error probability is $p^2$ (and more generally $p^t$ for $t$ channels) since if the same symbol is deleted in all channels, then it is not possible to recover its value and thus it will be deleted also in the output of the ML$^D$ decoder. This was already observed in~\cite{srinivasavaradhan2018maximum} and in their simulation results. Our main goal in this section is to calculate a tighter lower bound on $\perr(n,q,p)$. %We will also analyze these failure and error probabilities for the VT code\cite{VT} and the SVT code~\cite{7837631}.

In this section, we use the following additional notations.  For a word $\bfx \in \Sigma_q^*$, we denote by $\cL (\bfx)$ the number of runs in $\bfx$, and $\rho(\bfx) = (r_1, r_2, \ldots, r_{\cL(\bfx)})$ denotes the \emph{run-length profile of $\bfx$}, which is a vector of length $\cL(\bfx)$, in which the $i$-th entry corresponds to the length of the $i$-th run of $\bfx$ (for $1 \le i \le \cL(\bfx)$). Similarly, we define $\cA (\bfx)$ as the number of (maximal) alternating segments in $\bfx$, and $\omega(\bfx) = (a_1, a_2, \ldots, a_{\cA (\bfx)}) $ is the \emph{alternating-length profile of $\bfx$}, which is a length-$\cA(\bfx)$ vector, in which the $i$-th entry corresponds to the length of the $i$-th maximal alternating segment (for $1 \le i \le \cA(\bfx)$).  

The lower bound $p^2$ on $\perr(n,q,p)$ is not tight since if symbols from the same run are deleted, then the outputs of the two channels of this run are the same, and it is impossible to detect that this run experienced a deletion in both of its copies. The expected normalized distance due to deletions within runs is denoted by $\ensuremath{\mathsf{P_{err}^{run}}}(\del(p),\Sigma_q^n,\cD_{\ML^D},d)$ or in short $\prun(n,q,p)$ and the next lemma gives a lower bound on this probability.

\begin{lemma} \label{lm:2ch_run_del}
For the deletion channel $\del(p)$, it holds that 
%It holds that The average deletion probability of the ML decoder's output because of the runs is at least 
$$ \prun(n,q,p)  \ge \frac{1}{q^n}\cdot \frac{1}{n} \left( q(1-(1-p)^{n})^2+ \sum_{r=1}^{n-1}  (q-1)q^{n-r-1} (2q+(n-r-1)(q-1)) (1-(1-p)^{r})^2 \right) \triangleq \cP_{\textmd{run}} (n,q,p).$$
Furthermore, when $n$ approaches infinity, we have that
$$ \lim_{n\to \infty} \cP_{\textmd{run}} (n,q,p) = \frac{\left(q-1\right)}{q}+\frac{2\left(q-1\right)^{2}\left(p-1\right)}{q\left(p+q-1\right)}+\frac{\left(q-1\right)^{2}\left(p-1\right)^{2}}{q\left(q-\left(p-1\right)^{2}\right)}\triangleq \cP_{\textmd{run}} (q,p).$$
Finally, when $n$ approaches infinity and $p$ approaches zero, it holds that $\cP_{\textmd{run}} (q,p)  \approx  \frac{q+1}{q-1}p^2,$ i.e., $$\lim_{p \to 0} \frac{\cP_{\textmd{run}} (q,p)}{\frac{q+1}{q-1}p^2} =1.$$
\end{lemma}
\begin{IEEEproof}
The lower bound is given by considering the case in which both channel outputs experience a single deletion in the same run. First, we note that if both channel outputs, $\bfy_1$ and $\bfy_2$, experienced the same number of deletions in each run, then $\bfy_1=\bfy_2= \cS\cC\cS(\bfy_1, \bfy_2)$. Thus, in this case $\cD_{\ML^D} (\bfy_1, \bfy_2) = \bfy_1$ and $d_L(\cD_{\ML^D} (\bfy_1, \bfy_2), \bfx)$ is the number of deletions that occurred in each output, where $\bfx$ is the transmitted word. 
Assume $\bfx$ has a run of length $r \in \N$. The probability that both channel outputs have experienced at least one deletion in this run is given by $(1-(1-p)^r)^2 $. In this case, the distance $d_L(\cD_{\ML^D} (\bfy_1, \bfy_2), \bfx)$ increases by at least 1 as a result of the deletions in this run. 

Our goal is to calculate a lower bound on the expected normalized distance of a given word $\bfx$ and we do that by considering the increase in the normalized Levenshtein distance as a result of only deletions in the same run. We denote this value by $\prun(\bfx,q,p)$ and its calculation is given below. We also denote by $d_{\textrm{run}}((\bfy_1,\bfy_2),\bfx)$ the number of runs in $\bfx$ in which both $\bfy_1$ and $\bfy_2$ had at least one deletion. Assume that the run-length profile of $\bfx$ is $\rho(\bfx) = (r_1, r_2, \ldots, r_{\cL(\bfx)})$. By definition we have that
\begin{align*}
\prun(\bfx,q,p) &\geq  \sum_{\bfy_1, \bfy_2 :\cD(\bfy_1, \bfy_2) \neq  \bfx} \frac{d_{\textrm{run}}((\bfy_1,\bfy_2),\bfx)}{|\bfx|}\cdot \pr_\ch\{  \bfy_1 \textmd{ rec. }| \bfx \textmd{ trans.}\} \cdot \pr_\ch\{  \bfy_2 \textmd{ rec. }| \bfx \textmd{ trans.}\}.
\end{align*}
In order to calculate $d_{\textrm{run}}((\bfy_1,\bfy_2),\bfx)$, we consider each run of $\bfx$ independently and if both $\bfy_1$ and $\bfy_2$ experienced at least one deletion in a given run, then the value of $d_{\textrm{run}}((\bfy_1,\bfy_2),\bfx)$ increases by at least one.
Therefore, we get that 
\begin{align*}
\prun(\bfx,q,p) &\geq  \frac{1}{n} \sum_{i=1}^{\cL(\bfx)} \sum_{\substack{\bfy_1, \bfy_2 : \\  \cD(\bfy_1, \bfy_2) \neq  \bfx} } 1 \cdot \pr_\ch\{  \bfy_1 \textmd{ and }\bfy_2 \textmd{ had at least one deletion in the $i$-th run} |  \bfx \textmd{ trans.} \} p(\bfy_1 | \bfx) p(\bfy_2 | \bfx) &\\
& = \frac{1}{n} \sum_{i=1}^{\cL(\bfx)} 
 \pr_\ch\{  \bfy_1 \textmd{ and }\bfy_2 \textmd{ had at least one deletion in the $i$-th run} |  \bfx \textmd{ trans.} 
 \}  \\
 & = \frac{1}{n} \sum_{i=1}^{\cL(\bfx)} 
 \pr_\ch\{  \bfy_1  \textmd{ had at least one deletion in the $i$-th run} |  \bfx \textmd{ trans.}  \}  \\   
 & \hspace{10ex}  \cdot \pr_\ch\{  \bfy_2  \textmd{ had at least one deletion in the $i$-th run} |  \bfx \textmd{ trans.} 
 \}  \\
& = \frac{1}{n}\sum_{i=1}^{\cL(\bfx)} 1 \cdot (1-(1-p)^{r_i})^2 %= \frac{1}{n}\sum_{i=1}^{\cL(\bfx)}(1-(1-p)^{r_i})^2.& 
\end{align*}
Now let us consider $\ensuremath{\mathsf{P_{err}^{run}}}(\del(p),\Sigma_q^n,\cD_{\ML^D},d)$, which is the expected normalized distance due to runs,
\begin{align*}
\ensuremath{\mathsf{P_{err}^{run}}}(\del(p),\Sigma_q^n,\cD_{\ML^D},d) &= \frac{1}{q^n}\sum_{\bfx\in\Sigma_q^n}\prun(\bfx,q,p) \\
& \ge  \frac{1}{q^n} \cdot \frac{1}{n} \sum_{\bfx \in \Sigma_q^n} \sum_{i=1}^{\cL(\bfx)} (1-(1-p)^{r_i})^2.
\end{align*}
Next, for an integer $1\le r \le n$, let us denote by $R_{q,n}(r)$, the total number of runs of length $r$ occurring in all possible words of length $n$ over $\Sigma_q$. Thus, from the above discussion, we have that, 
\begin{align*}
\ensuremath{\mathsf{P_{err}^{run}}}(\del(p),\Sigma_q^n,\cD_{\ML^D},d) & \ge \frac{1}{q^n}\cdot \frac{1}{n} \sum_{r=1}^n  R_{q,n}(r) (1-(1-p)^{r})^2.
\end{align*}
The value of $R_{q,n}(r)$ can be calculated in a similar way as was done for alternating sequnces in~\cite{BEY23}. 
For $r=n$, this number is given by $q$. Additionally, for $1 \le r < n$, let us consider the number of words (over $\Sigma_q^n)$ with run of length $r$ that start in the $i$-th position for $1 \le i \le n$. 
For $i =1$ or $i=n-r+1$ this number is given by the selection of the symbol of the run, the symbol that follows (or precedes) the run, and the remaining $n-r-1$ symbols, which are not limited. Therefore, in total, the number is given by $q(q-1)q^{n-r-1}$. 
For $2 \le i \le n-r$, the number of words with run of length $r$ that starts in the $i$-th position is given by the selection of the symbol in the run, the selection of the preceding and the following symbol, and the selection of the remaining $n-r-2$ symbols. Thus, this number is given by $q(q-1)^2q^{n-r-2}$. Hence, in total we have that,   
\begin{align*}
R_{q,n}(r) &= 2 \cdot q(q-1)q^{n-r-1} + \sum_{i=2}^{n-r} q(q-1)^2q^{n-r-2}
\\& = (q-1)q^{n-r-1} (2q+(n-r-1)(q-1)),
\end{align*}
and as a result we get that
\begin{align*}
    &\ensuremath{\mathsf{P_{err}^{run}}}(\del(p),\Sigma_q^n,\cD_{\ML^D},d)  \ge \frac{1}{q^n}\cdot \frac{1}{n} \left( q(1-(1-p)^{n})^2+ \sum_{r=1}^{n-1}  (q-1)q^{n-r-1} (2q+(n-r-1)(q-1)) (1-(1-p)^{r})^2 \right).
    %\\ &= \frac{1}{q^n}\frac{1}{n}p^2(qn^2)+\frac{p^2}{q^{2n}} \left( (q-1)(\frac{q+1}{q})^{n-1} (nq-n+2+2q\frac{n-1}{q+1} +\frac{(q-1)(n-1)(nq-4q-2)}{(q+1)^2} -n(\frac{1}{q+1})^{n-2} \right)
\end{align*}
Let us simplify the expression as follows
\begin{align*}
 &\frac{1}{q^n}\cdot \frac{1}{n} \left( q(1-(1-p)^{n})^2+ \sum_{r=1}^{n-1}  (q-1)q^{n-r-1} (2q+(n-r-1)(q-1)) (1-(1-p)^{r})^2 \right)  
 \\ &  = \frac{1}{q^n}\cdot \frac{1}{n} \left( q(1-(1-p)^{n})^2+ \sum_{r=1}^{n-1}  (q-1)q^{n-r-1} \left(q(n-r+1)-(n-r-1)\right) \left(1-2(1-p)^{r} +(1-p)^{2r}\right)
 \right).
\end{align*}
To further simplify  $\sum_{r=1}^{n-1}  (q-1)q^{n-r-1} \left(q(n-r+1)-(n-r-1))\right) \left(1-2(1-p)^{r} +(1-p)^{2r}\right)$ we break it into six expressions, and the following equations can be verified
\begin{small}
\begin{align*}
 &\cS_1 \triangleq \sum_{r=1}^{n-1}  (q-1)q^{n-r} (n-r+1) = \frac{nq^{n+1}-(n+1)q^n-q^2+2q}{q-1}
\\ &\cS_2 \triangleq \sum_{r=1}^{n-1}  (q-1)q^{n-r} (n-r+1) (-2(1-p)^r) = \frac{2\left(q-1\right) \left(n\left(p-1\right)q^{n+1}+\left(n+1\right)\left(p-1\right)^{2}q^{n}+q^{2}\left(1-p\right)^{n}-2q\left(1-p\right)^{n+1}\right)}{\left(p+q-1\right)^{2}}
\\ &\cS_3 \triangleq \sum_{r=1}^{n-1}  (q\hspace{-0.1ex}-\hspace{-0.1ex}1)q^{n-r} (n\hspace{-0.1ex}-\hspace{-0.1ex}r+1) ((1\hspace{-0.1ex}-\hspace{-0.1ex}p)^{2r}) = \frac{(1\hspace{-0.1ex}-\hspace{-0.1ex}q)\left(\hspace{-0.1ex}-\hspace{-0.1ex}n\left(p\hspace{-0.1ex}-\hspace{-0.1ex}1\right)^{2}q^{n+1}+\left(n+1\right)\left(p\hspace{-0.1ex}-\hspace{-0.1ex}1\right)^{4}q^{n}+q^{2}\left(1\hspace{-0.1ex}-\hspace{-0.1ex}p\right)^{2n}\hspace{-0.1ex}-\hspace{-0.1ex}2q\left(1\hspace{-0.1ex}-\hspace{-0.1ex}p\right)^{2n+2)}\right)}{\left(p^{2}-2p-q+1\right)^{2}}
\\ &\cS_4 \triangleq -\sum_{r=1}^{n-1}  (q-1)q^{n-r-1} (n-r-1) = -\frac{\left(n-2\right)q^{n+1}-\left(n-1\right)q^{n}+q^{2}}{\left(q-1\right)q}
\\ &\cS_5 \triangleq -\sum_{r=1}^{n-1}  (q-1)q^{n-r-1} (n-r-1) (-2(1-p)^r)= \frac{2\left(q-1\right)\left(-\left(n-2\right)\left(p-1\right)q^{n+1}-\left(n-1\right)\left(p-1\right)^{2}q^{n}+q^{2}\left(1-p\right)^{n}\right)}{q\left(p+q-1\right)^{2}}
\\ &\cS_6 \triangleq -\sum_{r=1}^{n-1}  (q-1)q^{n-r-1} (n-r-1) (1-p)^{2r}= -\frac{\left(q-1\right)\left(\left(n-2\right)\left(p-1\right)^{2}q^{n+1}-\left(n-1\right)\left(p-1\right)^{4}q^{n}+q^{2}\left(1-p\right)^{2n}\right)}{q\left(p^{2}-2p-q+1\right)^{2}}.
\end{align*}
\end{small}

Now we have that, 
\begin{align*}
    \sum_{r=1}^{n-1}  (q-1)q^{n-r-1} \left(q(n-r+1)-(n-r-1))\right) \left(1-2(1-p)^{r} +(1-p)^{2r}\right) = \sum_{i=1}^6\cS_i.
\end{align*}
Thus, it can be deduced that, 
\begin{align*}
    &\perr(\del(p),\Sigma_q^n, \cD_{\ML^D},d)  \ge \frac{1}{q^n}\cdot \frac{1}{n} \left( q(1-(1-p)^n)^2+\sum_{i=1}^6\cS_i \right).
\end{align*}

Let us consider the case in which $n$ approaches infinity. In this case, we have that
\begin{align*}
\cP_{\textmd{run}} (q,p) \triangleq &\lim_{n \to \infty} \frac{1}{q^n}\cdot \frac{1}{n} \left( q(1-(1-p)^n)^2+\sum_{i=1}^6\cS_i \right) \\ =& 1+ \frac{2(q-1)(p-1)(q+p-1)}{(p+q-1)^2} + \frac{(q-1)(p-1)^2(q-(p-1)^2)}{(p^2-2p-q+1)^2} + \frac{1-q}{(q-1)q} -\frac{2(p-1)(q-1)(q+p-1)}{q(p+q-1)^2}
\\ & +\frac{(q-1)(p-1)^2((p-1)^2-q)}{q(p^2-2p-q+1)^2}
\\ & = \frac{\left(q-1\right)}{q}+\frac{2\left(q-1\right)^{2}\left(p-1\right)}{q\left(p+q-1\right)}+\frac{\left(q-1\right)^{2}\left(p-1\right)^{2}}{q\left(q-\left(p-1\right)^{2}\right)}.
\end{align*}
Finally, we consider the case where $n$ approaches infinity, and the probability $p$ vanishes to zero. In this case, the expected normalized distance due to runs approaches $\frac{(q+1)}{(q-1)}p^2$. The proof follows from the below equations that can be shown by algebraic manipulations. 
\begin{align*}
    \lim_{p \to 0 }\frac{\cP_{\textmd{run}} (q,p) }{\frac{(q+1)}{(q-1)}p^2} &=\lim_{p \to 0 } \left(\frac{\left(q-1\right)^{2}\left(-p-q+1+2qp\right)}{q\left(q+1\right)p^{2}\left(p+q-1\right)}+\frac{\left(q-1\right)^{3}\left(p-1\right)^{2}}{q\left(q-\left(p-1\right)^{2}\right)\left(q+1\right)p^{2}}\right) 
    \\& =\lim_{p \to 0 } \frac{pq^2+pq-pq^3-p+q^4+1-2q^2}{-p^3q^2-2p^3q-p^3-p^2q^3+p^2q^2+5p^2q+3p^2+3pq^3+3pq^2-3pq-3p+q^4-2q^2+1}
    \\ & = \frac{q^4-2q^2+1}{q^4-2q^2+1} =1.
\end{align*}

 \end{IEEEproof}
%Another proof works as follows. Assume that the close deletions in the two copies are in positions $i$ and $i+j$. Then, there is a run deletion if there is a run between positions $i$ and $i+j$, which happens with probability $p1/q^j$. Hence, the deletion probability because of the runs is $$p^2\left(1 + 2\sum_{j=1}^\infty \frac{1}{q^j}\right)  = p^2\left(1 + \frac{2}{q-1}\right) = \frac{q+1}{q-1}p^2.$$

However, runs are not the only source of errors in the output of the ML$^D$ decoder. For example, assume the $i$-th and the $(i+1)$-st symbols are deleted from the first and the second channel output, respectively. If the transmitted word $\bfx$ is of the form $\bfx = (x_1,\ldots,x_{i-1},0,1,x_{i+2},\ldots,x_n)$, then the two channels' outputs are $\bfy_1 = (x_1,\ldots,x_{i-1},0,x_{i+2},\ldots,x_n)$ and $\bfy_2 = (x_1,\ldots,x_{i-1},1,x_{i+2},\ldots,x_n)$. However, these two outputs could also be received upon deletions exactly in the same positions if the transmitted word was $\bfx' = (x_1,\ldots,x_{i-1},1,0,x_{i+2},\ldots,x_n)$. Hence, the ML$^D$ decoder can output the correct word only in one of these two cases. Longer alternating sequences cause the same problem as well and the \emph{occurrence} probability of this event, denoted by $\mathsf{P_{err}^{alt}}(\del(p),\Sigma_q^n, \cD_{\ML^D},d)$, or $\palt(n,q,p)$ in short, will be bounded from below in the next lemma. 
%This scenario extended to any alternating sequence, as will be shown in the next lemma. The error probability of this event is denoted by $\palt(q,p)$
\begin{lemma} \label{lm:2ch_alt_del}
For the deletion channel $\del(p)$, it holds that 
%It holds that The average deletion probability of the ML decoder's output because of the runs is at least 
\begin{small}
\begin{align*}
\palt(n,q,p)&  \ge  \frac{1}{q^n}\cdot \frac{1}{n} \left( q(q-1) (1-(1-p)^{n})(1-(1-p)^{n-1}) \right.
\\& \left. + \sum_{a=2}^{n-1}  (2(q-1)^2 q^{n-a}+(n-a-1)(q-1)^3q^{n-a-1}) (1-(1-p)^{a}-(1-p)^{a-1}+(1-p)^{2a-1}) \right) \triangleq \cP_{\textmd{alt}} (n, q, p).
\end{align*}
\end{small}

\noindent Furthermore, when $n$ approaches infinity, we have that 
$$\lim_{n \to \infty} \cP_{\textmd{alt}} (n,q, p) = \frac{(q-1)^2}{q^2} + \frac{(q-1)^3(p-1)(2-p)}{q^2(p+q-1)} + \frac{(q-1)^3(p-1)^3}{q^2(p^2-2p-q+1)}\triangleq \cP_{\textmd{alt}} (q, p).$$
Finally, when $n$ approaches infinity and $p$ approaches zero, it holds that $\cP_{\textmd{alt}} (q, p) \approx 2p^2,$ i.e., 
$$\lim_{p \to 0} \frac{\cP_{\textmd{alt}} (q,p)}{2p^2}=1.$$
\end{lemma} 
\begin{IEEEproof}
The lower bound is given by considering the case in which both channel outputs experience at least a single deletion in the same alternating sequence (but in different symbols within it). First, we note that if the same symbol is deleted in both channel outputs, this is considered a deletion in the same run, and therefore, the contribution to the expected normalized distance is covered by Lemma~\ref{lm:2ch_run_del}. Next, we consider the case in which both channel outputs, $\bfy_1$ and $\bfy_2$, experience a single deletion in each alternating sequence (in different symbols within the sequence). For simplicity in the analysis, we assume that the alternating sequences do not overlap; that is, each symbol in 
$\bfx$ belongs to at most one alternating sequence. In this case, in any of the erroneous alternating sequences, the decoder cannot distinguish between the alternating sequence and the alternating sequence with the opposite order of symbols. That is, the same channel outputs $\bfy_1$ and $\bfy_2$, can be obtained by applying the same deletions on the channel input, in which any of the erroneous alternating sequence  $ABAB \ldots$ is replaced with $BABA\ldots$ when  $A,B\in \Sigma_q$  are any two distinct symbols in the alphabet. In this scenario, the decoder $\cD_{\ML^D}$, which selects the word that maximizes the embedding number, must choose between two equally likely possibilities for each erroneous alternating sequence. Since $\cC=\Sigma_q^n$,  the probability of the decoder selecting the incorrect alternating sequence is 
$0.5$ for each such sequence due to symmetry in the likelihood of both options. In any such error event, the decoder returns the word where the erroneous alternating sequence appears in the opposite order. This event increases the Levenshtein distance $d_L(\cD_{\ML^D}(\bfy_1, \bfy_2), x)$ by $2$ for each Since this occurs with probability $0.5$, on average, such deletions in the same alternating sequence increase the Levenshtein distance by 1 per sequence.

%In this case, the distance $d_L(\cD_{\ML^D} (\bfy_1, \bfy_2), \bfx)$ increases by at least 1 as a result of the deletions in this run. 
Assume there is a deletion in the first channel in the $i$-th position and the closest deletion in the second channel is $j>0$ positions apart, i.e., either in position $i-j$ or $i+j$. W.l.o.g. assume it is in the $(i+j)$-th position and $\bfx_{[i,i+j]}$ is an alternating sequence $ABAB\cdots$. Then, the 
%Assume that the close deletions in the two copies are in positions $i$ and $i+j$ and $\bfx_{[i:j]}$ is an alternating sequence. 
same outputs from the two channels could be received if the transmitted word was the same as $\bfx$ but with the opposite order of the symbols of the alternating sequence, that is, the symbols of the word in the positions of $[i,i+j]$ are $BABA\cdots$, and let us denote this word by ${\bar{\bfx}}_{[i,i+j]}$. 

Our goal is to calculate a lower bound on the expected normalized distance of $\bfx$ by considering the increase of the normalized distance which results from deletions in the same alternating sequence. We denote this value by $\palt(\bfx,q,p)$. Following the notations from the previous paragraph, in this case, the Levenshtein distance of the decoder's output and the transmitted word is either $0$ if the decoder output is the correct word ($\bfx$), or $2$ (if the decoder output is ${\bar{\bfx}}_{[i,i+j]}$). Since we assume all the words over $\cC = \Sigma_q^n$ are equally transmitted, by averaging these two cases we get that any alternating sequence contributes $1$ to the Levenshtein distance. Let us denote by ${d_{\textrm{alt}}((\bfy_1,\bfy_2),\bfx)}$ the number of alternating sequences in which both $\bfy_1$ and $\bfy_2$ had at least one deletion (in different symbols). We recall that $\omega(\bfx) = (a_1, \ldots, a_{\cA(\bfx)})$ denotes the alternate length profile of $\bfx$.  By definition, we have that
\begin{align*}
\palt(\bfx,q,p) &\ge \sum_{\bfy_1, \bfy_2 :\cD(\bfy_1, \bfy_2) \neq  \bfx} \frac{d_{\textrm{alt}}((\bfy_1,\bfy_2),\bfx)}{|\bfx|}\cdot \pr_\ch\{  \bfy_1 \textmd{ rec. }| \bfx \textmd{ trans.}\} \pr_\ch\{  \bfy_2 \textmd{ rec. }| \bfx \textmd{ trans.}\}.
\end{align*}

To calculate ${d_{\textrm{alt}}((\bfy_1,\bfy_2),\bfx)}$, we consider each alternating sequence independently and we note that in each such  alternating sequence, if both $\bfy_1$ and $\bfy_2$ had at least one deletion, then ${d_{\textrm{alt}}((\bfy_1,\bfy_2),\bfx)}$ increase on average by at least 1. We also note that for an alternating sequence of length $a>1$, the probability that both channel outputs had at least one deletion in two distinct symbols is given by $(1-(1-p)^{a})(1-(1-p)^{a-1})$. Thus, we have that, 

\vspace{-2ex}\begin{small}
\begin{align*}
\palt(\bfx,q,p) &\geq  \frac{1}{n}\sum_{i=1}^{\cA(\bfx)} \sum_{\bfy_1, \bfy_2} 1 \cdot \pr_\ch\{  \bfy_1 \textmd{ and }\bfy_2 \textmd{ had at least one distinct deletion in the $i$-th  alternating sequence} | \bfx \textmd{ trans.} \} p(\bfy_1| \bfx) p(\bfy_2| \bfx) &\\
& = \frac{1}{n}\sum_{i=1}^{\cA(\bfx)} \pr_\ch\{  \bfy_1 \textmd{ and }\bfy_2 \textmd{ had at least one distinct deletion in the $i$-th  alternating sequence} | \bfx \textmd{ trans.} \} \\
& = \frac{1}{n}\sum_{i=1}^{\cA(\bfx)} \pr_\ch\{  \bfy_1 \textmd{ had at least one distinct deletion in the $i$-th  alternating sequence} | \bfx \textmd{ trans.} \} \\ & \hspace{10ex} \cdot \pr_\ch\{  \bfy_2 \textmd{ had at least one distinct deletion in the $i$-th  alternating sequence} | \bfx \textmd{ trans.} \}\\
&= \frac{1}{n}\sum_{i=1}^{\cA(\bfx)} 1 \cdot (1-(1-p)^{a_i})(1-(1-p)^{a_i-1}).& 
\end{align*}
\end{small}

Now, let us consider the expected normalized distance due to alternating segments. 
\begin{align*}
\ensuremath{\mathsf{P_{err}^{alt}}}(\del(p),&\Sigma_q^n,\cD_{\ML^D},d) = \frac{1}{q^n}\sum_{\bfx\in\Sigma_q^n}\palt(\bfx,q,p) \\
& \ge  \frac{1}{q^n} \cdot \frac{1}{n} \sum_{\bfx \in \Sigma_q^n} \sum_{i=1}^{\cA(\bfx)} (1-(1-p)^{a_i})(1-(1-p)^{a_i-1})
\\& = \frac{1}{q^n} \cdot \frac{1}{n} \sum_{\bfx \in \Sigma_q^n} \sum_{i=1}^{\cA(\bfx)} (1-(1-p)^{a_i}-(1-p)^{a_i-1}+ (1-p)^{2a_i-1}).
\end{align*}
Next, for an integer $1\le a \le n$, let us denote by $A_{q,n}(a)$, the number of alternating sequences of length $a$ occurring in all possible words of length $n$ over $\Sigma_q$. Thus, from the above discussion, we have that
\begin{align*}
\ensuremath{\mathsf{P_{err}^{alt}}}(\del(p),\Sigma_q^n, \cD_{\ML^D},d) & \ge \frac{1}{q^n}\cdot \frac{1}{n} \sum_{a=1}^n  A_{q,n}(a) (1-(1-p)^{a_i})(1-(1-p)^{a_i-1}).
\end{align*}
The value of $A_{q,n}(a)$ was calculated in~\cite{BEY23}, where it was shown that $A_{q,n}(1) = 2q^{n-1}+(n-2)q^{n-2}$, $A_{q,n}(n)=q(q-1)$, and for $2\le a\le n-1$, 
$$A_{q,n}(a)=2(q-1)^2 q^{n-a}+(n-a-1)(q-1)^3q^{n-a-1}.$$
Note that it is enough to  consider $a \ge 2$, since when $a=1$ the alternate sequence is in also a run of length one, and was considered in Lemma~\ref{lm:2ch_run_del}. Thus, we have that, 
\begin{align*}
&\ensuremath{\mathsf{P_{err}^{alt}}}(\del(p),\Sigma_q^n, \cD_{\ML^D},d)  \ge \frac{1}{q^n}\cdot \frac{1}{n} \sum_{a=2}^n  A_{q,n}(a) (1-(1-p)^{a})(1-(1-p)^{a-1})
\\ & = \frac{1}{q^n}\cdot \frac{1}{n} \Bigl(q(q-1) (1-(1-p)^{n})(1-(1-p)^{n-1})  
\\&  +\sum_{a=2}^{n-1}  (2(q-1)^2 q^{n-a}+(n-a-1)(q-1)^3q^{n-a-1}) (1-(1-p)^{a})(1-(1-p)^{a-1}) \Bigr)
\\ & = \frac{1}{q^n}\cdot \frac{1}{n} \Bigl(q(q-1) (1-(1-p)^{n})(1-(1-p)^{n-1})   
\\&  + \sum_{a=2}^{n-1}  (2(q-1)^2 q^{n-a}+(n-a-1)(q-1)^3q^{n-a-1}) (1-(1-p)^{a}-(1-p)^{a-1}+(1-p)^{2a-1}) \Bigr).
\end{align*}
To further simplify $\sum_{a=2}^{n-1}  (2(q-1)^2 q^{n-a}+(n-a-1)(q-1)^3q^{n-a-1}) (1-(1-p)^{a}-(1-p)^{a-1}+(1-p)^{2a-1})$ we break it into $8$ expressions, as can be seen below. 
\begin{align*}
 &\cS_1 \triangleq \sum_{a=2}^{n-1}  2(q-1)^2 q^{n-a} = \frac{2(q-1)(q^n-q^2)}{q}
\\& \cS_2 \triangleq \sum_{a=2}^{n-1}  (2(q-1)^2 q^{n-a}) (-(1-p)^{a}) = \frac{2(q-1)^2(-p^2q^n+2pq^n+q^2(1-p)^{n}-q^n)}{q(p+q-1)}
\\& \cS_3 \triangleq \sum_{a=2}^{n-1}  (2(q-1)^2 q^{n-a}) (-(1-p)^{a-1}) = \frac{2(q-1)^2(p^2q^n-2pq^n-q^2(1-p)^{n}+q^n)}{(p-1)q(p+q-1)}
\\& \cS_4 \triangleq \sum_{a=2}^{n-1}  (2(q-1)^2 q^{n-a}) ((1-p)^{2a-1}) = \frac{2\left(q-1\right)^{2}\left(\left(p-1\right)^{4}q^{n}-q^{2}\left(1-p\right)^{2n}\right)}{\left(p-1\right)q\left(p^{2}-2p-q+1\right)} % \frac{2(q-1)^2(p^4q^n-4p^3q^n+6p^2q^n-4pq^n -q^2(1-p)^{2n}+q^n)}{(p-1)q(p^2-2p-q+1)}
\\ & \cS_5 \triangleq \sum_{a=2}^{n-1} ((n-a-1)(q-1)^3q^{n-a-1}) = \frac{(q-1)\left( (n-3)q^{n+1}-(n-2)q^n+q^3\right)}{q^2}
\\& \cS_6 \triangleq \sum_{a=2}^{n-1} ((n-a-1)(q-1)^3q^{n-a-1}) (-(1-p)^{a}) = - \frac{(q-1)^3\left((n-3)(p-1)^2q^{n+1}+(n-2)(p-1)^3q^n+q^3(1-p)^n\right)}{q^2(p+q-1)^2}
\\& \cS_7 \triangleq \sum_{a=2}^{n-1} ((n-a-1)(q-1)^3q^{n-a-1}) (-(1-p)^{a-1}) = \frac{(q-1)^3\left((n-3)(p-1)^2q^{n+1}+(n-2)(p-1)^3q^n+q^3(1-p)^n\right)}{(p-1)q^2(p+q-1)^2}
\\& \cS_8 \triangleq \sum_{a=2}^{n-1} ((n-a-1)(q-1)^3q^{n-a-1}) ((1-p)^{2a-1}) = -\frac{(q-1)^3\left((n-3)(p-1)^4q^{n+1}-(n-2)(p-1)^6q^n+q^3(1-p)^{2n}\right)}{(p-1)q^2(p^2-2p-q+1)^2}.
\end{align*}
Thus, we have that, 
\begin{align*}
   &\ensuremath{\mathsf{P_{err}^{alt}}}(\del(p),\Sigma_q^n, \cD_{\ML^D},d) \ge  \frac{1}{q^n}\cdot \frac{1}{n} \left(q(q-1) (1-(1-p)^{n})(1-(1-p)^{n-1})+ \sum_{i=1}^{8}  \cS_i \right).
\end{align*}
Now let us consider the case in which $n$ approaches infinity. 
\begin{align*}
   \cP_{\textmd{alt}} (q, p) &= \lim_{n \to \infty} \frac{1}{q^n}\cdot \frac{1}{n} \left(q(q-1) (1-(1-p)^{n})(1-(1-p)^{n-1})+ \sum_{i=1}^{8}  \cS_i \right)
   \\ & = \frac{(q-1)^2}{q^2} - \frac{(q-1)^3(p-1)^2q}{q^2(p+q-1)^2}- \frac{(q-1)^3(p-1)^3}{q^2(p+q-1)^2} +  \frac{(q-1)^3(p-1)q}{q^2(p+q-1)^2} + \frac{(q-1)^3(p-1)^2}{q^2(p+q-1)^2}
   \\& - \frac{(q-1)^3(p-1)^3q}{q^2(p^2-2p-q+1)^2}
    + \frac{(q-1)^3(p-1)^5}{q^2(p^2-2p-q+1)^2}
    \\ & = \frac{(q-1)^2}{q^2} - \frac{(q-1)^3(p-1)^2}{q^2(p+q-1)}
    + \frac{(q-1)^3(p-1)}{q^2(p+q-1)} + \frac{(q-1)^3(p-1)^3}{q^2(p^2-2p-q+1)}
    \\ & = \frac{(q-1)^2}{q^2} + \frac{(q-1)^3(p-1)(2-p)}{q^2(p+q-1)}
     + \frac{(q-1)^3(p-1)^3}{q^2(p^2-2p-q+1)}.
    %\\& = \frac{(q-1)^2(p-2)}{2(p+q-1)(p^2-2p+1-q)}
\end{align*}
Finally, we consider the case when $n$ approaches infinity, and $p$ approaches zero. In this case, the expected normalized distance due to alternating sequences approaches $2p^2$ as can be seen below
\begin{align*}
    \lim_{p \to 0 } \frac{\cP_{\textmd{alt}} (q, p)}{2p^2} =  \lim_{p \to 0 } \frac{(q-1)^2(p-2)}{2(p+q-1)(p^2-2p+1-q)} =\frac{-2(q-1)^2}{2(q-1)(1-q)} = 1.
\end{align*}
%Therefore, the occurrence probability of this event is at least 
%$$2p^2\cdot \sum_{j=1}^\infty \frac{q-1}{q}\cdot \frac{1}{q^{j-1}} = 2p^2,$$
%where  $\frac{q-1}{q}\cdot \frac{1}{q^{j-1}}$ is the probability that $\bfx_{[i,i+j]}$ is any alternating sequence and the multiplication by 2 takes into account the cases of deletion in either position $i-j$ or $i+j$ \textcolor{blue}{(i.e., switching the first and the second channel)}.
\end{IEEEproof}

The results in  Lemma~\ref{lm:2ch_run_del} and Lemma~\ref{lm:2ch_alt_del} both present lower bounds on the probabilities $\prun(n,q,p)$ and $\palt(n,q,p)$ respectively. These results indeed provide lower bounds since they actually neglect cases in which one of the channel outputs experiences more than a single deletion in one of the runs/alternating segments. These error events do increase the normalized distance of the decoder, but their probability is in the order of $p^3$.
%, for the simplicity of our discussion we denote by $f_{n,q}$ a function of $n$ and $q$, such that this probability eqauls $f_{n,q} \cdot p^3$}. 
As a conclusion from Lemma~\ref{lm:2ch_run_del} and Lemma~\ref{lm:2ch_alt_del}, we can give a lower bound on the expected normalized distance for the case of two deletion channels. This lower bound is obtained by considering the sum of the Levenshtein normalized distance due to deletions in the same runs (Lemma~\ref{lm:2ch_run_del}), the Levenshtein normalized distance due to alternating sequence errors (Lemma~\ref{lm:2ch_alt_del}), and additional errors which are in the order of $p^3$. This result is summarized in the next theorem. 
\begin{theorem}
    For the deletion channel $\del(p)$, the expected normalized distance of the ML$^D$ decoder for the case of two channel-outputs is bounded from below by
    \begin{align*}
        %\perr(q,p) \ge \prun(q,p) + \palt(q,p)+O(p^3) = \frac{3q-1}{q-1}p^2+O(p^3).
\perr(n,q,p) &\ge  \prun(n,q,p) + \palt(n,q,p) 
\\ & = \cP_{\textmd{run}} (n, q, p) + \cP_{\textmd{alt}} (n, q, p)  \triangleq \cP_{\textmd{err}} (n,q,p). 
    \end{align*}
When $n$ approaches infinity, let
\begin{align*}
\lim_{n \to \infty }\cP_{\textmd{err}}  (n,q,p) \triangleq \cP_{\textmd{err}} (q,p).
\end{align*}
It holds that, when $n$ approaches infinity and $p$ approaches zero, it holds that 
$\lim_{p \to \infty}  \cP_{\textmd{err}} (q,p) \approx \frac{3q-1}{q-1}p^2,$
i.e., 
$$\lim_{p \to \infty} \frac{ \cP_{\textmd{err}} (q,p)}{\frac{3q-1}{q-1}p^2} =1.$$
\end{theorem}
\begin{IEEEproof}
    The theorem follows by considering the expected normalized distance that increases due to errors within runs, errors within alternating sequences, and the lower bounds given in Lemma~\ref{lm:2ch_run_del} and Lemma~\ref{lm:2ch_alt_del}. Note that while both types of errors can occur within the same sequence $\bfx$, they affect distinct regions, as one applies to runs and the other to alternating sequences. 
    \end{IEEEproof}

\begin{figure}[h!]
		\centering
		\subfigure{\includegraphics[width=85mm]{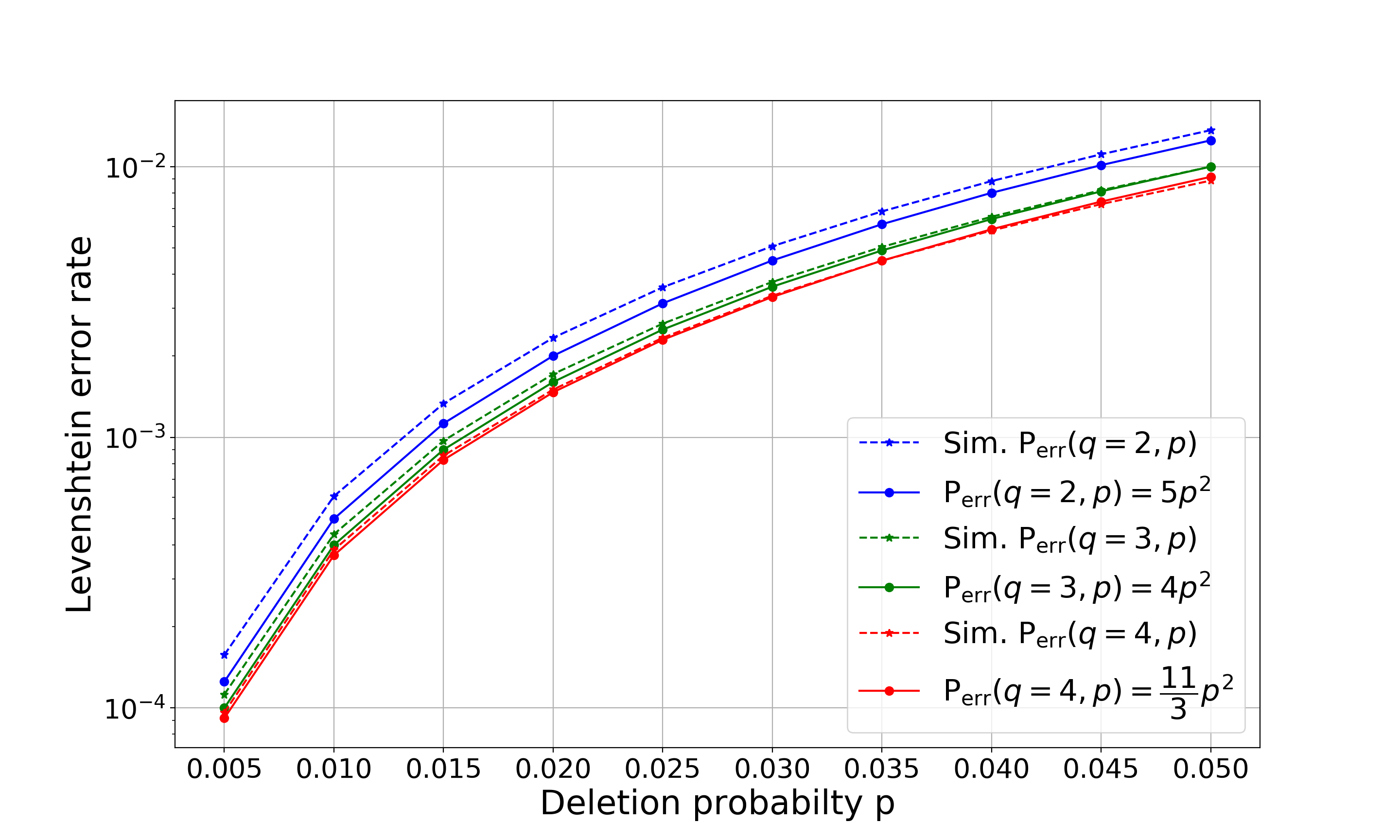}}
		\caption{The Levenshtein error rate as a function of the deletion probability $p$. The Levenshtein error rate is the average Levenshtein distance between the decoder's output and the transmitted simulated word, normalized by the transmitted word's length.  This value is an approximation of the expected normalized distance. } %These simulations verify Theorem~\ref{th:2ch}.}
		 \label{fig:LER-Deletion}
	\end{figure}

We verified the theoretical results presented in this section by computer simulations. These simulations were performed over words of length $n=450$ which were used to create two noisy outputs given a fixed deletion probability $p\in[0.005,0.05]$. Then, the two outputs were decoded by the ML$^D$ decoder as described earlier in this section. Finally, we calculated the Levenshtein error rate of the decoded word. Fig.~\ref{fig:LER-Deletion} plots the results of the Levenshtein error rate, which is the average Levenshtein distance between the decoder's output and the transmitted simulated word, normalized by the transmitted word's length. This value evaluates the expected normalized distance. Fig.~\ref{fig:LER-Deletion} confirms the approximation of the probability  $\perr(q,p)$ for $q\in \{ 2,3,4\}$ and when $p$ approaches zero. These probabilities are given by $\perr(2,p)=5p^2$,  $\perr(3,p)=4p^2$, and $\perr(4,p)=\frac{11}{3}p^2$. It can be seen that for larger values of $p$ (i.e., when $p$ is not approaching zero), the lower bound given for $\cP_{\textmd{err}} (q,p)$ is not applicable.  %Similarly, in Fig.~\ref{fig:runAltErrorsDel} we present the error probabilities $\palt(q,p)$ and  $\prun(q,p)$ separately, along with the corresponding value as calculated in Lemma~\ref{lm:2ch_run_del},~\ref{lm:2ch_alt_del}, respectively. \textcolor{red}{Lastly, in Fig.~\ref{fig:failuerRate}, we simulated the ML$^D$ decoder for the VT codes and the SVT codes and calculated the failure rate $\pfail(q,p)$. The implementation of the VT code was taken from~\cite{DBLP:journals/corr/abs-1906-07887}, and we modified this implementation for SVT codes. - I think This part should be removed.} 

Complexity wise, it is well known that the time complexity to calculate the SCS length and the embedding numbers of two sequences are both quadratic with the sequences' lengths. However, the number of SCSs can grow exponentially~\cite{itoga1981string, elzinga2008algorithms}.  Thus, given a set of SCSs of size $L$, the complexity of the ML$^D$ decoder for $t=2$ will be $O(Ln^2)$. The main idea behind these algorithms uses dynamic programming in order to calculate the SCS length and the embedding numbers for all prefixes of the given words. However, when calculating for example the SCS for $\bfy_1$ and $\bfy_2$ it is already known that $\mathsf{SCS}(\bfy_1,\bfy_2) \leq n$. Hence, it is not hard to observe that (see e.g.~\cite{apostolico1992fast}) many paths corresponding to prefixes which their length difference is greater than $d_1+d_2$ can be eliminated, where $d_1,d_2$ is the number of deletions in $\bfy_1,\bfy_2$, respectively. In particular, when $d_1$ and $d_2$ are fixed, then the time complexity is linear. In our simulations we used this improvement when implementing the ML$^D$ decoder. Other improvements and algorithms of the ML decoder are discussed in~\cite{srinivasavaradhan2018maximum,srinivasavaradhan2019symbolwise}.

\section{Conclusion}\label{sec:conc}

In this paper, we first  studied the ML$^*$ decoder of the 1-deletion and 2-deletion channels and then studied the problem of estimating the expected normalized distance of two deletion channels when the code is the entire space. It should be noted that we also characterized the ML$^*$ decoder for the $1\text{-}\mathsf{Ins}$ channel, where exactly 1 symbol is inserted into the transmitted word. When the code is the entire space, the ML$^*$ decoder of the $1\text{-}\mathsf{Ins}$ channel in almost all of the cases simply returns the channel outputs. In cases where the channel outputs contain an extremely long run (more then half of the word), the ML$^*$ decoder shortens it by one symbol. These results were proved by Ra\"{i}ssa Nataf and Tomer Tsachor for alphabet of size $q=2$~\cite{NT21}, and for any $q>2$ by Or Steiner and Michael Makhlevich~\cite{SM21}. While the results in the paper provide a significant contribution in the area of codes for insertions and deletions and sequence reconstruction, there are still several interesting problems which are left open. Some of them are summarized as follows.
\begin{enumerate}
\item Study the non-identical channels case. For example two deletion channels with different probabilities $p_1$ and $p_2$.
\item Study the expected normalized distance for more than two channels, both for insertions and deletions.
\item Study channels which introduce insertions, deletions, and substitutions.
\item Design coding schemes as well as complexity-efficient algorithms for the ML decoder in each case.
\end{enumerate}

\appendices

\section{}\label{app:A}
\begin{customclaim}{\ref{cl:tau}.}
For all $n\geq 1$ it holds that $\tau((\Sigma_2)^n) \leq 2\log(n)$. %\log(n)+2$.
\end{customclaim}
\begin{IEEEproof}
For $1\leq r\leq n$, let $N(r)$ denote the number of words in $\Sigma_2^n$ which the length of their maximal run is $r$.
Note that $N(r) \leq n 2^{n-r-1}$. This holds since we can set the location of the maximal run to start at some index $i$, which has less than $n$ options. There are two options for the bit value in the maximal run, the two bits before and after the run are fixed and have to opposite to the bit value in the run, and the rest of the bits can be arbitrary. Then, for $\ell(n)\in \N$, it holds that 
\begin{align*}
\tau((\Sigma_2)^n) & =  \frac{\sum_{r=1}^{n}r N(r)}{2^n} = \frac{\sum_{r=1}^{\ell(n)}r N(r)}{2^n} + \frac{\sum_{r=\ell(n)+1}^{n}r N(r)}{2^n} & \\
&\leq \frac{\sum_{r=1}^{\ell(n)}\ell(n) N(r)}{2^n} + \frac{\sum_{r=\ell(n)+1}^{n}r n 2^{n-r-1}}{2^n} & \\
&= \frac{\ell(n)\sum_{r=1}^{\ell(n)}N(r)}{2^n} + \frac{n2^{n-1}\sum_{r=\ell(n)+1}^{n}r 2^{-r}}{2^n} & \\
&\leq \frac{\ell(n)2^n}{2^n} + \frac{n2^{n-1}\cdot n 2^{-\ell(n)-1}}{2^n} = \ell(n) + \frac{n^2}{2^{\ell(n)+2}}.&
\end{align*}
Finally, by setting $\ell(n) = \lceil 2\log(n)\rceil-2$ we get that
\begin{align*}
\tau((\Sigma_2)^n) &\leq \lceil 2\log(n)\rceil -2 + \frac{n^2}{2^{\lceil 2\log(n)\rceil}} & \\
& \leq \lceil 2\log(n)\rceil -1 \leq 2\log(n). & 
\end{align*}

\end{IEEEproof}

%\bibliographystyle{ieeetr}
%\bibliographystyle{abbrv}
%\bibliography{references}
%\input{TR_JOURNAL_Yotam_V3.bbl}

\end{document}